\documentclass[%
 amsmath,amssymb,
 aps, physrev,
twocolumn,superscriptaddress,showpacs,nofootinbibfloatfix,amsmath,amsfonts,amssymb,longbibliography,prl
]{revtex4-2}
\usepackage{amsmath,amssymb}
\usepackage{hyperref} 
\usepackage{braket}
\usepackage{xcolor}
\usepackage{tikz}
\usetikzlibrary{arrows.meta}
\usetikzlibrary{shapes, arrows, calc, positioning}
\usepackage[normalem]{ulem}
\usepackage{ulem}
\usepackage{csquotes}
\usepackage{cancel}
\usepackage{graphicx}% Include figure files
\usepackage{dcolumn}% Align table columns on decimal point
\usepackage{bm}% bold math
\usepackage{svg}
\newcommand*\circled[1]{\tikz[baseline=(char.base)]{
            \node[shape=circle,draw,inner sep=1pt] (char) {#1};}}
\begin{document}

%\preprint{APS}

\title{Renormalization-Inspired Effective Field Neural Networks for Scalable
    Modeling of Classical and Quantum Many-Body Systems
}% 

\author{Xi Liu}
\affiliation{%
 Department of Physics, Hong Kong University of Science and Technology, Clear Water Bay, Hong Kong SAR, China
}%
 %\email{xi.liu@u.nus.edu}
\author{Yujun Zhao}%
\affiliation{%
 Department of Physics, Hong Kong University of Science and Technology, Clear Water Bay, Hong Kong SAR, China
}%

\author{Chun Yu Wan}
 \affiliation{%
 Department of Physics, Hong Kong University of Science and Technology, Clear Water Bay, Hong Kong SAR, China
}%
\author{Yang Zhang}
 %\affiliation{%
 %Department of Physics, Massachusetts Institute of Technology, Cambridge, Massachusetts 02139, USA
%}%
\affiliation{Department of Physics and Astronomy, University of Tennessee, Knoxville, TN 37996, USA}
\affiliation{Min H. Kao Department of Electrical Engineering and Computer Science, University of Tennessee, Knoxville, Tennessee 37996, USA}

\author{Junwei Liu}
\email{liuj@ust.hk}
 \affiliation{%
 Department of Physics, Hong Kong University of Science and Technology, Clear Water Bay, Hong Kong SAR, China
}%

\date{\today}

\begin{abstract}
We introduce Effective Field Neural Networks (EFNNs), a new architecture based on continued functions---mathematical tools used in renormalization to handle divergent perturbative series. Our key insight is that neural networks can implement these continued functions directly, providing a principled approach to many-body interactions. Testing on three systems (a classical 3-spin infinite-
range model, a continuous classical Heisenberg spin system, and a quantum double exchange model), we find that EFNN outperforms standard deep networks, ResNet, and DenseNet. Most striking is EFNN's generalization: trained on 10×10 lattices, it accurately predicts behavior on systems up to 40×40 with no additional training---and the accuracy improves with system size, with a computational time speed-up of $10^{3}$ compared to ED for $40\times 40$ lattice. This demonstrates that EFNN captures the underlying physics rather than merely fitting data, making it valuable beyond many-body problems to any field where renormalization ideas apply.
\end{abstract}

\maketitle

The collective behaviors of many interacting particles---such as spins, molecules, and atoms---give rise to some of the most intriguing phenomena in condensed matter physics. However, theoretical and computational studies of many-body problems often confront the curse of dimensionality and one must devise strategies to circumvent it. Machine learning has emerged as a powerful tool for extracting effective features from high-dimensional datasets and tackling long-standing problems in condensed matter physics. Numerous machine learning techniques have been applied to a wide range of classical and quantum many-body systems, including classifying phases of matter with supervised~\cite{carrasquilla2017machine,ch2017machine,iakovlev2018supervised} and unsupervised machine learning~\cite{wang2016discovering,van2017learning,ch2018unsupervised,rodriguez2019identifying,arnold2021interpretable}, representing many-body quantum states~\cite{carleo2017solving,nomura2017restricted,choo2018symmetries,yoshioka2019constructing,carrasquilla2019reconstructing}, accelerating Monte Carlo simulations~\cite{liu2017self,liu2017self2,huang2017accelerated,kanwar2020equivariant}, fitting high-dimensional potential energy surfaces~\cite{behler2007generalized, rupp2012fast,bartok2013representing,behler2015constructing, marchand2020machine} and searching for spin glass ground states~\cite{fan2023searching}.

Standard deep neural networks (DNNs) are ineffective in solving many-body problems unless augmented with physical knowledge. For example, an 8×8 classical two-dimensional Ising model requires highly complex DNN structures ~\cite{mills2018deep}. Consequently, extensive research has focused on encoding physics into neural network architectures to enhance their effectiveness in physical applications, including incorporating physical laws and constraints~\cite{wu2018physics,raissi2019physics,brunton2020machine,lu2021deepxde, mattheakis2019physical, bogatskiy2020lorentz, yang2020physics, mills2019extensive, Batzner2022E3,XIE2024lm,10.1063/5.0142281sj,Zhong2023_general_xhj,GAO2021466Enhancing,10.1093/nsr/nwad128MAGUS,PhysRevB.109.144426Spin-dependent_graph,li2018neural}. These physics-encoded neural networks have found applications in computing effective Hamiltonians~\cite{Zhong2024xhj,zhang2024advancingnonadiabaticmoleculardynamics_xhj,PhysRevB.105.174422spin_ann}, property prediction~\cite{Zhong2024xhj,Zhong2023_general_xhj,PhysRevB.105.174422spin_ann,PhysRevB.108.235159charge,10.1063/5.0106617GPUMD,Xie2024LASP,Yang2024Attention,barrosoluque2024openmaterials2024omat24,10.1063/5.0142281reconstructions,PhysRevB.109.144426Spin-dependent_graph}, structure search~\cite{GAO2021466Enhancing,10.1093/nsr/nwad128MAGUS}, etc. These methods, although often outperforming standard DNNs, tend to rely on intuitive and simplified imitations of physical principles, and still remain elusive in revealing the underlying  many-body interactions.

In this work, we propose Effective Field Neural Networks (EFNNs), a new architecture inspired by field theory. The core motivation is to decompose many-body interactions into single quasi-particle representations governed by an emergent effective field. Unlike standard DNNs where adjacent layers rely only on previous layers, EFNNs incorporate initial features at every layer, forming a self-similar structure trained via recursive self-refining that captures many-body interactions through renormalization principles. Unlike approaches such as FermiNet~\cite{PhysRevResearch.2.033429FermiNet}, which directly parameterizes high-dimensional wave-functions without explicit separation of quasi-particles and fields, our method does not presuppose a fixed ansatz for the wavefunction. Instead, the quasi-particle and effective field representations are learned through a recursive self-refining process, mirroring the renormalization group (RG) framework in field theory, while leveraging the expressive power of deep neural networks. This recursive approach progressively refines the accuracy of both the quasi-particle and effective field descriptions.

\begin{figure}[tbp]
    \centering\includegraphics[width=0.9\linewidth]{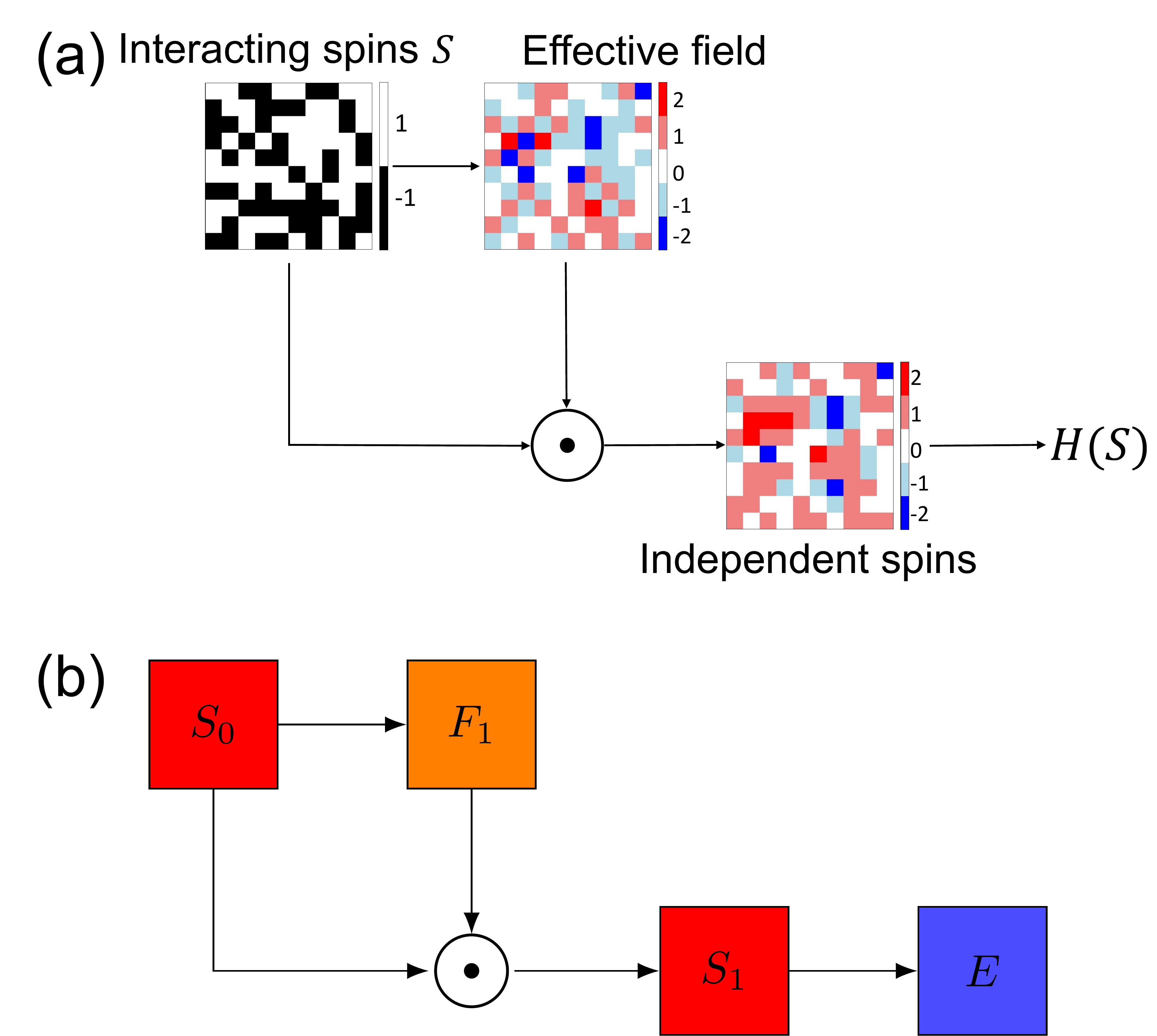}
    % First figure (Top figure) with scaling
       %\makebox[\textwidth][l]{\includegraphics[width=1\linewidth]{fig1/hadamard_fig1.pdf}
     %}
    \caption{Energy evaluation of a 2D Ising model reformulated as a neural network. (a) Calculation of the effective field by summing the interacting spins. Each interacting spin is multiplied by its corresponding effective field to obtain independent spin values, and the total energy is determined by summing up these independent spins. (b) Neural network representation of the energy evaluation process. $\odot$ represents the element-wise multiplication. }
    \label{fig:figure1}
\end{figure}

\begin{figure}[hbtp!]
    \centering\includegraphics[width=0.9\linewidth]{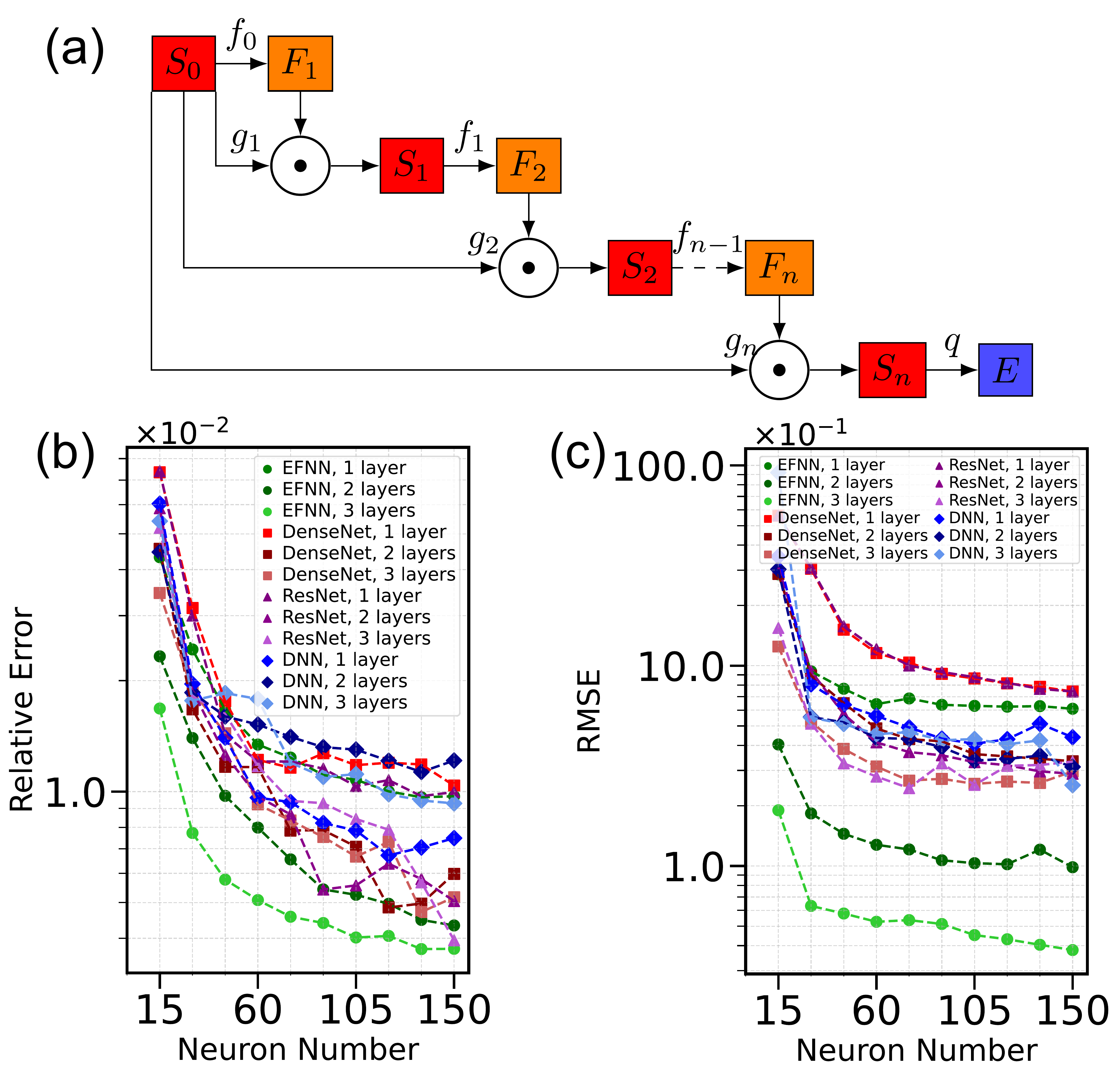}
     %\makebox[\textwidth][l]{\includegraphics[width=1\linewidth,trim=7 7 7 7,  % Adjust these values as needed
    %clip]{V20Figs/fig_efnn_vs_dnn/hadamard_spin3.pdf}
    %}
    \caption{Performance of EFNNs on a classical 3-spin infinite range model and classical Heisenberg spin system. (a) Architecture of the EFNN. (b) Performance of EFNNs, DenseNets, ResNets, DNNs on 3-spin infinite range model. (c) Performance of EFNNs, DenseNets, ResNets, DNNs on classical Heisenberg spin system.}
    \label{fig_spin3}
\end{figure}

We first illustrate the EFNN architecture by reformulating the classical 2D Ising model within the EFNN framework. The Hamiltonian of a classical 2D Ising model is defined as $H(S)=-J \sum_{\langle i ,j\rangle} s_{i} s_{j}$, where $\langle i, j\rangle$ denotes a summation over pairs of nearest-neighbor sites, $s_{i}=\pm 1$ represents the spin at site $i$, $J$ is the interaction strength between two spins, and $S$ is the collection of all spins. We define the  effective field on spin $s_{i}$ as $\phi_{i}(S)=-\frac{J}{2}\times (\textrm{sum of nearest neighbors})$, and the quasi-particle as $s_{i}\phi_{i}(S)$, yielding total energy $H(S)=\sum_{i}s_{i}\phi_{i}(S)$. As illustrated schematically in Fig.~\ref{fig:figure1}(a), the interacting spins are first mapped to an effective field, then the effective field element-wise multiplied with the interacting spins is mapped to independent spins. Finally, a summation is performed over the independent spins, and the total energy is obtained.  We reformulate this as a neural network structure (Fig.~\ref{fig:figure1}(b)), where $S_{0} = S$ represents the interacting spins. The effective field layer $F_{1}$ and the quasi-particle layer $S_{1}$ constitute a single field-particle (FP) layer. The evaluation of $S_{1}$ relies on both the effective layer $F_{1}$ and the interacting spin layer $S_{0}$, therefore a connection is required from $S_{0}$ to $S_{1}$, as well as from $F_{1}$ to $S_{1}$, distinguishing this architecture from a standard DNN. Finally, a summation is performed over $S_{1}$ to obtain the energy $E$.

In the classical 2D Ising model, both the effective field and quasi-particles can be exactly calculated, which, however, becomes infeasible for more complicated many-body interactions. Following the spirit of renormalization from field theory, we can recursively refine the evaluations of the effective field and the quasi-particles. 
This recursive structure, when combined with the multiplicative interactions and self-similar connectivity pattern, implements a continued function---a mathematical tool from renormalization theory for handling divergent perturbative series (see Supplementary Material~\cite{supplementary_material} for the complete derivation).  As shown in Fig.~\ref{fig_spin3}(a), we can extend the FP layer number in Fig.~\ref{fig:figure1}(b) to obtain the deep Effective Field Neural Networks. The  evaluation procedure is detailed as
\begin{align}
 F_{i}&=f_{i-1}(S_{i-1}),\, S_{i}=g_i(S_0)\odot F_{i}, \, E=q(S_n), \nonumber%i=1,\ldots,n.\label{eq1}
\end{align}
where the function $f_{i-1}$ maps the previous quasi-particle layer $S_{i-1}$ (for $i\geq 2$) or the initial layer $S_{0}$ (for $i=1$) to the effective field layer $F_{i}$. The input layer $S_{0}$ is first processed by $g_{i}$, then element-wise multiplied by $F_{i}$ to produce the quasi-particle layer $S_{i}$. Each FP layer consists of $F_{i}$ and $S_{i}$. The final quasi-particle layer $S_{n}$ generates the output energy $E$ through the function $q$, which sums all elements of the last layer. The functions $f_{i}$ are nonlinear (specifically, we use hyperbolic tangent functions) to implement renormalization by ensuring finite values, while $g_{j}$ are nonlinear to enhance expressiveness. Linear functions may suffice only for simpler models like the classical 2D Ising model with nearest-neighbor interactions.

We present three case studies to demonstrate the capability of EFNNs to capture many-body interactions. The first case study focuses on the classical 3-spin infinite-range model in 1D, with Hamiltonian:
\begin{align}
 H(S)&= -\sum_{i<j<k}J_{ijk}s_{i}s_{j}s_{k}
    \label{eqn_H_stochastic},
\end{align}
where the spins take binary values $\{0,1\}$, $J_{ijk}$ are constants, and the system comprises 15 lattice sites.  We train EFNNs to evaluate $H(S)$ in Eq.~(\ref{eqn_H_stochastic}), and compare its performance with those of DenseNets, ResNets, and DNNs. 

\begin{figure}
    \centering
\includegraphics[width=0.9\linewidth]{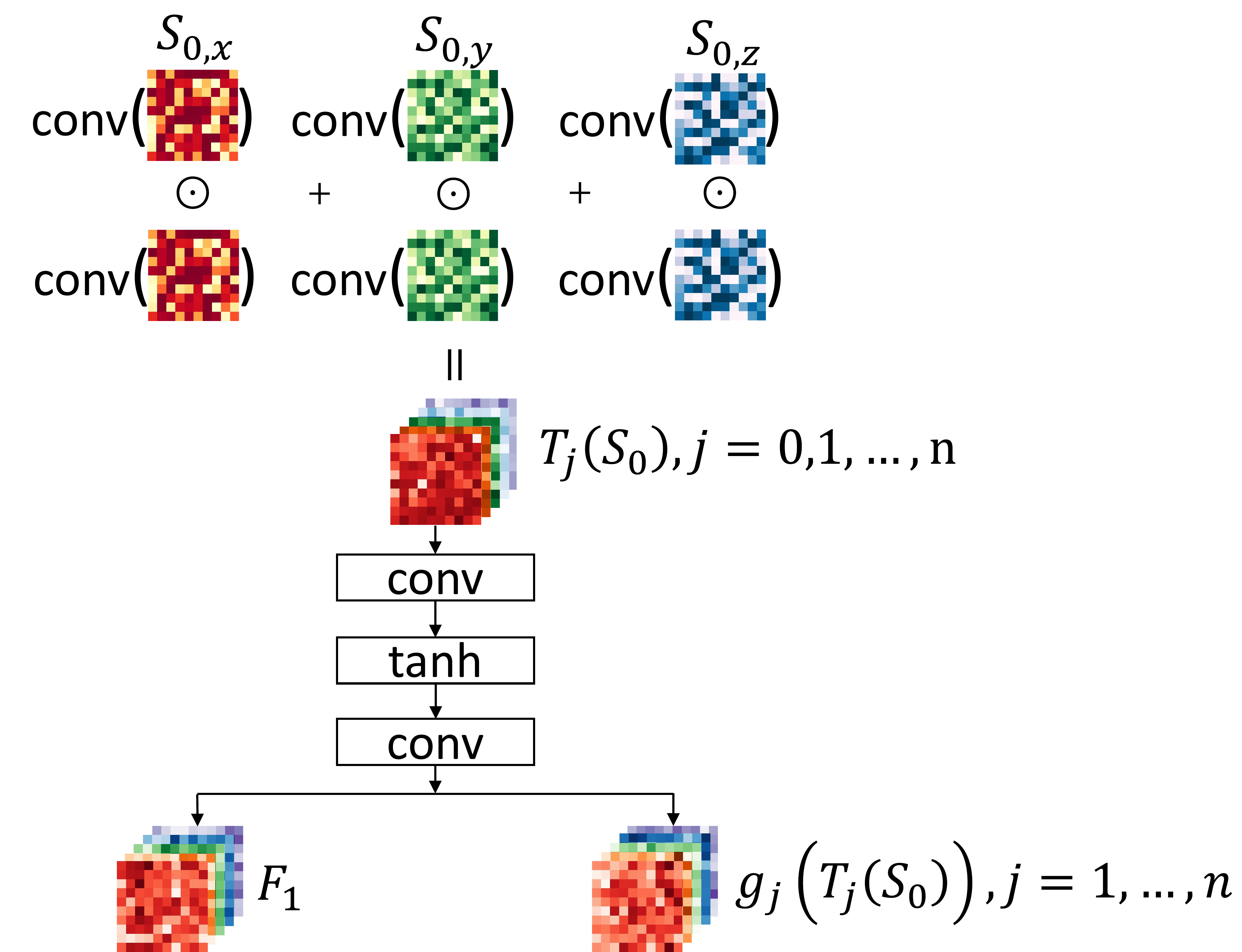}
    
    \caption{
    Symmetrization: Each of the three channels of $S_{0}$ ($S_{0,x}$, $S_{0,y}$, $S_{0,z}$) undergoes convolution and element-wise multiplication within the same channel. The resulting products are summed to form the symmetrization layer $T_{j}(S_{0})$. Subsequently, $T_{j}(S_{0})$ is passed through a sequence of layers---convolution, $\tanh$ activation, and another convolution}---to produce $F_{1}$ for $j=0$ or $g_{j}(T_{j}(S_{0}))$ for $j=1,\ldots,n$.
    \label{fig_conv_details}
\end{figure}

We generate 16000 training and 4000 test samples using Eq.~(\ref{eqn_H_stochastic}). The nonlinear mappings $f_{i}$ and $g_{j}$ use sequential linear layers with tanh activations. The network is trained with the Adam optimizer and a decaying learning rate. Performance is evaluated by the relative error, defined as the root mean square error (RMSE) on the test set divided by  the absolute value of the mean of $H(S)$. We vary the number of network layers from 1 to 3 and the number of neurons from 15 to 150. The performance of EFNNs, DenseNets, ResNets, and DNNs is presented in Fig.~\ref{fig_spin3}(b). The relative error decreases with more neurons for all networks, but EFNNs significantly outperform others as layers increase, with the relative error decreasing below $5\times 10^{-3}$ for EFNNs with 2 and 3 layers when the number of neurons $\geq 105$. The only exception is the ResNet with 3 layers and 150 neurons. In most cases, DenseNets, ResNets, and DNNs perform worse than 2-layer EFNNs.

To test the capability of EFNNs beyond discrete systems, our second case study examines a classical Heisenberg spin system, using the same three-body interaction model in Eq.~(\ref{eqn_H_stochastic}) with continuous spins, where $s_i$ takes continuous values in $[-1, 1]$ rather than discrete values ${0, 1}$. We use the RMSE of the total energy as the performance metric, with results shown in Fig.~\ref{fig_spin3}(c), since the average value of Eq.~(\ref{eqn_H_stochastic}) is 0. Compared to the discrete case, the configuration space is now continuous, yet EFNNs with 2 and 3 layers still outperform all other networks. Fig.~\ref{fig_spin3}(c) also shows the general trend that the error decreases with increasing number of neurons. The most accurate result is achieved by the 3-layer EFNN with 150 neurons, where the error of the total energy reduces to $4\times 10^{-2}$. These results demonstrate that EFNNs effectively capture many-body interactions through the renormalization procedure of their recursive continued-function structure; failing to incorporate this structure leads to inaccurate results. We show in the Supplementary Material~\cite{supplementary_material} that DNNs completely lack the structure of a continued function, while DenseNets and ResNets misrepresent a continued function.

We proceed to our third, more challenging case study: evaluating the Monte Carlo energy of the quantum double exchange model~\cite{zener1951interaction,anderson1955considerations,de1960effects} at finite temperature. The Hamiltonian of an $N\times N$ lattice reads
\begin{equation}
    H(S)=-t \sum_{\langle i, j\rangle, \alpha}\left(\hat{c}_{i \alpha}^{\dagger} \hat{c}_{j \alpha}+\text {h.c.}\right)-\frac{J}{2} \sum_{i, \alpha, \beta} \vec{s}_{i} \cdot \hat{c}_{i \alpha}^{\dagger} \vec{\sigma}_{\alpha \beta} \hat{c}_{i \beta}, \label{eqn_H_quantum}
\end{equation}
where ${\langle i ,j\rangle}$ denotes nearest neighbors, $\hat{c}_{i \alpha}$ is the electron annihilation operator with spin ${\alpha}$ at site $i$, $\vec{\sigma}$ are the Pauli matrices, and $\vec{s}_{i}\in\mathbb{S}^{2}$ is a 3D classical unit vector representing the local spin at site $i$, the classical spin interacts with the electrons through an on-site coupling, and $S$ is the collection of all classical spins $\vec{s}_{i}$. We set $N=10$, $J=16t$, $\mu=-8.3t$, $T=0.1t$, $t=1$.

At finite temperature $T$, the Monte Carlo energy  $E_{\text{MC}}(S) = -T \sum_n\log (1+e^{-E_{n}(S)/T})$ determines the acceptance ratio in Monte Carlo simulations, where $E_{n}(S)$ are the eigenvalues of $H(S)$ with chemical potential in Eq.~(\ref{eqn_H_quantum}). Computing the exact value of $E_{\text{MC}}(S)$ requires diagonalization of Eq.~(\ref{eqn_H_quantum}) at each Monte Carlo update step, incurring $O(N^{6})$ computational complexity per step. With millions of Monte Carlo steps required for convergence, this approach becomes computationally prohibitive. An intuitive solution is to approximate $E_{\text{MC}}(S)$ using a perturbative expansion, yielding an effective model with RKKY-type interactions~\cite{ruderman1954indirect,kasuya1956theory,yosida1957magnetic}. This effective model with 2-body interactions has proven sufficiently accurate to replace exact diagonalization in Determinant Quantum Monte Carlo (DQMC) simulations~\cite{liu2017femion}. The effective model including up to 4-body interactions takes the form
\begin{align}
    E_{\operatorname{eff}}(S)&\sim J_{0}+\sum_{ij}g_{ij}\vec{s}_{i}\cdot \vec{s}_{j}+\sum_{ijkl}g_{ijkl}(\vec{s}_{i}\cdot \vec{s}_{j})(\vec{s}_{k}\cdot \vec{s}_{l}).\label{eqn_quantum_eff}
\end{align} 
Here, the dot products include all pairs of spins on the lattice, including self-interactions. The perturbative approach faces a fundamental dilemma: truncating at low orders fails to capture the full physics, while including higher-order terms leads to numerical divergence due to the factorial growth in the number of terms. EFNN addresses this challenge through renormalization.

\begin{figure}
    \centering
     \hspace{-1cm}
     \includegraphics[width=1.1\linewidth]{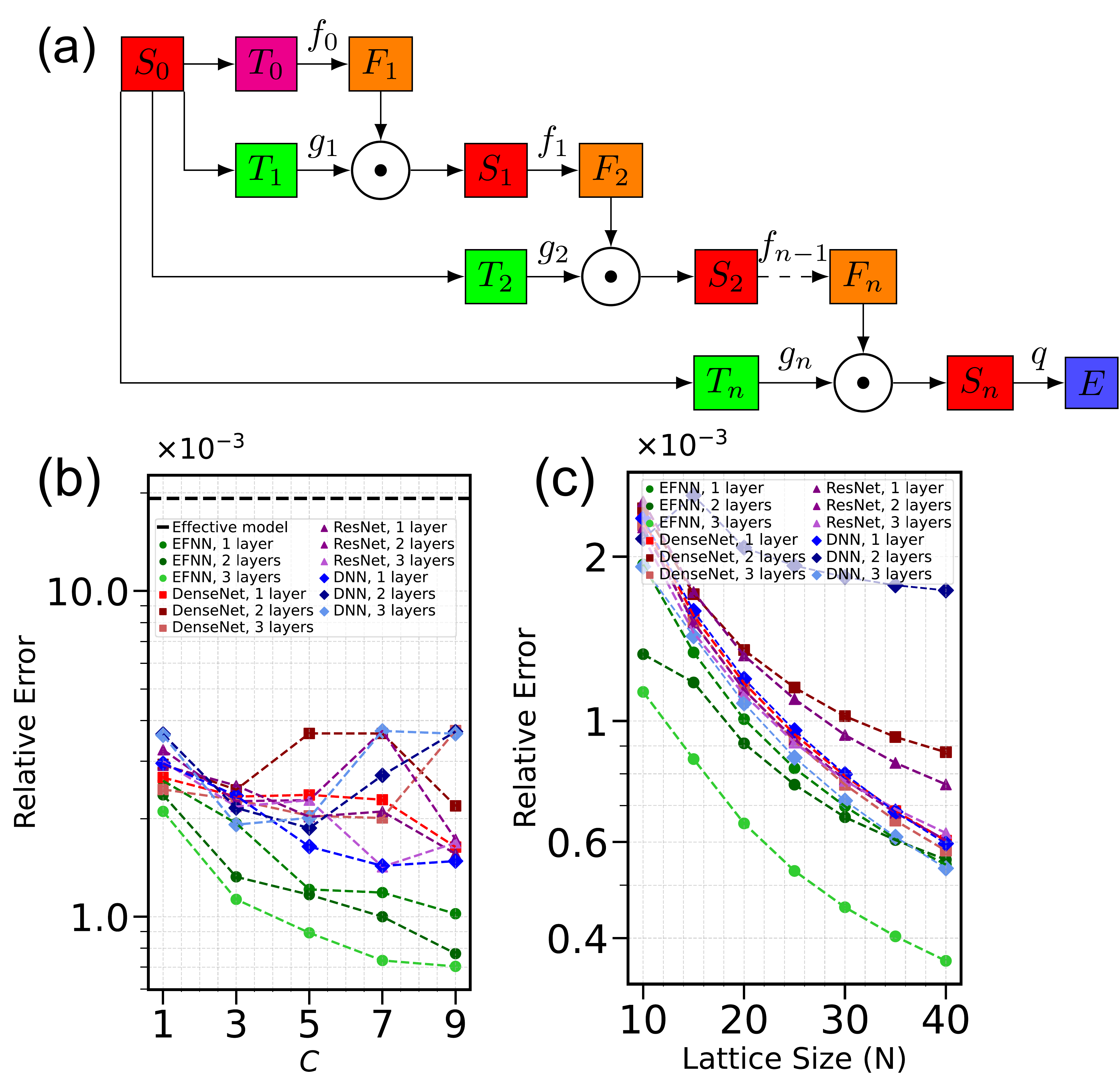}
     %\makebox[\textwidth][l]{\includegraphics[width=0.9\linewidth, %trim=15 10 10 10,  % Adjust these values as needed
    %clip]{V20Figs/fig_qt_efnn/hadamard_efnn_vs_eff.pdf}
    %}
    \caption{Relative error of EFNNs. (a) Architecture of EFNN incorporating symmetrization layers $T_{j}$, $j=0,1,\ldots,n$. (b)  Performance of EFNNs, DenseNets, ResNets, DNNs and effective model on a test set with lattice size $N=10$, for channel numbers $C=1,3,5,7,9$. (c)  Generalization performance of networks trained on $10\times 10$ lattices ($C=3$) when applied to larger systems.}
    \label{fig_efnn_vs_eff}
\end{figure}

The spin interactions in Eq.~(\ref{eqn_quantum_eff}) are exclusively expressed as dot products to satisfy the $\operatorname{O}(3)$ symmetry of Eq.~(\ref{eqn_H_quantum}). Inspired by this symmetry, we incorporate symmetrization layers into the EFNN architecture, depicted in Fig.~\ref{fig_conv_details}. The symmetrization is achieved by summing dot products of each channel of $S_{0}$ after convolution:
\begin{align}
    T_{j}(S_{0})&=\sum_{k=x,y,z}\operatorname{conv}(S_{0,k})\odot \operatorname{conv}(S_{0,k}),\,\,j=0,1,\ldots,n.\label{eqn_efnn_sym_Tj}\nonumber
\end{align}
Here, each $S_{0,k}\in\mathbb{R}^{N\times N}$ is convolved into $\mathbb{R}^{C\times N \times N}$.  Supplementary Material~\cite{supplementary_material} details the convolution process and proves that $T_{j}$ remains invariant under $\operatorname{O}(3)$ transformations. As shown in Fig.~\ref{fig_efnn_vs_eff}(a), in the first FP layer, $F_{1}$ and $S_{1}$ are initialized as
\begin{align}
    F_{1}&=f_{0}(T_{0}(S_{0})),\,\, S_{1}=g_{1}(T_{1}(S_{0}))\odot F_{1}. \nonumber
\end{align}
For subsequent layers $j=2,\ldots,n$, the effective field and quasi-particle layers are
\begin{align}
    F_{j}&=f_{j-1}(S_{j-1}), \,\, S_{j}=g_{j}(T_{j}(S_{0}))\odot F_{j}. \nonumber
\end{align}
The mappings $f_{j}$ and $g_{k}$ consist of a convolutional layer, a $\tanh$ activation, and another convolutional layer. For the mapping $q(S_{n})=E$, the final quasi-particle layer $S_{n}$ is convolved  into a single-channel matrix, and an element-wise summation is performed to produce the scalar energy prediction $E$, see Supplementary Material~\cite{supplementary_material}. The layer dimensions are as follows: input layer $S_{0}\in\mathbb{R}^{3\times N\times N}$, intermediate layers $F_{j}$, $S_{j}\in\mathbb{R}^{C\times N\times N}$, for $j=1,\ldots,n$, and output $E=q(S_{n})\in\mathbb{R}$.  For comparison, we also incorporate symmetrization layers and final summation $q$ into DenseNet, ResNet, and DNN architectures, detailed in Supplementary Material~\cite{supplementary_material}.

We generate a total of $2\times 10^{6}$ data points and split them into 80\% for training and 20\% for testing. These data are obtained by exact diagonalization of the Hamiltonian in Eq.~(\ref{eqn_H_quantum}) with chemical potential. For the effective model in Eq.~(\ref{eqn_quantum_eff}), we perform linear regression to obtain the coefficients.  While the full effective model would require 12758826 parameters, we reduce this to 213201 by restricting interactions within each $5\times 5$ tile. The EFNNs, DenseNets, ResNets, and DNNs with symmetrization are trained using networks with $C=1,3,5,7,9$ channels and $n=1,2,3$ layers, with parameter counts ranging from 200 to 30000. As shown in Fig.~\ref{fig_efnn_vs_eff}(b), the effective model achieves a relative error of approximately $2\times 10^{-2}$, whereas the neural networks achieve relative errors on the order of $10^{-3}$---approximately six times better using significantly fewer parameters. Among the neural networks, EFNNs with 1, 2, and 3 layers consistently outperform all other architectures, with some overlap observed between 1-layer EFNNs and 3-layer DNNs, as well as 1-, 2-, and 3-layer DenseNets at $C=1,3$. For EFNNs, the error systematically decreases as $C$ increases and as the number of layers increases from 1 to 3.  This superior performance with fewer parameters shows that the EFNNs possess inherent renormalization capabilities beyond simple parameter fitting.

We next test the generalization capability of these neural networks trained on $10\times 10$ lattices by evaluating their performance on systems up to $40\times 40$. We select networks with $C=3$ channels as a representative configuration and evaluate EFNN, DenseNet, ResNet, and DNN architectures across this range of lattice sizes. The convolution operations with padding under periodic boundary conditions (PBC) effectively capture interactions between localized and neighboring spins. As shown in Fig.~\ref{fig_efnn_vs_eff}(c), all neural network architectures show improved performance on larger lattices, with relative errors decreasing as the system size increases from $10\times 10$ to $40\times 40$. The EFNNs exhibit the best extrapolation capability and renormalization ability: the 3-layer EFNN achieves the lowest errors across all lattice sizes, reaching a remarkable relative error of approximately $4 \times 10^{-4}$ at $N=40$. The 2-layer EFNN mostly outperforms the 1-layer EFNN, while among non-EFNN architectures, only the 3-layer DNN has comparable performance to the 1-layer EFNN. All other configurations consistently produce higher errors, confirming EFNN's superior ability to capture scale-invariant features. Notably, as demonstrated in Supplementary Material~\cite{supplementary_material}, for $N=40$, EFNN achieves $10^{3}$ speed-up compared to ED. 

\begin{figure}[tb!]
    \centering \includegraphics[width=1\linewidth]{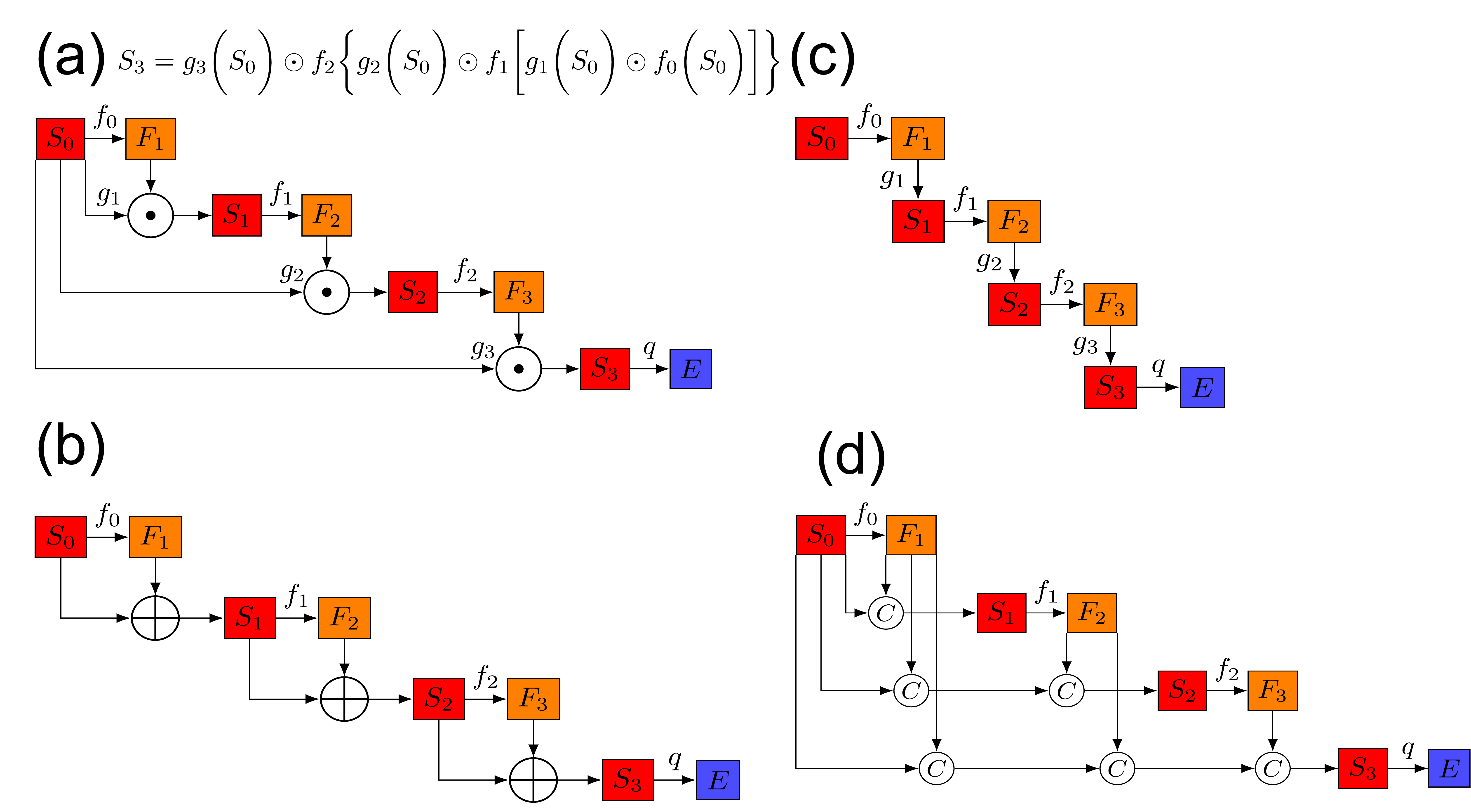}
       %\makebox[\textwidth][l]{\includegraphics[width=1\linewidth]{V20Figs/fig_pade/hadamard_pade.pdf}
     %}
\caption{Comparison of neural network structures. (a) In an EFNN, the initial layer $S_{0}$ is recursively integrated into every quasi-particle layer. After a mapping, $S_{0}$ is multiplied with subsequent layers, forming a continued-function representation (see the equation above the figure). (b) A typical architecture of ResNet, only one skip connection starts from $S_{0}$ and summation is used, this is not a correct continued-function structure. (c) Removing the $S_{0}$ connections transforms an EFNN into a standard DNN, with no continued-function structure. (d) A typical architecture of DenseNet,  \protect\circled{$C$} is concatenation operation, this is not a correct continued-function structure.  }
\label{fig_continued_fraction}
    \end{figure}

We give some discussions and conclude this paper. In all three case studies---the classical 3-spin infinite-range model, continuous classical Heisenberg system, and quantum double exchange model---EFNNs consistently outperform DenseNets, ResNets, DNNs, and the effective model, achieving the lowest errors. Particularly striking is EFNN's ability to generalize: trained on $10\times 10$ lattices, it accurately predicted energies for systems as large as $40\times 40$, while DenseNets, ResNets, and standard DNNs struggle with this extrapolation task. This superior performance isn't accidental---it stems from EFNN's mathematical foundation in continued functions. While conventional architectures like DenseNet and ResNet were designed through trial and error with various architectural innovations, EFNN's structure directly implements continued functions, a mathematical framework that naturally captures many-body interactions up to infinite order. More importantly, for a perturbative expansion like Eq.~(\ref{eqn_quantum_eff}), EFNN selects the most relevant terms via training, and expresses these terms in nonlinear continued function structure.  Fig.~\ref{fig_continued_fraction} compares these structures, demonstrating that only EFNN (Fig.~\ref{fig_continued_fraction}(a)) possesses the correct continued function structure. The Supplementary Material~\cite{supplementary_material} provides a detailed mathematical explanation of why EFNN implements continued functions while ResNet and DenseNet do not. A special case of continued functions is the Pad\'{e} approximant, which takes the form of a continued fraction and is ubiquitously used throughout physics and numerical methods for its rapid convergence, strong extrapolation capability, and compact representation~\cite{Baker_Graves-Morris_1996}. Continued functions generalize continued fractions by replacing the inversion operation with generic nonlinear functions. Here, we use tanh activation functions instead of inverses, which makes EFNN trainable through standard backpropagation---something impossible with traditional Padé approximants due to gradient singularities. This simple substitution preserves the mathematical power of continued functions while enabling practical neural network training. Notably,  EFNN actually implements self-similar approximation~\cite{Yukalov2019_Interplay}, a powerful mathematical framework that has been successfully applied across diverse fields including high energy physics~\cite{Abhignan2021SimpleTools,YUKALOV1994553_higher_orders,YUKALOV1993573_Self_similar,YUKALOV1999219_Perturbation,QUIGG1979167_quarkonium}, condensed matter physics~\cite{PhysRevE.55.6552_bootstrap,Gluzman2017_Critical}, quantum mechanics~\cite{Yukalov2017_epj,Yukalov_2017_bose,YUKALOV1999219_Perturbation,YUKALOV1996336_Temporal,Yukalova1992_Dubna,math3020510,Yukalov2019_Interplay}, quantum field theory~\cite{Yukalov:2015eh_multicomponent,Yukalov2007_Calculation,Yukalov2019_Interplay,yukalov2000equationstatequantumchromodynamics}, and applied mathematics~\cite{YUKALOVA20083074_equations,PhysRevE.55.6552_bootstrap,YUKALOV_2000_MPLB,Yukalov_Gluzman_1998_Resummation_MPLB,Renormalization_group_october,boom_crashes_1998,weighted_fixed_points_1999,Yukalov2001_market_analyse,Reconstructing_generalized_exp_laws_2003,yukalov2000cooperativeelectromagneticeffects,YUKALOV200191_nonequilibrium_phenomena}. Previously, applying self-similar approximation required tedious manual calculations to obtain order by order matching; EFNN changes this by automating the process through neural network optimization. EFNN's self-similar structure also naturally comes from renormalization group equation, as detailed in the Supplementary Material~\cite{supplementary_material}.

In conclusion, we have developed EFNN, a neural network architecture inspired by field theory that captures high-order interactions in both classical and quantum models. The key to EFNN's success lies in how it recursively refines effective fields and constructs quasi-particles to encode complex many-body effects.  This partitioning into field and quasi-particle layers provides clear physical interpretation while maintaining computational efficiency.  As we showed, the connections from the true particle layer $S_0$ to the quasi-particle layers create a continued function structure with renormalization properties similar to Pad\'{e} approximants.   What EFNN actually implements is self-similar approximation---a powerful mathematical technique that has been widely used in physics but previously required manual calculations for each problem. Once properly trained, this architecture learns effective representations that accurately predict energies even for systems with complicated many-body interactions. Our tests on three diverse systems---the classical 3-spin infinite-range model, continuous classical Heisenberg spin system, and quantum double exchange model---demonstrate that EFNN consistently outperforms DenseNet, ResNet, and standard DNNs. Most remarkably, EFNNs trained on small systems improve accuracy on larger ones, opening possibilities for training on expensive DFT data from small lattices then inferring on much larger systems. Looking ahead, we plan to develop theoretical error bounds and apply EFNN to other challenging problems in physics.

\section*{DATA AVAILABILITY}
The data that support the findings of this study are openly available in Zenodo at Ref.~\cite{efnn2026data}.

\textit{Acknowledgments}---This work is supported by the Hong Kong Research Grants Council (16306220) and National Key R$\&$D Program of China (2021YFA1401500). Y.Z. is supported by the Max Planck Partner lab on quantum materials from Max Planck Institute Chemical Physics of Solids.

%\textit{Acknowledgments}---This work is supported by the Hong Kong Research Grants Council (16306220), National Key R$\&$D Program of China (2021YFA1401500), and National Natural Science Foundation of China (12022416).

%\nocite{*}
%TC:ignore
%\printbibliography
\bibliography{reference}% 

\begin{thebibliography}{86}%
\makeatletter
\providecommand \@ifxundefined [1]{%
 \@ifx{#1\undefined}
}%
\providecommand \@ifnum [1]{%
 \ifnum #1\expandafter \@firstoftwo
 \else \expandafter \@secondoftwo
 \fi
}%
\providecommand \@ifx [1]{%
 \ifx #1\expandafter \@firstoftwo
 \else \expandafter \@secondoftwo
 \fi
}%
\providecommand \natexlab [1]{#1}%
\providecommand \enquote  [1]{``#1''}%
\providecommand \bibnamefont  [1]{#1}%
\providecommand \bibfnamefont [1]{#1}%
\providecommand \citenamefont [1]{#1}%
\providecommand \href@noop [0]{\@secondoftwo}%
\providecommand \href [0]{\begingroup \@sanitize@url \@href}%
\providecommand \@href[1]{\@@startlink{#1}\@@href}%
\providecommand \@@href[1]{\endgroup#1\@@endlink}%
\providecommand \@sanitize@url [0]{\catcode `\\12\catcode `\$12\catcode
  `\&12\catcode `\#12\catcode `\^12\catcode `\_12\catcode `\%12\relax}%
\providecommand \@@startlink[1]{}%
\providecommand \@@endlink[0]{}%
\providecommand \url  [0]{\begingroup\@sanitize@url \@url }%
\providecommand \@url [1]{\endgroup\@href {#1}{\urlprefix }}%
\providecommand \urlprefix  [0]{URL }%
\providecommand \Eprint [0]{\href }%
\providecommand \doibase [0]{https://doi.org/}%
\providecommand \selectlanguage [0]{\@gobble}%
\providecommand \bibinfo  [0]{\@secondoftwo}%
\providecommand \bibfield  [0]{\@secondoftwo}%
\providecommand \translation [1]{[#1]}%
\providecommand \BibitemOpen [0]{}%
\providecommand \bibitemStop [0]{}%
\providecommand \bibitemNoStop [0]{.\EOS\space}%
\providecommand \EOS [0]{\spacefactor3000\relax}%
\providecommand \BibitemShut  [1]{\csname bibitem#1\endcsname}%
\let\auto@bib@innerbib\@empty
%</preamble>
\bibitem [{\citenamefont {Carrasquilla}\ and\ \citenamefont
  {Melko}(2017)}]{carrasquilla2017machine}%
  \BibitemOpen
  \bibfield  {author} {\bibinfo {author} {\bibfnamefont {J.}~\bibnamefont
  {Carrasquilla}}\ and\ \bibinfo {author} {\bibfnamefont {R.~G.}\ \bibnamefont
  {Melko}},\ }\bibfield  {title} {\bibinfo {title} {{Machine learning phases of
  matter}},\ }\href {https://doi.org/10.1038/nphys4035} {\bibfield  {journal}
  {\bibinfo  {journal} {Nature Physics}\ }\textbf {\bibinfo {volume} {13}},\
  \bibinfo {pages} {431} (\bibinfo {year} {2017})}\BibitemShut {NoStop}%
\bibitem [{\citenamefont {Ch'ng}\ \emph {et~al.}(2017)\citenamefont {Ch'ng},
  \citenamefont {Carrasquilla}, \citenamefont {Melko},\ and\ \citenamefont
  {Khatami}}]{ch2017machine}%
  \BibitemOpen
  \bibfield  {author} {\bibinfo {author} {\bibfnamefont {K.}~\bibnamefont
  {Ch'ng}}, \bibinfo {author} {\bibfnamefont {J.}~\bibnamefont {Carrasquilla}},
  \bibinfo {author} {\bibfnamefont {R.~G.}\ \bibnamefont {Melko}},\ and\
  \bibinfo {author} {\bibfnamefont {E.}~\bibnamefont {Khatami}},\ }\bibfield
  {title} {\bibinfo {title} {{Machine Learning Phases of Strongly Correlated
  Fermions}},\ }\href {https://doi.org/10.1103/PhysRevX.7.031038} {\bibfield
  {journal} {\bibinfo  {journal} {Phys. Rev. X}\ }\textbf {\bibinfo {volume}
  {7}},\ \bibinfo {pages} {031038} (\bibinfo {year} {2017})}\BibitemShut
  {NoStop}%
\bibitem [{\citenamefont {Iakovlev}\ \emph {et~al.}(2018)\citenamefont
  {Iakovlev}, \citenamefont {Sotnikov},\ and\ \citenamefont
  {Mazurenko}}]{iakovlev2018supervised}%
  \BibitemOpen
  \bibfield  {author} {\bibinfo {author} {\bibfnamefont {I.~A.}\ \bibnamefont
  {Iakovlev}}, \bibinfo {author} {\bibfnamefont {O.~M.}\ \bibnamefont
  {Sotnikov}},\ and\ \bibinfo {author} {\bibfnamefont {V.~V.}\ \bibnamefont
  {Mazurenko}},\ }\bibfield  {title} {\bibinfo {title} {{Supervised learning
  approach for recognizing magnetic skyrmion phases}},\ }\href
  {https://doi.org/10.1103/PhysRevB.98.174411} {\bibfield  {journal} {\bibinfo
  {journal} {Phys. Rev. B}\ }\textbf {\bibinfo {volume} {98}},\ \bibinfo
  {pages} {174411} (\bibinfo {year} {2018})}\BibitemShut {NoStop}%
\bibitem [{\citenamefont {Wang}(2016)}]{wang2016discovering}%
  \BibitemOpen
  \bibfield  {author} {\bibinfo {author} {\bibfnamefont {L.}~\bibnamefont
  {Wang}},\ }\bibfield  {title} {\bibinfo {title} {{Discovering phase
  transitions with unsupervised learning}},\ }\href
  {https://doi.org/10.1103/PhysRevB.94.195105} {\bibfield  {journal} {\bibinfo
  {journal} {Phys. Rev. B}\ }\textbf {\bibinfo {volume} {94}},\ \bibinfo
  {pages} {195105} (\bibinfo {year} {2016})}\BibitemShut {NoStop}%
\bibitem [{\citenamefont {van Nieuwenburg}\ \emph {et~al.}(2017)\citenamefont
  {van Nieuwenburg}, \citenamefont {Liu},\ and\ \citenamefont
  {Huber}}]{van2017learning}%
  \BibitemOpen
  \bibfield  {author} {\bibinfo {author} {\bibfnamefont {E.~P.~L.}\
  \bibnamefont {van Nieuwenburg}}, \bibinfo {author} {\bibfnamefont {Y.-H.}\
  \bibnamefont {Liu}},\ and\ \bibinfo {author} {\bibfnamefont {S.~D.}\
  \bibnamefont {Huber}},\ }\bibfield  {title} {\bibinfo {title} {{Learning
  phase transitions by confusion}},\ }\href {https://doi.org/10.1038/nphys4037}
  {\bibfield  {journal} {\bibinfo  {journal} {Nature Physics}\ }\textbf
  {\bibinfo {volume} {13}},\ \bibinfo {pages} {435} (\bibinfo {year}
  {2017})}\BibitemShut {NoStop}%
\bibitem [{\citenamefont {Ch'ng}\ \emph {et~al.}(2018)\citenamefont {Ch'ng},
  \citenamefont {Vazquez},\ and\ \citenamefont {Khatami}}]{ch2018unsupervised}%
  \BibitemOpen
  \bibfield  {author} {\bibinfo {author} {\bibfnamefont {K.}~\bibnamefont
  {Ch'ng}}, \bibinfo {author} {\bibfnamefont {N.}~\bibnamefont {Vazquez}},\
  and\ \bibinfo {author} {\bibfnamefont {E.}~\bibnamefont {Khatami}},\
  }\bibfield  {title} {\bibinfo {title} {{Unsupervised machine learning account
  of magnetic transitions in the Hubbard model}},\ }\href
  {https://doi.org/10.1103/PhysRevE.97.013306} {\bibfield  {journal} {\bibinfo
  {journal} {Phys. Rev. E}\ }\textbf {\bibinfo {volume} {97}},\ \bibinfo
  {pages} {013306} (\bibinfo {year} {2018})}\BibitemShut {NoStop}%
\bibitem [{\citenamefont {Rodriguez-Nieva}\ and\ \citenamefont
  {Scheurer}(2019)}]{rodriguez2019identifying}%
  \BibitemOpen
  \bibfield  {author} {\bibinfo {author} {\bibfnamefont {J.~F.}\ \bibnamefont
  {Rodriguez-Nieva}}\ and\ \bibinfo {author} {\bibfnamefont {M.~S.}\
  \bibnamefont {Scheurer}},\ }\bibfield  {title} {\bibinfo {title}
  {{Identifying topological order through unsupervised machine learning}},\
  }\href {https://doi.org/10.1038/s41567-019-0512-x} {\bibfield  {journal}
  {\bibinfo  {journal} {Nature Physics}\ }\textbf {\bibinfo {volume} {15}},\
  \bibinfo {pages} {790} (\bibinfo {year} {2019})}\BibitemShut {NoStop}%
\bibitem [{\citenamefont {Arnold}\ \emph {et~al.}(2021)\citenamefont {Arnold},
  \citenamefont {Sch\"afer}, \citenamefont {\ifmmode~\check{Z}\else
  \v{Z}\fi{}onda},\ and\ \citenamefont {Lode}}]{arnold2021interpretable}%
  \BibitemOpen
  \bibfield  {author} {\bibinfo {author} {\bibfnamefont {J.}~\bibnamefont
  {Arnold}}, \bibinfo {author} {\bibfnamefont {F.}~\bibnamefont {Sch\"afer}},
  \bibinfo {author} {\bibfnamefont {M.}~\bibnamefont {\ifmmode~\check{Z}\else
  \v{Z}\fi{}onda}},\ and\ \bibinfo {author} {\bibfnamefont {A.~U.~J.}\
  \bibnamefont {Lode}},\ }\bibfield  {title} {\bibinfo {title} {{Interpretable
  and unsupervised phase classification}},\ }\href
  {https://doi.org/10.1103/PhysRevResearch.3.033052} {\bibfield  {journal}
  {\bibinfo  {journal} {Phys. Rev. Res.}\ }\textbf {\bibinfo {volume} {3}},\
  \bibinfo {pages} {033052} (\bibinfo {year} {2021})}\BibitemShut {NoStop}%
\bibitem [{\citenamefont {Carleo}\ and\ \citenamefont
  {Troyer}(2017)}]{carleo2017solving}%
  \BibitemOpen
  \bibfield  {author} {\bibinfo {author} {\bibfnamefont {G.}~\bibnamefont
  {Carleo}}\ and\ \bibinfo {author} {\bibfnamefont {M.}~\bibnamefont
  {Troyer}},\ }\bibfield  {title} {\bibinfo {title} {{Solving the quantum
  many-body problem with artificial neural networks}},\ }\href
  {https://doi.org/10.1126/science.aag2302} {\bibfield  {journal} {\bibinfo
  {journal} {Science}\ }\textbf {\bibinfo {volume} {355}},\ \bibinfo {pages}
  {602} (\bibinfo {year} {2017})}\BibitemShut {NoStop}%
\bibitem [{\citenamefont {Nomura}\ \emph {et~al.}(2017)\citenamefont {Nomura},
  \citenamefont {Darmawan}, \citenamefont {Yamaji},\ and\ \citenamefont
  {Imada}}]{nomura2017restricted}%
  \BibitemOpen
  \bibfield  {author} {\bibinfo {author} {\bibfnamefont {Y.}~\bibnamefont
  {Nomura}}, \bibinfo {author} {\bibfnamefont {A.~S.}\ \bibnamefont
  {Darmawan}}, \bibinfo {author} {\bibfnamefont {Y.}~\bibnamefont {Yamaji}},\
  and\ \bibinfo {author} {\bibfnamefont {M.}~\bibnamefont {Imada}},\ }\bibfield
   {title} {\bibinfo {title} {{Restricted Boltzmann machine learning for
  solving strongly correlated quantum systems}},\ }\href
  {https://doi.org/10.1103/PhysRevB.96.205152} {\bibfield  {journal} {\bibinfo
  {journal} {Phys. Rev. B}\ }\textbf {\bibinfo {volume} {96}},\ \bibinfo
  {pages} {205152} (\bibinfo {year} {2017})}\BibitemShut {NoStop}%
\bibitem [{\citenamefont {Choo}\ \emph {et~al.}(2018)\citenamefont {Choo},
  \citenamefont {Carleo}, \citenamefont {Regnault},\ and\ \citenamefont
  {Neupert}}]{choo2018symmetries}%
  \BibitemOpen
  \bibfield  {author} {\bibinfo {author} {\bibfnamefont {K.}~\bibnamefont
  {Choo}}, \bibinfo {author} {\bibfnamefont {G.}~\bibnamefont {Carleo}},
  \bibinfo {author} {\bibfnamefont {N.}~\bibnamefont {Regnault}},\ and\
  \bibinfo {author} {\bibfnamefont {T.}~\bibnamefont {Neupert}},\ }\bibfield
  {title} {\bibinfo {title} {{Symmetries and Many-Body Excitations with
  Neural-Network Quantum States}},\ }\href
  {https://doi.org/10.1103/PhysRevLett.121.167204} {\bibfield  {journal}
  {\bibinfo  {journal} {Phys. Rev. Lett.}\ }\textbf {\bibinfo {volume} {121}},\
  \bibinfo {pages} {167204} (\bibinfo {year} {2018})}\BibitemShut {NoStop}%
\bibitem [{\citenamefont {Yoshioka}\ and\ \citenamefont
  {Hamazaki}(2019)}]{yoshioka2019constructing}%
  \BibitemOpen
  \bibfield  {author} {\bibinfo {author} {\bibfnamefont {N.}~\bibnamefont
  {Yoshioka}}\ and\ \bibinfo {author} {\bibfnamefont {R.}~\bibnamefont
  {Hamazaki}},\ }\bibfield  {title} {\bibinfo {title} {{Constructing neural
  stationary states for open quantum many-body systems}},\ }\href
  {https://doi.org/10.1103/PhysRevB.99.214306} {\bibfield  {journal} {\bibinfo
  {journal} {Phys. Rev. B}\ }\textbf {\bibinfo {volume} {99}},\ \bibinfo
  {pages} {214306} (\bibinfo {year} {2019})}\BibitemShut {NoStop}%
\bibitem [{\citenamefont {Carrasquilla}\ \emph {et~al.}(2019)\citenamefont
  {Carrasquilla}, \citenamefont {Torlai}, \citenamefont {Melko},\ and\
  \citenamefont {Aolita}}]{carrasquilla2019reconstructing}%
  \BibitemOpen
  \bibfield  {author} {\bibinfo {author} {\bibfnamefont {J.}~\bibnamefont
  {Carrasquilla}}, \bibinfo {author} {\bibfnamefont {G.}~\bibnamefont
  {Torlai}}, \bibinfo {author} {\bibfnamefont {R.~G.}\ \bibnamefont {Melko}},\
  and\ \bibinfo {author} {\bibfnamefont {L.}~\bibnamefont {Aolita}},\
  }\bibfield  {title} {\bibinfo {title} {{Reconstructing quantum states with
  generative models}},\ }\href {https://doi.org/10.1038/s42256-019-0028-1}
  {\bibfield  {journal} {\bibinfo  {journal} {Nature Machine Intelligence}\
  }\textbf {\bibinfo {volume} {1}},\ \bibinfo {pages} {155} (\bibinfo {year}
  {2019})}\BibitemShut {NoStop}%
\bibitem [{\citenamefont {Liu}\ \emph {et~al.}(2017{\natexlab{a}})\citenamefont
  {Liu}, \citenamefont {Qi}, \citenamefont {Meng},\ and\ \citenamefont
  {Fu}}]{liu2017self}%
  \BibitemOpen
  \bibfield  {author} {\bibinfo {author} {\bibfnamefont {J.}~\bibnamefont
  {Liu}}, \bibinfo {author} {\bibfnamefont {Y.}~\bibnamefont {Qi}}, \bibinfo
  {author} {\bibfnamefont {Z.~Y.}\ \bibnamefont {Meng}},\ and\ \bibinfo
  {author} {\bibfnamefont {L.}~\bibnamefont {Fu}},\ }\bibfield  {title}
  {\bibinfo {title} {{Self-learning Monte Carlo method}},\ }\href
  {https://doi.org/10.1103/PhysRevB.95.041101} {\bibfield  {journal} {\bibinfo
  {journal} {Phys. Rev. B}\ }\textbf {\bibinfo {volume} {95}},\ \bibinfo
  {pages} {041101} (\bibinfo {year} {2017}{\natexlab{a}})}\BibitemShut
  {NoStop}%
\bibitem [{\citenamefont {Liu}\ \emph {et~al.}(2017{\natexlab{b}})\citenamefont
  {Liu}, \citenamefont {Shen}, \citenamefont {Qi}, \citenamefont {Meng},\ and\
  \citenamefont {Fu}}]{liu2017self2}%
  \BibitemOpen
  \bibfield  {author} {\bibinfo {author} {\bibfnamefont {J.}~\bibnamefont
  {Liu}}, \bibinfo {author} {\bibfnamefont {H.}~\bibnamefont {Shen}}, \bibinfo
  {author} {\bibfnamefont {Y.}~\bibnamefont {Qi}}, \bibinfo {author}
  {\bibfnamefont {Z.~Y.}\ \bibnamefont {Meng}},\ and\ \bibinfo {author}
  {\bibfnamefont {L.}~\bibnamefont {Fu}},\ }\bibfield  {title} {\bibinfo
  {title} {{Self-learning Monte Carlo method and cumulative update in fermion
  systems}},\ }\href {https://doi.org/10.1103/PhysRevB.95.241104} {\bibfield
  {journal} {\bibinfo  {journal} {Phys. Rev. B}\ }\textbf {\bibinfo {volume}
  {95}},\ \bibinfo {pages} {241104} (\bibinfo {year}
  {2017}{\natexlab{b}})}\BibitemShut {NoStop}%
\bibitem [{\citenamefont {Huang}\ and\ \citenamefont
  {Wang}(2017)}]{huang2017accelerated}%
  \BibitemOpen
  \bibfield  {author} {\bibinfo {author} {\bibfnamefont {L.}~\bibnamefont
  {Huang}}\ and\ \bibinfo {author} {\bibfnamefont {L.}~\bibnamefont {Wang}},\
  }\bibfield  {title} {\bibinfo {title} {{Accelerated Monte Carlo simulations
  with restricted Boltzmann machines}},\ }\href
  {https://doi.org/10.1103/PhysRevB.95.035105} {\bibfield  {journal} {\bibinfo
  {journal} {Phys. Rev. B}\ }\textbf {\bibinfo {volume} {95}},\ \bibinfo
  {pages} {035105} (\bibinfo {year} {2017})}\BibitemShut {NoStop}%
\bibitem [{\citenamefont {Kanwar}\ \emph {et~al.}(2020)\citenamefont {Kanwar},
  \citenamefont {Albergo}, \citenamefont {Boyda}, \citenamefont {Cranmer},
  \citenamefont {Hackett}, \citenamefont {Racani\`ere}, \citenamefont
  {Rezende},\ and\ \citenamefont {Shanahan}}]{kanwar2020equivariant}%
  \BibitemOpen
  \bibfield  {author} {\bibinfo {author} {\bibfnamefont {G.}~\bibnamefont
  {Kanwar}}, \bibinfo {author} {\bibfnamefont {M.~S.}\ \bibnamefont {Albergo}},
  \bibinfo {author} {\bibfnamefont {D.}~\bibnamefont {Boyda}}, \bibinfo
  {author} {\bibfnamefont {K.}~\bibnamefont {Cranmer}}, \bibinfo {author}
  {\bibfnamefont {D.~C.}\ \bibnamefont {Hackett}}, \bibinfo {author}
  {\bibfnamefont {S.}~\bibnamefont {Racani\`ere}}, \bibinfo {author}
  {\bibfnamefont {D.~J.}\ \bibnamefont {Rezende}},\ and\ \bibinfo {author}
  {\bibfnamefont {P.~E.}\ \bibnamefont {Shanahan}},\ }\bibfield  {title}
  {\bibinfo {title} {{Equivariant Flow-Based Sampling for Lattice Gauge
  Theory}},\ }\href {https://doi.org/10.1103/PhysRevLett.125.121601} {\bibfield
   {journal} {\bibinfo  {journal} {Phys. Rev. Lett.}\ }\textbf {\bibinfo
  {volume} {125}},\ \bibinfo {pages} {121601} (\bibinfo {year}
  {2020})}\BibitemShut {NoStop}%
\bibitem [{\citenamefont {Behler}\ and\ \citenamefont
  {Parrinello}(2007)}]{behler2007generalized}%
  \BibitemOpen
  \bibfield  {author} {\bibinfo {author} {\bibfnamefont {J.}~\bibnamefont
  {Behler}}\ and\ \bibinfo {author} {\bibfnamefont {M.}~\bibnamefont
  {Parrinello}},\ }\bibfield  {title} {\bibinfo {title} {{Generalized
  Neural-Network Representation of High-Dimensional Potential-Energy
  Surfaces}},\ }\href {https://doi.org/10.1103/PhysRevLett.98.146401}
  {\bibfield  {journal} {\bibinfo  {journal} {Phys. Rev. Lett.}\ }\textbf
  {\bibinfo {volume} {98}},\ \bibinfo {pages} {146401} (\bibinfo {year}
  {2007})}\BibitemShut {NoStop}%
\bibitem [{\citenamefont {Rupp}\ \emph {et~al.}(2012)\citenamefont {Rupp},
  \citenamefont {Tkatchenko}, \citenamefont {M\"uller},\ and\ \citenamefont
  {von Lilienfeld}}]{rupp2012fast}%
  \BibitemOpen
  \bibfield  {author} {\bibinfo {author} {\bibfnamefont {M.}~\bibnamefont
  {Rupp}}, \bibinfo {author} {\bibfnamefont {A.}~\bibnamefont {Tkatchenko}},
  \bibinfo {author} {\bibfnamefont {K.-R.}\ \bibnamefont {M\"uller}},\ and\
  \bibinfo {author} {\bibfnamefont {O.~A.}\ \bibnamefont {von Lilienfeld}},\
  }\bibfield  {title} {\bibinfo {title} {{Fast and Accurate Modeling of
  Molecular Atomization Energies with Machine Learning}},\ }\href
  {https://doi.org/10.1103/PhysRevLett.108.058301} {\bibfield  {journal}
  {\bibinfo  {journal} {Phys. Rev. Lett.}\ }\textbf {\bibinfo {volume} {108}},\
  \bibinfo {pages} {058301} (\bibinfo {year} {2012})}\BibitemShut {NoStop}%
\bibitem [{\citenamefont {Bart\'ok}\ \emph {et~al.}(2013)\citenamefont
  {Bart\'ok}, \citenamefont {Kondor},\ and\ \citenamefont
  {Cs\'anyi}}]{bartok2013representing}%
  \BibitemOpen
  \bibfield  {author} {\bibinfo {author} {\bibfnamefont {A.~P.}\ \bibnamefont
  {Bart\'ok}}, \bibinfo {author} {\bibfnamefont {R.}~\bibnamefont {Kondor}},\
  and\ \bibinfo {author} {\bibfnamefont {G.}~\bibnamefont {Cs\'anyi}},\
  }\bibfield  {title} {\bibinfo {title} {{On representing chemical
  environments}},\ }\href {https://doi.org/10.1103/PhysRevB.87.184115}
  {\bibfield  {journal} {\bibinfo  {journal} {Phys. Rev. B}\ }\textbf {\bibinfo
  {volume} {87}},\ \bibinfo {pages} {184115} (\bibinfo {year}
  {2013})}\BibitemShut {NoStop}%
\bibitem [{\citenamefont {Behler}(2015)}]{behler2015constructing}%
  \BibitemOpen
  \bibfield  {author} {\bibinfo {author} {\bibfnamefont {J.}~\bibnamefont
  {Behler}},\ }\bibfield  {title} {\bibinfo {title} {{Constructing
  high-dimensional neural network potentials: A tutorial review}},\ }\href
  {https://doi.org/https://doi.org/10.1002/qua.24890} {\bibfield  {journal}
  {\bibinfo  {journal} {International Journal of Quantum Chemistry}\ }\textbf
  {\bibinfo {volume} {115}},\ \bibinfo {pages} {1032} (\bibinfo {year}
  {2015})}\BibitemShut {NoStop}%
\bibitem [{\citenamefont {Marchand}\ \emph {et~al.}(2020)\citenamefont
  {Marchand}, \citenamefont {Jain}, \citenamefont {Glensk},\ and\ \citenamefont
  {Curtin}}]{marchand2020machine}%
  \BibitemOpen
  \bibfield  {author} {\bibinfo {author} {\bibfnamefont {D.}~\bibnamefont
  {Marchand}}, \bibinfo {author} {\bibfnamefont {A.}~\bibnamefont {Jain}},
  \bibinfo {author} {\bibfnamefont {A.}~\bibnamefont {Glensk}},\ and\ \bibinfo
  {author} {\bibfnamefont {W.~A.}\ \bibnamefont {Curtin}},\ }\bibfield  {title}
  {\bibinfo {title} {{Machine learning for metallurgy {I}. {A} neural-network
  potential for {A}l-{C}u}},\ }\href
  {https://doi.org/10.1103/PhysRevMaterials.4.103601} {\bibfield  {journal}
  {\bibinfo  {journal} {Phys. Rev. Mater.}\ }\textbf {\bibinfo {volume} {4}},\
  \bibinfo {pages} {103601} (\bibinfo {year} {2020})}\BibitemShut {NoStop}%
\bibitem [{\citenamefont {Fan}\ \emph {et~al.}(2023)\citenamefont {Fan},
  \citenamefont {Shen}, \citenamefont {Nussinov}, \citenamefont {Liu},
  \citenamefont {Sun},\ and\ \citenamefont {Liu}}]{fan2023searching}%
  \BibitemOpen
  \bibfield  {author} {\bibinfo {author} {\bibfnamefont {C.}~\bibnamefont
  {Fan}}, \bibinfo {author} {\bibfnamefont {M.}~\bibnamefont {Shen}}, \bibinfo
  {author} {\bibfnamefont {Z.}~\bibnamefont {Nussinov}}, \bibinfo {author}
  {\bibfnamefont {Z.}~\bibnamefont {Liu}}, \bibinfo {author} {\bibfnamefont
  {Y.}~\bibnamefont {Sun}},\ and\ \bibinfo {author} {\bibfnamefont {Y.-Y.}\
  \bibnamefont {Liu}},\ }\bibfield  {title} {\bibinfo {title} {{Searching for
  spin glass ground states through deep reinforcement learning}},\ }\href
  {https://doi.org/10.1038/s41467-023-36363-w} {\bibfield  {journal} {\bibinfo
  {journal} {Nature Communications}\ }\textbf {\bibinfo {volume} {14}},\
  \bibinfo {pages} {725} (\bibinfo {year} {2023})}\BibitemShut {NoStop}%
\bibitem [{\citenamefont {Mills}\ and\ \citenamefont
  {Tamblyn}(2018)}]{mills2018deep}%
  \BibitemOpen
  \bibfield  {author} {\bibinfo {author} {\bibfnamefont {K.}~\bibnamefont
  {Mills}}\ and\ \bibinfo {author} {\bibfnamefont {I.}~\bibnamefont
  {Tamblyn}},\ }\bibfield  {title} {\bibinfo {title} {{Deep neural networks for
  direct, featureless learning through observation: The case of two-dimensional
  spin models}},\ }\href {https://doi.org/10.1103/PhysRevE.97.032119}
  {\bibfield  {journal} {\bibinfo  {journal} {Phys. Rev. E}\ }\textbf {\bibinfo
  {volume} {97}},\ \bibinfo {pages} {032119} (\bibinfo {year}
  {2018})}\BibitemShut {NoStop}%
\bibitem [{\citenamefont {Wu}\ \emph {et~al.}(2018)\citenamefont {Wu},
  \citenamefont {Xiao},\ and\ \citenamefont {Paterson}}]{wu2018physics}%
  \BibitemOpen
  \bibfield  {author} {\bibinfo {author} {\bibfnamefont {J.-L.}\ \bibnamefont
  {Wu}}, \bibinfo {author} {\bibfnamefont {H.}~\bibnamefont {Xiao}},\ and\
  \bibinfo {author} {\bibfnamefont {E.}~\bibnamefont {Paterson}},\ }\bibfield
  {title} {\bibinfo {title} {{Physics-informed machine learning approach for
  augmenting turbulence models: A comprehensive framework}},\ }\href
  {https://doi.org/10.1103/PhysRevFluids.3.074602} {\bibfield  {journal}
  {\bibinfo  {journal} {Phys. Rev. Fluids}\ }\textbf {\bibinfo {volume} {3}},\
  \bibinfo {pages} {074602} (\bibinfo {year} {2018})}\BibitemShut {NoStop}%
\bibitem [{\citenamefont {Raissi}\ \emph {et~al.}(2019)\citenamefont {Raissi},
  \citenamefont {Perdikaris},\ and\ \citenamefont
  {Karniadakis}}]{raissi2019physics}%
  \BibitemOpen
  \bibfield  {author} {\bibinfo {author} {\bibfnamefont {M.}~\bibnamefont
  {Raissi}}, \bibinfo {author} {\bibfnamefont {P.}~\bibnamefont {Perdikaris}},\
  and\ \bibinfo {author} {\bibfnamefont {G.}~\bibnamefont {Karniadakis}},\
  }\bibfield  {title} {\bibinfo {title} {{Physics-informed neural networks: A
  deep learning framework for solving forward and inverse problems involving
  nonlinear partial differential equations}},\ }\href
  {https://doi.org/https://doi.org/10.1016/j.jcp.2018.10.045} {\bibfield
  {journal} {\bibinfo  {journal} {Journal of Computational Physics}\ }\textbf
  {\bibinfo {volume} {378}},\ \bibinfo {pages} {686} (\bibinfo {year}
  {2019})}\BibitemShut {NoStop}%
\bibitem [{\citenamefont {Brunton}\ \emph {et~al.}(2020)\citenamefont
  {Brunton}, \citenamefont {Noack},\ and\ \citenamefont
  {Koumoutsakos}}]{brunton2020machine}%
  \BibitemOpen
  \bibfield  {author} {\bibinfo {author} {\bibfnamefont {S.~L.}\ \bibnamefont
  {Brunton}}, \bibinfo {author} {\bibfnamefont {B.~R.}\ \bibnamefont {Noack}},\
  and\ \bibinfo {author} {\bibfnamefont {P.}~\bibnamefont {Koumoutsakos}},\
  }\bibfield  {title} {\bibinfo {title} {{Machine Learning for Fluid
  Mechanics}},\ }\href
  {https://doi.org/https://doi.org/10.1146/annurev-fluid-010719-060214}
  {\bibfield  {journal} {\bibinfo  {journal} {Annual Review of Fluid
  Mechanics}\ }\textbf {\bibinfo {volume} {52}},\ \bibinfo {pages} {477}
  (\bibinfo {year} {2020})}\BibitemShut {NoStop}%
\bibitem [{\citenamefont {Lu}\ \emph {et~al.}(2021)\citenamefont {Lu},
  \citenamefont {Meng}, \citenamefont {Mao},\ and\ \citenamefont
  {Karniadakis}}]{lu2021deepxde}%
  \BibitemOpen
  \bibfield  {author} {\bibinfo {author} {\bibfnamefont {L.}~\bibnamefont
  {Lu}}, \bibinfo {author} {\bibfnamefont {X.}~\bibnamefont {Meng}}, \bibinfo
  {author} {\bibfnamefont {Z.}~\bibnamefont {Mao}},\ and\ \bibinfo {author}
  {\bibfnamefont {G.~E.}\ \bibnamefont {Karniadakis}},\ }\bibfield  {title}
  {\bibinfo {title} {{DeepXDE: A Deep Learning Library for Solving Differential
  Equations}},\ }\href {https://doi.org/10.1137/19M1274067} {\bibfield
  {journal} {\bibinfo  {journal} {SIAM Review}\ }\textbf {\bibinfo {volume}
  {63}},\ \bibinfo {pages} {208} (\bibinfo {year} {2021})}\BibitemShut
  {NoStop}%
\bibitem [{\citenamefont {Mattheakis}\ \emph {et~al.}(2020)\citenamefont
  {Mattheakis}, \citenamefont {Protopapas}, \citenamefont {Sondak},
  \citenamefont {Giovanni},\ and\ \citenamefont
  {Kaxiras}}]{mattheakis2019physical}%
  \BibitemOpen
  \bibfield  {author} {\bibinfo {author} {\bibfnamefont {M.}~\bibnamefont
  {Mattheakis}}, \bibinfo {author} {\bibfnamefont {P.}~\bibnamefont
  {Protopapas}}, \bibinfo {author} {\bibfnamefont {D.}~\bibnamefont {Sondak}},
  \bibinfo {author} {\bibfnamefont {M.~D.}\ \bibnamefont {Giovanni}},\ and\
  \bibinfo {author} {\bibfnamefont {E.}~\bibnamefont {Kaxiras}},\ }\href
  {https://arxiv.org/abs/1904.08991} {\bibinfo {title} {{Physical Symmetries
  Embedded in Neural Networks}}} (\bibinfo {year} {2020})\BibitemShut {NoStop}%
\bibitem [{\citenamefont {Bogatskiy}\ \emph {et~al.}(2020)\citenamefont
  {Bogatskiy}, \citenamefont {Anderson}, \citenamefont {Offermann},
  \citenamefont {Roussi}, \citenamefont {Miller},\ and\ \citenamefont
  {Kondor}}]{bogatskiy2020lorentz}%
  \BibitemOpen
  \bibfield  {author} {\bibinfo {author} {\bibfnamefont {A.}~\bibnamefont
  {Bogatskiy}}, \bibinfo {author} {\bibfnamefont {B.}~\bibnamefont {Anderson}},
  \bibinfo {author} {\bibfnamefont {J.}~\bibnamefont {Offermann}}, \bibinfo
  {author} {\bibfnamefont {M.}~\bibnamefont {Roussi}}, \bibinfo {author}
  {\bibfnamefont {D.}~\bibnamefont {Miller}},\ and\ \bibinfo {author}
  {\bibfnamefont {R.}~\bibnamefont {Kondor}},\ }\bibfield  {title} {\bibinfo
  {title} {{L}orentz group equivariant neural network for particle physics},\
  }in\ \href {https://proceedings.mlr.press/v119/bogatskiy20a.html} {\emph
  {\bibinfo {booktitle} {Proceedings of the 37th International Conference on
  Machine Learning}}},\ \bibinfo {series} {Proceedings of Machine Learning
  Research}, Vol.\ \bibinfo {volume} {119}\ (\bibinfo  {publisher} {PMLR},\
  \bibinfo {year} {2020})\ pp.\ \bibinfo {pages} {992--1002}\BibitemShut
  {NoStop}%
\bibitem [{\citenamefont {Yang}\ \emph {et~al.}(2020)\citenamefont {Yang},
  \citenamefont {Zhang},\ and\ \citenamefont {Karniadakis}}]{yang2020physics}%
  \BibitemOpen
  \bibfield  {author} {\bibinfo {author} {\bibfnamefont {L.}~\bibnamefont
  {Yang}}, \bibinfo {author} {\bibfnamefont {D.}~\bibnamefont {Zhang}},\ and\
  \bibinfo {author} {\bibfnamefont {G.~E.}\ \bibnamefont {Karniadakis}},\
  }\bibfield  {title} {\bibinfo {title} {{Physics-Informed Generative
  Adversarial Networks for Stochastic Differential Equations}},\ }\href
  {https://doi.org/10.1137/18M1225409} {\bibfield  {journal} {\bibinfo
  {journal} {SIAM Journal on Scientific Computing}\ }\textbf {\bibinfo {volume}
  {42}},\ \bibinfo {pages} {A292} (\bibinfo {year} {2020})}\BibitemShut
  {NoStop}%
\bibitem [{\citenamefont {Mills}\ \emph {et~al.}(2019)\citenamefont {Mills},
  \citenamefont {Ryczko}, \citenamefont {Luchak}, \citenamefont {Domurad},
  \citenamefont {Beeler},\ and\ \citenamefont {Tamblyn}}]{mills2019extensive}%
  \BibitemOpen
  \bibfield  {author} {\bibinfo {author} {\bibfnamefont {K.}~\bibnamefont
  {Mills}}, \bibinfo {author} {\bibfnamefont {K.}~\bibnamefont {Ryczko}},
  \bibinfo {author} {\bibfnamefont {I.}~\bibnamefont {Luchak}}, \bibinfo
  {author} {\bibfnamefont {A.}~\bibnamefont {Domurad}}, \bibinfo {author}
  {\bibfnamefont {C.}~\bibnamefont {Beeler}},\ and\ \bibinfo {author}
  {\bibfnamefont {I.}~\bibnamefont {Tamblyn}},\ }\bibfield  {title} {\bibinfo
  {title} {{Extensive deep neural networks for transferring small scale
  learning to large scale systems}},\ }\href
  {https://doi.org/10.1039/C8SC04578J} {\bibfield  {journal} {\bibinfo
  {journal} {Chem. Sci.}\ }\textbf {\bibinfo {volume} {10}},\ \bibinfo {pages}
  {4129} (\bibinfo {year} {2019})}\BibitemShut {NoStop}%
\bibitem [{\citenamefont {Batzner}\ \emph {et~al.}(2022)\citenamefont
  {Batzner}, \citenamefont {Musaelian}, \citenamefont {Sun}, \citenamefont
  {Geiger}, \citenamefont {Mailoa}, \citenamefont {Kornbluth}, \citenamefont
  {Molinari}, \citenamefont {Smidt},\ and\ \citenamefont
  {Kozinsky}}]{Batzner2022E3}%
  \BibitemOpen
  \bibfield  {author} {\bibinfo {author} {\bibfnamefont {S.}~\bibnamefont
  {Batzner}}, \bibinfo {author} {\bibfnamefont {A.}~\bibnamefont {Musaelian}},
  \bibinfo {author} {\bibfnamefont {L.}~\bibnamefont {Sun}}, \bibinfo {author}
  {\bibfnamefont {M.}~\bibnamefont {Geiger}}, \bibinfo {author} {\bibfnamefont
  {J.~P.}\ \bibnamefont {Mailoa}}, \bibinfo {author} {\bibfnamefont
  {M.}~\bibnamefont {Kornbluth}}, \bibinfo {author} {\bibfnamefont
  {N.}~\bibnamefont {Molinari}}, \bibinfo {author} {\bibfnamefont {T.~E.}\
  \bibnamefont {Smidt}},\ and\ \bibinfo {author} {\bibfnamefont
  {B.}~\bibnamefont {Kozinsky}},\ }\bibfield  {title} {\bibinfo {title}
  {{E(3)-equivariant graph neural networks for data-efficient and accurate
  interatomic potentials}},\ }\href
  {https://doi.org/10.1038/s41467-022-29939-5} {\bibfield  {journal} {\bibinfo
  {journal} {Nature Communications}\ }\textbf {\bibinfo {volume} {13}},\
  \bibinfo {pages} {2453} (\bibinfo {year} {2022})}\BibitemShut {NoStop}%
\bibitem [{\citenamefont {Xie}\ \emph {et~al.}(2024{\natexlab{a}})\citenamefont
  {Xie}, \citenamefont {Lu}, \citenamefont {Meng},\ and\ \citenamefont
  {Liu}}]{XIE2024lm}%
  \BibitemOpen
  \bibfield  {author} {\bibinfo {author} {\bibfnamefont {F.}~\bibnamefont
  {Xie}}, \bibinfo {author} {\bibfnamefont {T.}~\bibnamefont {Lu}}, \bibinfo
  {author} {\bibfnamefont {S.}~\bibnamefont {Meng}},\ and\ \bibinfo {author}
  {\bibfnamefont {M.}~\bibnamefont {Liu}},\ }\bibfield  {title} {\bibinfo
  {title} {{GPTFF: A high-accuracy out-of-the-box universal AI force field for
  arbitrary inorganic materials}},\ }\href
  {https://doi.org/https://doi.org/10.1016/j.scib.2024.08.039} {\bibfield
  {journal} {\bibinfo  {journal} {Science Bulletin}\ }\textbf {\bibinfo
  {volume} {69}},\ \bibinfo {pages} {3525} (\bibinfo {year}
  {2024}{\natexlab{a}})}\BibitemShut {NoStop}%
\bibitem [{\citenamefont {Han}\ \emph {et~al.}(2023{\natexlab{a}})\citenamefont
  {Han}, \citenamefont {Wang}, \citenamefont {Ding}, \citenamefont {Gao},
  \citenamefont {Pan}, \citenamefont {Jia},\ and\ \citenamefont
  {Sun}}]{10.1063/5.0142281sj}%
  \BibitemOpen
  \bibfield  {author} {\bibinfo {author} {\bibfnamefont {Y.}~\bibnamefont
  {Han}}, \bibinfo {author} {\bibfnamefont {J.}~\bibnamefont {Wang}}, \bibinfo
  {author} {\bibfnamefont {C.}~\bibnamefont {Ding}}, \bibinfo {author}
  {\bibfnamefont {H.}~\bibnamefont {Gao}}, \bibinfo {author} {\bibfnamefont
  {S.}~\bibnamefont {Pan}}, \bibinfo {author} {\bibfnamefont {Q.}~\bibnamefont
  {Jia}},\ and\ \bibinfo {author} {\bibfnamefont {J.}~\bibnamefont {Sun}},\
  }\bibfield  {title} {\bibinfo {title} {{Prediction of surface reconstructions
  using MAGUS}},\ }\href {https://doi.org/10.1063/5.0142281} {\bibfield
  {journal} {\bibinfo  {journal} {The Journal of Chemical Physics}\ }\textbf
  {\bibinfo {volume} {158}},\ \bibinfo {pages} {174109} (\bibinfo {year}
  {2023}{\natexlab{a}})}\BibitemShut {NoStop}%
\bibitem [{\citenamefont {Zhong}\ \emph {et~al.}(2023)\citenamefont {Zhong},
  \citenamefont {Yu}, \citenamefont {Gong},\ and\ \citenamefont
  {Xiang}}]{Zhong2023_general_xhj}%
  \BibitemOpen
  \bibfield  {author} {\bibinfo {author} {\bibfnamefont {Y.}~\bibnamefont
  {Zhong}}, \bibinfo {author} {\bibfnamefont {H.}~\bibnamefont {Yu}}, \bibinfo
  {author} {\bibfnamefont {X.}~\bibnamefont {Gong}},\ and\ \bibinfo {author}
  {\bibfnamefont {H.}~\bibnamefont {Xiang}},\ }\bibfield  {title} {\bibinfo
  {title} {{A General Tensor Prediction Framework Based on Graph Neural
  Networks}},\ }\href {https://doi.org/10.1021/acs.jpclett.3c01200} {\bibfield
  {journal} {\bibinfo  {journal} {The Journal of Physical Chemistry Letters}\
  }\textbf {\bibinfo {volume} {14}},\ \bibinfo {pages} {6339} (\bibinfo {year}
  {2023})}\BibitemShut {NoStop}%
\bibitem [{\citenamefont {Gao}\ \emph {et~al.}(2021)\citenamefont {Gao},
  \citenamefont {Wang}, \citenamefont {Han},\ and\ \citenamefont
  {Sun}}]{GAO2021466Enhancing}%
  \BibitemOpen
  \bibfield  {author} {\bibinfo {author} {\bibfnamefont {H.}~\bibnamefont
  {Gao}}, \bibinfo {author} {\bibfnamefont {J.}~\bibnamefont {Wang}}, \bibinfo
  {author} {\bibfnamefont {Y.}~\bibnamefont {Han}},\ and\ \bibinfo {author}
  {\bibfnamefont {J.}~\bibnamefont {Sun}},\ }\bibfield  {title} {\bibinfo
  {title} {{Enhancing crystal structure prediction by decomposition and
  evolution schemes based on graph theory}},\ }\href
  {https://doi.org/https://doi.org/10.1016/j.fmre.2021.06.005} {\bibfield
  {journal} {\bibinfo  {journal} {Fundamental Research}\ }\textbf {\bibinfo
  {volume} {1}},\ \bibinfo {pages} {466} (\bibinfo {year} {2021})}\BibitemShut
  {NoStop}%
\bibitem [{\citenamefont {Wang}\ \emph {et~al.}(2023)\citenamefont {Wang},
  \citenamefont {Gao}, \citenamefont {Han}, \citenamefont {Ding}, \citenamefont
  {Pan}, \citenamefont {Wang}, \citenamefont {Jia}, \citenamefont {Wang},
  \citenamefont {Xing},\ and\ \citenamefont {Sun}}]{10.1093/nsr/nwad128MAGUS}%
  \BibitemOpen
  \bibfield  {author} {\bibinfo {author} {\bibfnamefont {J.}~\bibnamefont
  {Wang}}, \bibinfo {author} {\bibfnamefont {H.}~\bibnamefont {Gao}}, \bibinfo
  {author} {\bibfnamefont {Y.}~\bibnamefont {Han}}, \bibinfo {author}
  {\bibfnamefont {C.}~\bibnamefont {Ding}}, \bibinfo {author} {\bibfnamefont
  {S.}~\bibnamefont {Pan}}, \bibinfo {author} {\bibfnamefont {Y.}~\bibnamefont
  {Wang}}, \bibinfo {author} {\bibfnamefont {Q.}~\bibnamefont {Jia}}, \bibinfo
  {author} {\bibfnamefont {H.-T.}\ \bibnamefont {Wang}}, \bibinfo {author}
  {\bibfnamefont {D.}~\bibnamefont {Xing}},\ and\ \bibinfo {author}
  {\bibfnamefont {J.}~\bibnamefont {Sun}},\ }\bibfield  {title} {\bibinfo
  {title} {{MAGUS: machine learning and graph theory assisted universal
  structure searcher}},\ }\href {https://doi.org/10.1093/nsr/nwad128}
  {\bibfield  {journal} {\bibinfo  {journal} {National Science Review}\
  }\textbf {\bibinfo {volume} {10}},\ \bibinfo {pages} {nwad128} (\bibinfo
  {year} {2023})}\BibitemShut {NoStop}%
\bibitem [{\citenamefont {Yu}\ \emph {et~al.}(2024)\citenamefont {Yu},
  \citenamefont {Zhong}, \citenamefont {Hong}, \citenamefont {Xu},
  \citenamefont {Ren}, \citenamefont {Gong},\ and\ \citenamefont
  {Xiang}}]{PhysRevB.109.144426Spin-dependent_graph}%
  \BibitemOpen
  \bibfield  {author} {\bibinfo {author} {\bibfnamefont {H.}~\bibnamefont
  {Yu}}, \bibinfo {author} {\bibfnamefont {Y.}~\bibnamefont {Zhong}}, \bibinfo
  {author} {\bibfnamefont {L.}~\bibnamefont {Hong}}, \bibinfo {author}
  {\bibfnamefont {C.}~\bibnamefont {Xu}}, \bibinfo {author} {\bibfnamefont
  {W.}~\bibnamefont {Ren}}, \bibinfo {author} {\bibfnamefont {X.}~\bibnamefont
  {Gong}},\ and\ \bibinfo {author} {\bibfnamefont {H.}~\bibnamefont {Xiang}},\
  }\bibfield  {title} {\bibinfo {title} {{Spin-dependent graph neural network
  potential for magnetic materials}},\ }\href
  {https://doi.org/10.1103/PhysRevB.109.144426} {\bibfield  {journal} {\bibinfo
   {journal} {Phys. Rev. B}\ }\textbf {\bibinfo {volume} {109}},\ \bibinfo
  {pages} {144426} (\bibinfo {year} {2024})}\BibitemShut {NoStop}%
\bibitem [{\citenamefont {Li}\ and\ \citenamefont {Wang}(2018)}]{li2018neural}%
  \BibitemOpen
  \bibfield  {author} {\bibinfo {author} {\bibfnamefont {S.-H.}\ \bibnamefont
  {Li}}\ and\ \bibinfo {author} {\bibfnamefont {L.}~\bibnamefont {Wang}},\
  }\bibfield  {title} {\bibinfo {title} {{Neural Network Renormalization
  Group}},\ }\href {https://doi.org/10.1103/PhysRevLett.121.260601} {\bibfield
  {journal} {\bibinfo  {journal} {Phys. Rev. Lett.}\ }\textbf {\bibinfo
  {volume} {121}},\ \bibinfo {pages} {260601} (\bibinfo {year}
  {2018})}\BibitemShut {NoStop}%
\bibitem [{\citenamefont {Zhong}\ \emph {et~al.}(2024)\citenamefont {Zhong},
  \citenamefont {Liu}, \citenamefont {Zhang}, \citenamefont {Tao},
  \citenamefont {Sun}, \citenamefont {Chu}, \citenamefont {Gong}, \citenamefont
  {Yang},\ and\ \citenamefont {Xiang}}]{Zhong2024xhj}%
  \BibitemOpen
  \bibfield  {author} {\bibinfo {author} {\bibfnamefont {Y.}~\bibnamefont
  {Zhong}}, \bibinfo {author} {\bibfnamefont {S.}~\bibnamefont {Liu}}, \bibinfo
  {author} {\bibfnamefont {B.}~\bibnamefont {Zhang}}, \bibinfo {author}
  {\bibfnamefont {Z.}~\bibnamefont {Tao}}, \bibinfo {author} {\bibfnamefont
  {Y.}~\bibnamefont {Sun}}, \bibinfo {author} {\bibfnamefont {W.}~\bibnamefont
  {Chu}}, \bibinfo {author} {\bibfnamefont {X.-G.}\ \bibnamefont {Gong}},
  \bibinfo {author} {\bibfnamefont {J.-H.}\ \bibnamefont {Yang}},\ and\
  \bibinfo {author} {\bibfnamefont {H.}~\bibnamefont {Xiang}},\ }\bibfield
  {title} {\bibinfo {title} {{Accelerating the calculation of electron--phonon
  coupling strength with machine learning}},\ }\href
  {https://doi.org/10.1038/s43588-024-00668-7} {\bibfield  {journal} {\bibinfo
  {journal} {Nature Computational Science}\ }\textbf {\bibinfo {volume} {4}},\
  \bibinfo {pages} {615} (\bibinfo {year} {2024})}\BibitemShut {NoStop}%
\bibitem [{\citenamefont {Zhang}\ \emph {et~al.}(2024)\citenamefont {Zhang},
  \citenamefont {Zhong}, \citenamefont {Tao}, \citenamefont {Qing},
  \citenamefont {Shang}, \citenamefont {Lan}, \citenamefont {Prezhdo},
  \citenamefont {Gong}, \citenamefont {Chu},\ and\ \citenamefont
  {Xiang}}]{zhang2024advancingnonadiabaticmoleculardynamics_xhj}%
  \BibitemOpen
  \bibfield  {author} {\bibinfo {author} {\bibfnamefont {C.}~\bibnamefont
  {Zhang}}, \bibinfo {author} {\bibfnamefont {Y.}~\bibnamefont {Zhong}},
  \bibinfo {author} {\bibfnamefont {Z.-G.}\ \bibnamefont {Tao}}, \bibinfo
  {author} {\bibfnamefont {X.}~\bibnamefont {Qing}}, \bibinfo {author}
  {\bibfnamefont {H.}~\bibnamefont {Shang}}, \bibinfo {author} {\bibfnamefont
  {Z.}~\bibnamefont {Lan}}, \bibinfo {author} {\bibfnamefont {O.~V.}\
  \bibnamefont {Prezhdo}}, \bibinfo {author} {\bibfnamefont {X.-G.}\
  \bibnamefont {Gong}}, \bibinfo {author} {\bibfnamefont {W.}~\bibnamefont
  {Chu}},\ and\ \bibinfo {author} {\bibfnamefont {H.}~\bibnamefont {Xiang}},\
  }\href {https://arxiv.org/abs/2408.06654} {\bibinfo {title} {Advancing
  nonadiabatic molecular dynamics simulations for solids: Achieving supreme
  accuracy and efficiency with machine learning}} (\bibinfo {year}
  {2024})\BibitemShut {NoStop}%
\bibitem [{\citenamefont {Yu}\ \emph {et~al.}(2022)\citenamefont {Yu},
  \citenamefont {Xu}, \citenamefont {Li}, \citenamefont {Lou}, \citenamefont
  {Bellaiche}, \citenamefont {Hu}, \citenamefont {Gong},\ and\ \citenamefont
  {Xiang}}]{PhysRevB.105.174422spin_ann}%
  \BibitemOpen
  \bibfield  {author} {\bibinfo {author} {\bibfnamefont {H.}~\bibnamefont
  {Yu}}, \bibinfo {author} {\bibfnamefont {C.}~\bibnamefont {Xu}}, \bibinfo
  {author} {\bibfnamefont {X.}~\bibnamefont {Li}}, \bibinfo {author}
  {\bibfnamefont {F.}~\bibnamefont {Lou}}, \bibinfo {author} {\bibfnamefont
  {L.}~\bibnamefont {Bellaiche}}, \bibinfo {author} {\bibfnamefont
  {Z.}~\bibnamefont {Hu}}, \bibinfo {author} {\bibfnamefont {X.}~\bibnamefont
  {Gong}},\ and\ \bibinfo {author} {\bibfnamefont {H.}~\bibnamefont {Xiang}},\
  }\bibfield  {title} {\bibinfo {title} {{Complex spin Hamiltonian represented
  by an artificial neural network}},\ }\href
  {https://doi.org/10.1103/PhysRevB.105.174422} {\bibfield  {journal} {\bibinfo
   {journal} {Phys. Rev. B}\ }\textbf {\bibinfo {volume} {105}},\ \bibinfo
  {pages} {174422} (\bibinfo {year} {2022})}\BibitemShut {NoStop}%
\bibitem [{\citenamefont {Lv}\ \emph {et~al.}(2023)\citenamefont {Lv},
  \citenamefont {Zhong}, \citenamefont {Liang}, \citenamefont {Li},
  \citenamefont {Huang},\ and\ \citenamefont
  {Zheng}}]{PhysRevB.108.235159charge}%
  \BibitemOpen
  \bibfield  {author} {\bibinfo {author} {\bibfnamefont {T.}~\bibnamefont
  {Lv}}, \bibinfo {author} {\bibfnamefont {Z.}~\bibnamefont {Zhong}}, \bibinfo
  {author} {\bibfnamefont {Y.}~\bibnamefont {Liang}}, \bibinfo {author}
  {\bibfnamefont {F.}~\bibnamefont {Li}}, \bibinfo {author} {\bibfnamefont
  {J.}~\bibnamefont {Huang}},\ and\ \bibinfo {author} {\bibfnamefont
  {R.}~\bibnamefont {Zheng}},\ }\bibfield  {title} {\bibinfo {title} {{Deep
  Charge: Deep learning model of electron density from a one-shot density
  functional theory calculation}},\ }\href
  {https://doi.org/10.1103/PhysRevB.108.235159} {\bibfield  {journal} {\bibinfo
   {journal} {Phys. Rev. B}\ }\textbf {\bibinfo {volume} {108}},\ \bibinfo
  {pages} {235159} (\bibinfo {year} {2023})}\BibitemShut {NoStop}%
\bibitem [{\citenamefont {Fan}\ \emph {et~al.}(2022)\citenamefont {Fan},
  \citenamefont {Wang}, \citenamefont {Ying}, \citenamefont {Song},
  \citenamefont {Wang}, \citenamefont {Wang}, \citenamefont {Zeng},
  \citenamefont {Xu}, \citenamefont {Lindgren}, \citenamefont {Rahm},
  \citenamefont {Gabourie}, \citenamefont {Liu}, \citenamefont {Dong},
  \citenamefont {Wu}, \citenamefont {Chen}, \citenamefont {Zhong},
  \citenamefont {Sun}, \citenamefont {Erhart}, \citenamefont {Su},\ and\
  \citenamefont {Ala-Nissila}}]{10.1063/5.0106617GPUMD}%
  \BibitemOpen
  \bibfield  {author} {\bibinfo {author} {\bibfnamefont {Z.}~\bibnamefont
  {Fan}}, \bibinfo {author} {\bibfnamefont {Y.}~\bibnamefont {Wang}}, \bibinfo
  {author} {\bibfnamefont {P.}~\bibnamefont {Ying}}, \bibinfo {author}
  {\bibfnamefont {K.}~\bibnamefont {Song}}, \bibinfo {author} {\bibfnamefont
  {J.}~\bibnamefont {Wang}}, \bibinfo {author} {\bibfnamefont {Y.}~\bibnamefont
  {Wang}}, \bibinfo {author} {\bibfnamefont {Z.}~\bibnamefont {Zeng}}, \bibinfo
  {author} {\bibfnamefont {K.}~\bibnamefont {Xu}}, \bibinfo {author}
  {\bibfnamefont {E.}~\bibnamefont {Lindgren}}, \bibinfo {author}
  {\bibfnamefont {J.~M.}\ \bibnamefont {Rahm}}, \bibinfo {author}
  {\bibfnamefont {A.~J.}\ \bibnamefont {Gabourie}}, \bibinfo {author}
  {\bibfnamefont {J.}~\bibnamefont {Liu}}, \bibinfo {author} {\bibfnamefont
  {H.}~\bibnamefont {Dong}}, \bibinfo {author} {\bibfnamefont {J.}~\bibnamefont
  {Wu}}, \bibinfo {author} {\bibfnamefont {Y.}~\bibnamefont {Chen}}, \bibinfo
  {author} {\bibfnamefont {Z.}~\bibnamefont {Zhong}}, \bibinfo {author}
  {\bibfnamefont {J.}~\bibnamefont {Sun}}, \bibinfo {author} {\bibfnamefont
  {P.}~\bibnamefont {Erhart}}, \bibinfo {author} {\bibfnamefont
  {Y.}~\bibnamefont {Su}},\ and\ \bibinfo {author} {\bibfnamefont
  {T.}~\bibnamefont {Ala-Nissila}},\ }\bibfield  {title} {\bibinfo {title}
  {{GPUMD: A package for constructing accurate machine-learned potentials and
  performing highly efficient atomistic simulations}},\ }\href
  {https://doi.org/10.1063/5.0106617} {\bibfield  {journal} {\bibinfo
  {journal} {The Journal of Chemical Physics}\ }\textbf {\bibinfo {volume}
  {157}},\ \bibinfo {pages} {114801} (\bibinfo {year} {2022})}\BibitemShut
  {NoStop}%
\bibitem [{\citenamefont {Xie}\ \emph {et~al.}(2024{\natexlab{b}})\citenamefont
  {Xie}, \citenamefont {Yang}, \citenamefont {Chen}, \citenamefont {Shi},
  \citenamefont {Kang}, \citenamefont {Ma}, \citenamefont {Li}, \citenamefont
  {Shang},\ and\ \citenamefont {Liu}}]{Xie2024LASP}%
  \BibitemOpen
  \bibfield  {author} {\bibinfo {author} {\bibfnamefont {X.-T.}\ \bibnamefont
  {Xie}}, \bibinfo {author} {\bibfnamefont {Z.-X.}\ \bibnamefont {Yang}},
  \bibinfo {author} {\bibfnamefont {D.}~\bibnamefont {Chen}}, \bibinfo {author}
  {\bibfnamefont {Y.-F.}\ \bibnamefont {Shi}}, \bibinfo {author} {\bibfnamefont
  {P.-L.}\ \bibnamefont {Kang}}, \bibinfo {author} {\bibfnamefont
  {S.}~\bibnamefont {Ma}}, \bibinfo {author} {\bibfnamefont {Y.-F.}\
  \bibnamefont {Li}}, \bibinfo {author} {\bibfnamefont {C.}~\bibnamefont
  {Shang}},\ and\ \bibinfo {author} {\bibfnamefont {Z.-P.}\ \bibnamefont
  {Liu}},\ }\bibfield  {title} {\bibinfo {title} {{LASP to the Future of Atomic
  Simulation: Intelligence and Automation}},\ }\href
  {https://doi.org/10.1021/prechem.4c00060} {\bibfield  {journal} {\bibinfo
  {journal} {Precision Chemistry}\ }\textbf {\bibinfo {volume} {2}},\ \bibinfo
  {pages} {612} (\bibinfo {year} {2024}{\natexlab{b}})}\BibitemShut {NoStop}%
\bibitem [{\citenamefont {Yang}\ \emph {et~al.}(2024)\citenamefont {Yang},
  \citenamefont {Xie}, \citenamefont {Kang}, \citenamefont {Wang},
  \citenamefont {Shang},\ and\ \citenamefont {Liu}}]{Yang2024Attention}%
  \BibitemOpen
  \bibfield  {author} {\bibinfo {author} {\bibfnamefont {Z.-X.}\ \bibnamefont
  {Yang}}, \bibinfo {author} {\bibfnamefont {X.-T.}\ \bibnamefont {Xie}},
  \bibinfo {author} {\bibfnamefont {P.-L.}\ \bibnamefont {Kang}}, \bibinfo
  {author} {\bibfnamefont {Z.-X.}\ \bibnamefont {Wang}}, \bibinfo {author}
  {\bibfnamefont {C.}~\bibnamefont {Shang}},\ and\ \bibinfo {author}
  {\bibfnamefont {Z.-P.}\ \bibnamefont {Liu}},\ }\bibfield  {title} {\bibinfo
  {title} {{Many-Body Function Corrected Neural Network with Atomic Attention
  (MBNN-att) for Molecular Property Prediction}},\ }\href
  {https://doi.org/10.1021/acs.jctc.4c00660} {\bibfield  {journal} {\bibinfo
  {journal} {Journal of Chemical Theory and Computation}\ }\textbf {\bibinfo
  {volume} {20}},\ \bibinfo {pages} {6717} (\bibinfo {year}
  {2024})}\BibitemShut {NoStop}%
\bibitem [{\citenamefont {Barroso-Luque}\ \emph {et~al.}(2024)\citenamefont
  {Barroso-Luque}, \citenamefont {Shuaibi}, \citenamefont {Fu}, \citenamefont
  {Wood}, \citenamefont {Dzamba}, \citenamefont {Gao}, \citenamefont {Rizvi},
  \citenamefont {Zitnick},\ and\ \citenamefont
  {Ulissi}}]{barrosoluque2024openmaterials2024omat24}%
  \BibitemOpen
  \bibfield  {author} {\bibinfo {author} {\bibfnamefont {L.}~\bibnamefont
  {Barroso-Luque}}, \bibinfo {author} {\bibfnamefont {M.}~\bibnamefont
  {Shuaibi}}, \bibinfo {author} {\bibfnamefont {X.}~\bibnamefont {Fu}},
  \bibinfo {author} {\bibfnamefont {B.~M.}\ \bibnamefont {Wood}}, \bibinfo
  {author} {\bibfnamefont {M.}~\bibnamefont {Dzamba}}, \bibinfo {author}
  {\bibfnamefont {M.}~\bibnamefont {Gao}}, \bibinfo {author} {\bibfnamefont
  {A.}~\bibnamefont {Rizvi}}, \bibinfo {author} {\bibfnamefont {C.~L.}\
  \bibnamefont {Zitnick}},\ and\ \bibinfo {author} {\bibfnamefont {Z.~W.}\
  \bibnamefont {Ulissi}},\ }\href {https://arxiv.org/abs/2410.12771} {\bibinfo
  {title} {{Open Materials 2024 (OMat24) Inorganic Materials Dataset and
  Models}}} (\bibinfo {year} {2024})\BibitemShut {NoStop}%
\bibitem [{\citenamefont {Han}\ \emph {et~al.}(2023{\natexlab{b}})\citenamefont
  {Han}, \citenamefont {Wang}, \citenamefont {Ding}, \citenamefont {Gao},
  \citenamefont {Pan}, \citenamefont {Jia},\ and\ \citenamefont
  {Sun}}]{10.1063/5.0142281reconstructions}%
  \BibitemOpen
  \bibfield  {author} {\bibinfo {author} {\bibfnamefont {Y.}~\bibnamefont
  {Han}}, \bibinfo {author} {\bibfnamefont {J.}~\bibnamefont {Wang}}, \bibinfo
  {author} {\bibfnamefont {C.}~\bibnamefont {Ding}}, \bibinfo {author}
  {\bibfnamefont {H.}~\bibnamefont {Gao}}, \bibinfo {author} {\bibfnamefont
  {S.}~\bibnamefont {Pan}}, \bibinfo {author} {\bibfnamefont {Q.}~\bibnamefont
  {Jia}},\ and\ \bibinfo {author} {\bibfnamefont {J.}~\bibnamefont {Sun}},\
  }\bibfield  {title} {\bibinfo {title} {{Prediction of surface reconstructions
  using MAGUS}},\ }\href {https://doi.org/10.1063/5.0142281} {\bibfield
  {journal} {\bibinfo  {journal} {The Journal of Chemical Physics}\ }\textbf
  {\bibinfo {volume} {158}},\ \bibinfo {pages} {174109} (\bibinfo {year}
  {2023}{\natexlab{b}})}\BibitemShut {NoStop}%
\bibitem [{\citenamefont {Pfau}\ \emph {et~al.}(2020)\citenamefont {Pfau},
  \citenamefont {Spencer}, \citenamefont {Matthews},\ and\ \citenamefont
  {Foulkes}}]{PhysRevResearch.2.033429FermiNet}%
  \BibitemOpen
  \bibfield  {author} {\bibinfo {author} {\bibfnamefont {D.}~\bibnamefont
  {Pfau}}, \bibinfo {author} {\bibfnamefont {J.~S.}\ \bibnamefont {Spencer}},
  \bibinfo {author} {\bibfnamefont {A.~G. D.~G.}\ \bibnamefont {Matthews}},\
  and\ \bibinfo {author} {\bibfnamefont {W.~M.~C.}\ \bibnamefont {Foulkes}},\
  }\bibfield  {title} {\bibinfo {title} {{Ab initio solution of the
  many-electron Schr\"odinger equation with deep neural networks}},\ }\href
  {https://doi.org/10.1103/PhysRevResearch.2.033429} {\bibfield  {journal}
  {\bibinfo  {journal} {Phys. Rev. Res.}\ }\textbf {\bibinfo {volume} {2}},\
  \bibinfo {pages} {033429} (\bibinfo {year} {2020})}\BibitemShut {NoStop}%
\bibitem [{sup()}]{supplementary_material}%
  \BibitemOpen
  \href@noop {} {\bibinfo {title} {{Supplemental Material of Effective Field
  Neural Networks}}}\BibitemShut {NoStop}%
\bibitem [{\citenamefont {Zener}(1951)}]{zener1951interaction}%
  \BibitemOpen
  \bibfield  {author} {\bibinfo {author} {\bibfnamefont {C.}~\bibnamefont
  {Zener}},\ }\bibfield  {title} {\bibinfo {title} {{Interaction between the
  $d$-Shells in the Transition Metals. II. Ferromagnetic Compounds of Manganese
  with Perovskite Structure}},\ }\href {https://doi.org/10.1103/PhysRev.82.403}
  {\bibfield  {journal} {\bibinfo  {journal} {Phys. Rev.}\ }\textbf {\bibinfo
  {volume} {82}},\ \bibinfo {pages} {403} (\bibinfo {year} {1951})}\BibitemShut
  {NoStop}%
\bibitem [{\citenamefont {Anderson}\ and\ \citenamefont
  {Hasegawa}(1955)}]{anderson1955considerations}%
  \BibitemOpen
  \bibfield  {author} {\bibinfo {author} {\bibfnamefont {P.~W.}\ \bibnamefont
  {Anderson}}\ and\ \bibinfo {author} {\bibfnamefont {H.}~\bibnamefont
  {Hasegawa}},\ }\bibfield  {title} {\bibinfo {title} {{Considerations on
  Double Exchange}},\ }\href {https://doi.org/10.1103/PhysRev.100.675}
  {\bibfield  {journal} {\bibinfo  {journal} {Phys. Rev.}\ }\textbf {\bibinfo
  {volume} {100}},\ \bibinfo {pages} {675} (\bibinfo {year}
  {1955})}\BibitemShut {NoStop}%
\bibitem [{\citenamefont {de~Gennes}(1960)}]{de1960effects}%
  \BibitemOpen
  \bibfield  {author} {\bibinfo {author} {\bibfnamefont {P.~G.}\ \bibnamefont
  {de~Gennes}},\ }\bibfield  {title} {\bibinfo {title} {{Effects of Double
  Exchange in Magnetic Crystals}},\ }\href
  {https://doi.org/10.1103/PhysRev.118.141} {\bibfield  {journal} {\bibinfo
  {journal} {Phys. Rev.}\ }\textbf {\bibinfo {volume} {118}},\ \bibinfo {pages}
  {141} (\bibinfo {year} {1960})}\BibitemShut {NoStop}%
\bibitem [{\citenamefont {Ruderman}\ and\ \citenamefont
  {Kittel}(1954)}]{ruderman1954indirect}%
  \BibitemOpen
  \bibfield  {author} {\bibinfo {author} {\bibfnamefont {M.~A.}\ \bibnamefont
  {Ruderman}}\ and\ \bibinfo {author} {\bibfnamefont {C.}~\bibnamefont
  {Kittel}},\ }\bibfield  {title} {\bibinfo {title} {{Indirect Exchange
  Coupling of Nuclear Magnetic Moments by Conduction Electrons}},\ }\href
  {https://doi.org/10.1103/PhysRev.96.99} {\bibfield  {journal} {\bibinfo
  {journal} {Phys. Rev.}\ }\textbf {\bibinfo {volume} {96}},\ \bibinfo {pages}
  {99} (\bibinfo {year} {1954})}\BibitemShut {NoStop}%
\bibitem [{\citenamefont {Kasuya}(1956)}]{kasuya1956theory}%
  \BibitemOpen
  \bibfield  {author} {\bibinfo {author} {\bibfnamefont {T.}~\bibnamefont
  {Kasuya}},\ }\bibfield  {title} {\bibinfo {title} {{A Theory of Metallic
  Ferro- and Antiferromagnetism on Zener's Model}},\ }\href
  {https://doi.org/10.1143/PTP.16.45} {\bibfield  {journal} {\bibinfo
  {journal} {Progress of Theoretical Physics}\ }\textbf {\bibinfo {volume}
  {16}},\ \bibinfo {pages} {45} (\bibinfo {year} {1956})}\BibitemShut {NoStop}%
\bibitem [{\citenamefont {Yosida}(1957)}]{yosida1957magnetic}%
  \BibitemOpen
  \bibfield  {author} {\bibinfo {author} {\bibfnamefont {K.}~\bibnamefont
  {Yosida}},\ }\bibfield  {title} {\bibinfo {title} {{Magnetic Properties of
  Cu-Mn Alloys}},\ }\href {https://doi.org/10.1103/PhysRev.106.893} {\bibfield
  {journal} {\bibinfo  {journal} {Phys. Rev.}\ }\textbf {\bibinfo {volume}
  {106}},\ \bibinfo {pages} {893} (\bibinfo {year} {1957})}\BibitemShut
  {NoStop}%
\bibitem [{\citenamefont {Liu}\ \emph {et~al.}(2017{\natexlab{c}})\citenamefont
  {Liu}, \citenamefont {Shen}, \citenamefont {Qi}, \citenamefont {Meng},\ and\
  \citenamefont {Fu}}]{liu2017femion}%
  \BibitemOpen
  \bibfield  {author} {\bibinfo {author} {\bibfnamefont {J.}~\bibnamefont
  {Liu}}, \bibinfo {author} {\bibfnamefont {H.}~\bibnamefont {Shen}}, \bibinfo
  {author} {\bibfnamefont {Y.}~\bibnamefont {Qi}}, \bibinfo {author}
  {\bibfnamefont {Z.~Y.}\ \bibnamefont {Meng}},\ and\ \bibinfo {author}
  {\bibfnamefont {L.}~\bibnamefont {Fu}},\ }\bibfield  {title} {\bibinfo
  {title} {{Self-learning Monte Carlo method and cumulative update in fermion
  systems}},\ }\href {https://doi.org/10.1103/PhysRevB.95.241104} {\bibfield
  {journal} {\bibinfo  {journal} {Phys. Rev. B}\ }\textbf {\bibinfo {volume}
  {95}},\ \bibinfo {pages} {241104} (\bibinfo {year}
  {2017}{\natexlab{c}})}\BibitemShut {NoStop}%
\bibitem [{\citenamefont {Baker}\ and\ \citenamefont
  {Graves-Morris}(1996)}]{Baker_Graves-Morris_1996}%
  \BibitemOpen
  \bibfield  {author} {\bibinfo {author} {\bibfnamefont {G.~A.}\ \bibnamefont
  {Baker}}\ and\ \bibinfo {author} {\bibfnamefont {P.}~\bibnamefont
  {Graves-Morris}},\ }\href {https://doi.org/10.1017/CBO9780511530074} {\emph
  {\bibinfo {title} {{Padé Approximants}}}},\ \bibinfo {edition} {2nd}\ ed.\
  (\bibinfo  {publisher} {Cambridge University Press},\ \bibinfo {address}
  {Cambridge},\ \bibinfo {year} {1996})\BibitemShut {NoStop}%
\bibitem [{\citenamefont {Yukalov}(2019)}]{Yukalov2019_Interplay}%
  \BibitemOpen
  \bibfield  {author} {\bibinfo {author} {\bibfnamefont {V.~I.}\ \bibnamefont
  {Yukalov}},\ }\bibfield  {title} {\bibinfo {title} {{Interplay between
  Approximation Theory and Renormalization Group}},\ }\href
  {https://doi.org/10.1134/S1063779619020047} {\bibfield  {journal} {\bibinfo
  {journal} {Physics of Particles and Nuclei}\ }\textbf {\bibinfo {volume}
  {50}},\ \bibinfo {pages} {141} (\bibinfo {year} {2019})}\BibitemShut
  {NoStop}%
\bibitem [{\citenamefont {Abhignan}\ and\ \citenamefont
  {Sankaranarayanan}(2021)}]{Abhignan2021SimpleTools}%
  \BibitemOpen
  \bibfield  {author} {\bibinfo {author} {\bibfnamefont {V.}~\bibnamefont
  {Abhignan}}\ and\ \bibinfo {author} {\bibfnamefont {R.}~\bibnamefont
  {Sankaranarayanan}},\ }\bibfield  {title} {\bibinfo {title} {{Continued
  Functions and Perturbation Series: Simple Tools for Convergence of Diverging
  Series in O(n)-Symmetric $\phi^4$ Field Theory at Weak Coupling Limit}},\
  }\href {https://doi.org/10.1007/s10955-021-02719-z} {\bibfield  {journal}
  {\bibinfo  {journal} {Journal of Statistical Physics}\ }\textbf {\bibinfo
  {volume} {183}},\ \bibinfo {pages} {4} (\bibinfo {year} {2021})}\BibitemShut
  {NoStop}%
\bibitem [{\citenamefont {Yukalov}\ and\ \citenamefont
  {Yukalova}(1994)}]{YUKALOV1994553_higher_orders}%
  \BibitemOpen
  \bibfield  {author} {\bibinfo {author} {\bibfnamefont {V.}~\bibnamefont
  {Yukalov}}\ and\ \bibinfo {author} {\bibfnamefont {E.}~\bibnamefont
  {Yukalova}},\ }\bibfield  {title} {\bibinfo {title} {{Higher orders of
  self-similar approximations for thermodynamic potentials}},\ }\href
  {https://doi.org/10.1016/0378-4371(94)90324-7} {\bibfield  {journal}
  {\bibinfo  {journal} {Physica A: Statistical Mechanics and its Applications}\
  }\textbf {\bibinfo {volume} {206}},\ \bibinfo {pages} {553} (\bibinfo {year}
  {1994})}\BibitemShut {NoStop}%
\bibitem [{\citenamefont {Yukalov}\ and\ \citenamefont
  {Yukalova}(1993)}]{YUKALOV1993573_Self_similar}%
  \BibitemOpen
  \bibfield  {author} {\bibinfo {author} {\bibfnamefont {V.}~\bibnamefont
  {Yukalov}}\ and\ \bibinfo {author} {\bibfnamefont {E.}~\bibnamefont
  {Yukalova}},\ }\bibfield  {title} {\bibinfo {title} {{Self-similar
  approximations for thermodynamic potentials}},\ }\href
  {https://doi.org/10.1016/0378-4371(93)90241-U} {\bibfield  {journal}
  {\bibinfo  {journal} {Physica A: Statistical Mechanics and its Applications}\
  }\textbf {\bibinfo {volume} {198}},\ \bibinfo {pages} {573} (\bibinfo {year}
  {1993})}\BibitemShut {NoStop}%
\bibitem [{\citenamefont {Yukalov}\ and\ \citenamefont
  {Yukalova}(1999)}]{YUKALOV1999219_Perturbation}%
  \BibitemOpen
  \bibfield  {author} {\bibinfo {author} {\bibfnamefont {V.}~\bibnamefont
  {Yukalov}}\ and\ \bibinfo {author} {\bibfnamefont {E.}~\bibnamefont
  {Yukalova}},\ }\bibfield  {title} {\bibinfo {title} {{Self-Similar
  Perturbation Theory}},\ }\href {https://doi.org/10.1006/aphy.1999.5953}
  {\bibfield  {journal} {\bibinfo  {journal} {Annals of Physics}\ }\textbf
  {\bibinfo {volume} {277}},\ \bibinfo {pages} {219} (\bibinfo {year}
  {1999})}\BibitemShut {NoStop}%
\bibitem [{\citenamefont {Quigg}\ and\ \citenamefont
  {Rosner}(1979)}]{QUIGG1979167_quarkonium}%
  \BibitemOpen
  \bibfield  {author} {\bibinfo {author} {\bibfnamefont {C.}~\bibnamefont
  {Quigg}}\ and\ \bibinfo {author} {\bibfnamefont {J.~L.}\ \bibnamefont
  {Rosner}},\ }\bibfield  {title} {\bibinfo {title} {{Quantum mechanics with
  applications to quarkonium}},\ }\href
  {https://doi.org/10.1016/0370-1573(79)90095-4} {\bibfield  {journal}
  {\bibinfo  {journal} {Physics Reports}\ }\textbf {\bibinfo {volume} {56}},\
  \bibinfo {pages} {167} (\bibinfo {year} {1979})}\BibitemShut {NoStop}%
\bibitem [{\citenamefont {Yukalov}\ and\ \citenamefont
  {Gluzman}(1997)}]{PhysRevE.55.6552_bootstrap}%
  \BibitemOpen
  \bibfield  {author} {\bibinfo {author} {\bibfnamefont {V.~I.}\ \bibnamefont
  {Yukalov}}\ and\ \bibinfo {author} {\bibfnamefont {S.}~\bibnamefont
  {Gluzman}},\ }\bibfield  {title} {\bibinfo {title} {{Self-similar bootstrap
  of divergent series}},\ }\href {https://doi.org/10.1103/PhysRevE.55.6552}
  {\bibfield  {journal} {\bibinfo  {journal} {Physical Review E}\ }\textbf
  {\bibinfo {volume} {55}},\ \bibinfo {pages} {6552} (\bibinfo {year}
  {1997})}\BibitemShut {NoStop}%
\bibitem [{\citenamefont {Gluzman}\ and\ \citenamefont
  {Yukalov}(2017)}]{Gluzman2017_Critical}%
  \BibitemOpen
  \bibfield  {author} {\bibinfo {author} {\bibfnamefont {S.}~\bibnamefont
  {Gluzman}}\ and\ \bibinfo {author} {\bibfnamefont {V.~I.}\ \bibnamefont
  {Yukalov}},\ }\bibfield  {title} {\bibinfo {title} {{Critical indices from
  self-similar root approximants}},\ }\href
  {https://doi.org/10.1140/epjp/i2017-11820-2} {\bibfield  {journal} {\bibinfo
  {journal} {The European Physical Journal Plus}\ }\textbf {\bibinfo {volume}
  {132}},\ \bibinfo {pages} {535} (\bibinfo {year} {2017})}\BibitemShut
  {NoStop}%
\bibitem [{\citenamefont {Yukalov}\ and\ \citenamefont
  {Yukalova}(2017{\natexlab{a}})}]{Yukalov2017_epj}%
  \BibitemOpen
  \bibfield  {author} {\bibinfo {author} {\bibfnamefont {V.}~\bibnamefont
  {Yukalov}}\ and\ \bibinfo {author} {\bibfnamefont {E.}~\bibnamefont
  {Yukalova}},\ }\bibfield  {title} {\bibinfo {title} {{Critical temperature in
  weakly interacting multicomponent field theory}},\ }in\ \href
  {https://doi.org/10.1051/epjconf/201713803011} {\emph {\bibinfo {booktitle}
  {EPJ Web of Conferences}}},\ Vol.\ \bibinfo {volume} {138}\ (\bibinfo
  {publisher} {EDP Sciences},\ \bibinfo {year} {2017})\ p.\ \bibinfo {pages}
  {03011}\BibitemShut {NoStop}%
\bibitem [{\citenamefont {Yukalov}\ and\ \citenamefont
  {Yukalova}(2017{\natexlab{b}})}]{Yukalov_2017_bose}%
  \BibitemOpen
  \bibfield  {author} {\bibinfo {author} {\bibfnamefont {V.~I.}\ \bibnamefont
  {Yukalov}}\ and\ \bibinfo {author} {\bibfnamefont {E.~P.}\ \bibnamefont
  {Yukalova}},\ }\bibfield  {title} {\bibinfo {title} {{Bose--Einstein
  condensation temperature of weakly interacting atoms}},\ }\href
  {https://doi.org/10.1088/1612-202X/aa6eed} {\bibfield  {journal} {\bibinfo
  {journal} {Laser Physics Letters}\ }\textbf {\bibinfo {volume} {14}},\
  \bibinfo {pages} {073001} (\bibinfo {year} {2017}{\natexlab{b}})}\BibitemShut
  {NoStop}%
\bibitem [{\citenamefont {Yukalov}\ and\ \citenamefont
  {Yukalova}(1996)}]{YUKALOV1996336_Temporal}%
  \BibitemOpen
  \bibfield  {author} {\bibinfo {author} {\bibfnamefont {V.}~\bibnamefont
  {Yukalov}}\ and\ \bibinfo {author} {\bibfnamefont {E.}~\bibnamefont
  {Yukalova}},\ }\bibfield  {title} {\bibinfo {title} {{Temporal dynamics in
  perturbation theory}},\ }\href {https://doi.org/10.1016/0378-4371(95)00471-8}
  {\bibfield  {journal} {\bibinfo  {journal} {Physica A: Statistical Mechanics
  and its Applications}\ }\textbf {\bibinfo {volume} {225}},\ \bibinfo {pages}
  {336} (\bibinfo {year} {1996})}\BibitemShut {NoStop}%
\bibitem [{\citenamefont {Yukalova}\ and\ \citenamefont
  {Yukalov}(1992)}]{Yukalova1992_Dubna}%
  \BibitemOpen
  \bibfield  {author} {\bibinfo {author} {\bibfnamefont {E.~P.}\ \bibnamefont
  {Yukalova}}\ and\ \bibinfo {author} {\bibfnamefont {V.~I.}\ \bibnamefont
  {Yukalov}},\ }\href@noop {} {\emph {\bibinfo {title} {{Calculation of
  eigenvalues of {S}chr\"{o}dinger operators for arbitrary coupling}}}},\
  \bibinfo {type} {Tech. Rep.}\ (\bibinfo  {institution} {Joint Institute for
  Nuclear Research (JINR)},\ \bibinfo {address} {Dubna, Russia},\ \bibinfo
  {year} {1992})\BibitemShut {NoStop}%
\bibitem [{\citenamefont {Gluzman}\ and\ \citenamefont
  {Yukalov}(2015)}]{math3020510}%
  \BibitemOpen
  \bibfield  {author} {\bibinfo {author} {\bibfnamefont {S.}~\bibnamefont
  {Gluzman}}\ and\ \bibinfo {author} {\bibfnamefont {V.~I.}\ \bibnamefont
  {Yukalov}},\ }\bibfield  {title} {\bibinfo {title} {{Effective Summation and
  Interpolation of Series by Self-Similar Root Approximants}},\ }\href
  {https://doi.org/10.3390/math3020510} {\bibfield  {journal} {\bibinfo
  {journal} {Mathematics}\ }\textbf {\bibinfo {volume} {3}},\ \bibinfo {pages}
  {510} (\bibinfo {year} {2015})}\BibitemShut {NoStop}%
\bibitem [{\citenamefont {Yukalov}\ and\ \citenamefont
  {Yukalova}(2015)}]{Yukalov:2015eh_multicomponent}%
  \BibitemOpen
  \bibfield  {author} {\bibinfo {author} {\bibfnamefont {V.~I.}\ \bibnamefont
  {Yukalov}}\ and\ \bibinfo {author} {\bibfnamefont {E.}~\bibnamefont
  {Yukalova}},\ }\bibfield  {title} {\bibinfo {title} {{Phase transition in
  multicomponent field theory at finite temperature}},\ }in\ \href
  {https://doi.org/10.22323/1.225.0080} {\emph {\bibinfo {booktitle} {PoS}}},\
  Vol.\ \bibinfo {volume} {Baldin ISHEPP XXII}\ (\bibinfo  {publisher} {Sissa
  Medialab},\ \bibinfo {year} {2015})\ p.\ \bibinfo {pages} {080}\BibitemShut
  {NoStop}%
\bibitem [{\citenamefont {Yukalov}\ and\ \citenamefont
  {Yukalova}(2007)}]{Yukalov2007_Calculation}%
  \BibitemOpen
  \bibfield  {author} {\bibinfo {author} {\bibfnamefont {V.~I.}\ \bibnamefont
  {Yukalov}}\ and\ \bibinfo {author} {\bibfnamefont {E.~P.}\ \bibnamefont
  {Yukalova}},\ }\bibfield  {title} {\bibinfo {title} {{Calculation of critical
  exponents by self-similar factor approximants}},\ }\href
  {https://doi.org/10.1140/epjb/e2007-00044-4} {\bibfield  {journal} {\bibinfo
  {journal} {The European Physical Journal B}\ }\textbf {\bibinfo {volume}
  {55}},\ \bibinfo {pages} {93} (\bibinfo {year} {2007})}\BibitemShut {NoStop}%
\bibitem [{\citenamefont {Yukalov}\ and\ \citenamefont
  {Yukalova}(2000{\natexlab{a}})}]{yukalov2000equationstatequantumchromodynamics}%
  \BibitemOpen
  \bibfield  {author} {\bibinfo {author} {\bibfnamefont {V.~I.}\ \bibnamefont
  {Yukalov}}\ and\ \bibinfo {author} {\bibfnamefont {E.~P.}\ \bibnamefont
  {Yukalova}},\ }\href {https://arxiv.org/abs/hep-ph/0010028} {\bibinfo {title}
  {{Equation of State in Quantum Chromodynamics}}} (\bibinfo {year}
  {2000}{\natexlab{a}})\BibitemShut {NoStop}%
\bibitem [{\citenamefont {Yukalova}\ \emph {et~al.}(2008)\citenamefont
  {Yukalova}, \citenamefont {Yukalov},\ and\ \citenamefont
  {Gluzman}}]{YUKALOVA20083074_equations}%
  \BibitemOpen
  \bibfield  {author} {\bibinfo {author} {\bibfnamefont {E.}~\bibnamefont
  {Yukalova}}, \bibinfo {author} {\bibfnamefont {V.}~\bibnamefont {Yukalov}},\
  and\ \bibinfo {author} {\bibfnamefont {S.}~\bibnamefont {Gluzman}},\
  }\bibfield  {title} {\bibinfo {title} {{Self-similar factor approximants for
  evolution equations and boundary-value problems}},\ }\href
  {https://doi.org/10.1016/j.aop.2008.05.009} {\bibfield  {journal} {\bibinfo
  {journal} {Annals of Physics}\ }\textbf {\bibinfo {volume} {323}},\ \bibinfo
  {pages} {3074} (\bibinfo {year} {2008})}\BibitemShut {NoStop}%
\bibitem [{\citenamefont {Yukalov}(2000)}]{YUKALOV_2000_MPLB}%
  \BibitemOpen
  \bibfield  {author} {\bibinfo {author} {\bibfnamefont {V.}~\bibnamefont
  {Yukalov}},\ }\bibfield  {title} {\bibinfo {title} {{Self-similar
  extrapolation of asymptotic series and forecasting for time series}},\ }\href
  {https://doi.org/10.1142/S0217984900000999} {\bibfield  {journal} {\bibinfo
  {journal} {Modern Physics Letters B}\ }\textbf {\bibinfo {volume} {14}},\
  \bibinfo {pages} {791} (\bibinfo {year} {2000})}\BibitemShut {NoStop}%
\bibitem [{\citenamefont {Gluzman}\ and\ \citenamefont
  {Yukalov}(1998{\natexlab{a}})}]{Yukalov_Gluzman_1998_Resummation_MPLB}%
  \BibitemOpen
  \bibfield  {author} {\bibinfo {author} {\bibfnamefont {S.}~\bibnamefont
  {Gluzman}}\ and\ \bibinfo {author} {\bibfnamefont {V.}~\bibnamefont
  {Yukalov}},\ }\bibfield  {title} {\bibinfo {title} {{Resummation Methods for
  Analyzing Time Series}},\ }\href {https://doi.org/10.1142/S021798499800010X}
  {\bibfield  {journal} {\bibinfo  {journal} {Modern Physics Letters B}\
  }\textbf {\bibinfo {volume} {12}},\ \bibinfo {pages} {61} (\bibinfo {year}
  {1998}{\natexlab{a}})}\BibitemShut {NoStop}%
\bibitem [{\citenamefont {Gluzman}\ and\ \citenamefont
  {Yukalov}(1998{\natexlab{b}})}]{Renormalization_group_october}%
  \BibitemOpen
  \bibfield  {author} {\bibinfo {author} {\bibfnamefont {S.}~\bibnamefont
  {Gluzman}}\ and\ \bibinfo {author} {\bibfnamefont {V.}~\bibnamefont
  {Yukalov}},\ }\bibfield  {title} {\bibinfo {title} {{Renormalization Group
  Analysis of October Market Crashes}},\ }\href
  {https://doi.org/10.1142/S0217984998000111} {\bibfield  {journal} {\bibinfo
  {journal} {Modern Physics Letters B}\ }\textbf {\bibinfo {volume} {12}},\
  \bibinfo {pages} {75} (\bibinfo {year} {1998}{\natexlab{b}})}\BibitemShut
  {NoStop}%
\bibitem [{\citenamefont {Gluzman}\ and\ \citenamefont
  {Yukalov}(1998{\natexlab{c}})}]{boom_crashes_1998}%
  \BibitemOpen
  \bibfield  {author} {\bibinfo {author} {\bibfnamefont {S.}~\bibnamefont
  {Gluzman}}\ and\ \bibinfo {author} {\bibfnamefont {V.}~\bibnamefont
  {Yukalov}},\ }\href {http://arxiv.org/abs/cond-mat/9810092} {\bibinfo {title}
  {{Booms and Crashes in Self-Similar Markets}}} (\bibinfo {year}
  {1998}{\natexlab{c}})\BibitemShut {NoStop}%
\bibitem [{\citenamefont {Yukalov}\ and\ \citenamefont
  {Gluzman}(1999)}]{weighted_fixed_points_1999}%
  \BibitemOpen
  \bibfield  {author} {\bibinfo {author} {\bibfnamefont {V.}~\bibnamefont
  {Yukalov}}\ and\ \bibinfo {author} {\bibfnamefont {S.}~\bibnamefont
  {Gluzman}},\ }\bibfield  {title} {\bibinfo {title} {{Weighted Fixed Points in
  Self-Similar Analysis of Time Series}},\ }\href
  {https://doi.org/10.1142/S021797929900151X} {\bibfield  {journal} {\bibinfo
  {journal} {International Journal of Modern Physics B}\ }\textbf {\bibinfo
  {volume} {13}},\ \bibinfo {pages} {1463} (\bibinfo {year}
  {1999})}\BibitemShut {NoStop}%
\bibitem [{\citenamefont
  {Yukalov}(2001{\natexlab{a}})}]{Yukalov2001_market_analyse}%
  \BibitemOpen
  \bibfield  {author} {\bibinfo {author} {\bibfnamefont {V.}~\bibnamefont
  {Yukalov}},\ }\bibfield  {title} {\bibinfo {title} {{Self-Similar Approach to
  Market Analysis}},\ }\href {https://doi.org/10.1007/PL00011115} {\bibfield
  {journal} {\bibinfo  {journal} {The European Physical Journal B - Condensed
  Matter and Complex Systems}\ }\textbf {\bibinfo {volume} {20}},\ \bibinfo
  {pages} {609} (\bibinfo {year} {2001}{\natexlab{a}})}\BibitemShut {NoStop}%
\bibitem [{\citenamefont {Gluzman}\ \emph {et~al.}(2003)\citenamefont
  {Gluzman}, \citenamefont {Sornette},\ and\ \citenamefont
  {Yukalov}}]{Reconstructing_generalized_exp_laws_2003}%
  \BibitemOpen
  \bibfield  {author} {\bibinfo {author} {\bibfnamefont {S.}~\bibnamefont
  {Gluzman}}, \bibinfo {author} {\bibfnamefont {D.}~\bibnamefont {Sornette}},\
  and\ \bibinfo {author} {\bibfnamefont {V.}~\bibnamefont {Yukalov}},\
  }\bibfield  {title} {\bibinfo {title} {{Reconstructing Generalized
  Exponential Laws By Self-Similar Exponential Approximants}},\ }\href
  {https://doi.org/10.1142/S012918310300470X} {\bibfield  {journal} {\bibinfo
  {journal} {International Journal of Modern Physics C}\ }\textbf {\bibinfo
  {volume} {14}},\ \bibinfo {pages} {509} (\bibinfo {year} {2003})}\BibitemShut
  {NoStop}%
\bibitem [{\citenamefont {Yukalov}\ and\ \citenamefont
  {Yukalova}(2000{\natexlab{b}})}]{yukalov2000cooperativeelectromagneticeffects}%
  \BibitemOpen
  \bibfield  {author} {\bibinfo {author} {\bibfnamefont {V.~I.}\ \bibnamefont
  {Yukalov}}\ and\ \bibinfo {author} {\bibfnamefont {E.~P.}\ \bibnamefont
  {Yukalova}},\ }\href {https://arxiv.org/abs/cond-mat/0006159} {\bibinfo
  {title} {{Cooperative Electromagnetic Effects}}} (\bibinfo {year}
  {2000}{\natexlab{b}})\BibitemShut {NoStop}%
\bibitem [{\citenamefont
  {Yukalov}(2001{\natexlab{b}})}]{YUKALOV200191_nonequilibrium_phenomena}%
  \BibitemOpen
  \bibfield  {author} {\bibinfo {author} {\bibfnamefont {V.}~\bibnamefont
  {Yukalov}},\ }\bibfield  {title} {\bibinfo {title} {{Principle of Pattern
  Selection for Nonequilibrium Phenomena}},\ }\href
  {https://doi.org/10.1016/S0375-9601(01)00281-X} {\bibfield  {journal}
  {\bibinfo  {journal} {Physics Letters A}\ }\textbf {\bibinfo {volume}
  {284}},\ \bibinfo {pages} {91} (\bibinfo {year}
  {2001}{\natexlab{b}})}\BibitemShut {NoStop}%
\bibitem [{\citenamefont {Liu}\ \emph {et~al.}(2025)\citenamefont {Liu},
  \citenamefont {Zhao}, \citenamefont {Wan}, \citenamefont {Zhang},\ and\
  \citenamefont {Liu}}]{efnn2026data}%
  \BibitemOpen
  \bibfield  {author} {\bibinfo {author} {\bibfnamefont {X.}~\bibnamefont
  {Liu}}, \bibinfo {author} {\bibfnamefont {Y.}~\bibnamefont {Zhao}}, \bibinfo
  {author} {\bibfnamefont {C.~Y.}\ \bibnamefont {Wan}}, \bibinfo {author}
  {\bibfnamefont {Y.}~\bibnamefont {Zhang}},\ and\ \bibinfo {author}
  {\bibfnamefont {J.}~\bibnamefont {Liu}},\ }\href
  {https://doi.org/10.5281/zenodo.19081463} {\bibinfo {title} {Data for:
  Renormalization-inspired effective field neural networks for scalable
  modeling of classical and quantum many-body systems}} (\bibinfo {year}
  {2025})\BibitemShut {NoStop}%
\end{thebibliography}%


\begin{thebibliography}{8}%
\makeatletter
\providecommand \@ifxundefined [1]{%
 \@ifx{#1\undefined}
}%
\providecommand \@ifnum [1]{%
 \ifnum #1\expandafter \@firstoftwo
 \else \expandafter \@secondoftwo
 \fi
}%
\providecommand \@ifx [1]{%
 \ifx #1\expandafter \@firstoftwo
 \else \expandafter \@secondoftwo
 \fi
}%
\providecommand \natexlab [1]{#1}%
\providecommand \enquote  [1]{``#1''}%
\providecommand \bibnamefont  [1]{#1}%
\providecommand \bibfnamefont [1]{#1}%
\providecommand \citenamefont [1]{#1}%
\providecommand \href@noop [0]{\@secondoftwo}%
\providecommand \href [0]{\begingroup \@sanitize@url \@href}%
\providecommand \@href[1]{\@@startlink{#1}\@@href}%
\providecommand \@@href[1]{\endgroup#1\@@endlink}%
\providecommand \@sanitize@url [0]{\catcode `\\12\catcode `\$12\catcode
  `\&12\catcode `\#12\catcode `\^12\catcode `\_12\catcode `\%12\relax}%
\providecommand \@@startlink[1]{}%
\providecommand \@@endlink[0]{}%
\providecommand \url  [0]{\begingroup\@sanitize@url \@url }%
\providecommand \@url [1]{\endgroup\@href {#1}{\urlprefix }}%
\providecommand \urlprefix  [0]{URL }%
\providecommand \Eprint [0]{\href }%
\providecommand \doibase [0]{https://doi.org/}%
\providecommand \selectlanguage [0]{\@gobble}%
\providecommand \bibinfo  [0]{\@secondoftwo}%
\providecommand \bibfield  [0]{\@secondoftwo}%
\providecommand \translation [1]{[#1]}%
\providecommand \BibitemOpen [0]{}%
\providecommand \bibitemStop [0]{}%
\providecommand \bibitemNoStop [0]{.\EOS\space}%
\providecommand \EOS [0]{\spacefactor3000\relax}%
\providecommand \BibitemShut  [1]{\csname bibitem#1\endcsname}%
\let\auto@bib@innerbib\@empty
%</preamble>
\bibitem [{\citenamefont {Al-Mohy}\ and\ \citenamefont
  {Higham}(2010)}]{expm_Higham_2010}%
  \BibitemOpen
  \bibfield  {author} {\bibinfo {author} {\bibfnamefont {A.~H.}\ \bibnamefont
  {Al-Mohy}}\ and\ \bibinfo {author} {\bibfnamefont {N.~J.}\ \bibnamefont
  {Higham}},\ }\bibfield  {title} {\bibinfo {title} {{A New Scaling and
  Squaring Algorithm for the Matrix Exponential}},\ }\href
  {https://doi.org/10.1137/09074721X} {\bibfield  {journal} {\bibinfo
  {journal} {SIAM Journal on Matrix Analysis and Applications}\ }\textbf
  {\bibinfo {volume} {31}},\ \bibinfo {pages} {970} (\bibinfo {year}
  {2010})}\BibitemShut {NoStop}%
\bibitem [{\citenamefont {Bender}\ and\ \citenamefont
  {Orszag}(1999)}]{bender1999advanced}%
  \BibitemOpen
  \bibfield  {author} {\bibinfo {author} {\bibfnamefont {C.~M.}\ \bibnamefont
  {Bender}}\ and\ \bibinfo {author} {\bibfnamefont {S.~A.}\ \bibnamefont
  {Orszag}},\ }\href
  {https://doi.org/https://doi.org/10.1007/978-1-4757-3069-2} {\emph {\bibinfo
  {title} {{Advanced Mathematical Methods for Scientists and Engineers I}}}},\
  \bibinfo {edition} {1st}\ ed.\ (\bibinfo  {publisher} {Springer New York},\
  \bibinfo {address} {New York, NY},\ \bibinfo {year} {1999})\BibitemShut
  {NoStop}%
\bibitem [{\citenamefont {Yukalov}\ and\ \citenamefont
  {Yukalova}(1994)}]{YUKALOV1994553_higher_orders}%
  \BibitemOpen
  \bibfield  {author} {\bibinfo {author} {\bibfnamefont {V.}~\bibnamefont
  {Yukalov}}\ and\ \bibinfo {author} {\bibfnamefont {E.}~\bibnamefont
  {Yukalova}},\ }\bibfield  {title} {\bibinfo {title} {{Higher Orders of
  Self-Similar Approximations for Thermodynamic Potentials}},\ }\href
  {https://doi.org/10.1016/0378-4371(94)90324-7} {\bibfield  {journal}
  {\bibinfo  {journal} {Physica A: Statistical Mechanics and its Applications}\
  }\textbf {\bibinfo {volume} {206}},\ \bibinfo {pages} {553} (\bibinfo {year}
  {1994})}\BibitemShut {NoStop}%
\bibitem [{\citenamefont {Yukalov}\ and\ \citenamefont
  {Gluzman}(1997)}]{PhysRevE.55.6552_bootstrap}%
  \BibitemOpen
  \bibfield  {author} {\bibinfo {author} {\bibfnamefont {V.}~\bibnamefont
  {Yukalov}}\ and\ \bibinfo {author} {\bibfnamefont {S.}~\bibnamefont
  {Gluzman}},\ }\bibfield  {title} {\bibinfo {title} {{Self-Similar Bootstrap
  of Divergent Series}},\ }\href {https://doi.org/10.1103/PhysRevE.55.6552}
  {\bibfield  {journal} {\bibinfo  {journal} {Physical Review E}\ }\textbf
  {\bibinfo {volume} {55}},\ \bibinfo {pages} {6552} (\bibinfo {year}
  {1997})}\BibitemShut {NoStop}%
\bibitem [{\citenamefont {Yukalov}\ and\ \citenamefont
  {Gluzman}(1998)}]{PhysRevE.58.1359_exponential}%
  \BibitemOpen
  \bibfield  {author} {\bibinfo {author} {\bibfnamefont {V.}~\bibnamefont
  {Yukalov}}\ and\ \bibinfo {author} {\bibfnamefont {S.}~\bibnamefont
  {Gluzman}},\ }\bibfield  {title} {\bibinfo {title} {{Self-Similar Exponential
  Approximants}},\ }\href {https://doi.org/10.1103/PhysRevE.58.1359} {\bibfield
   {journal} {\bibinfo  {journal} {Physical Review E}\ }\textbf {\bibinfo
  {volume} {58}},\ \bibinfo {pages} {1359} (\bibinfo {year}
  {1998})}\BibitemShut {NoStop}%
\bibitem [{\citenamefont {Yukalov}\ and\ \citenamefont
  {Yukalova}(2002)}]{YUKALOV2002839_fractal}%
  \BibitemOpen
  \bibfield  {author} {\bibinfo {author} {\bibfnamefont {V.}~\bibnamefont
  {Yukalov}}\ and\ \bibinfo {author} {\bibfnamefont {E.}~\bibnamefont
  {Yukalova}},\ }\bibfield  {title} {\bibinfo {title} {{Self-Similar Structures
  and Fractal Transforms in Approximation Theory}},\ }\href
  {https://doi.org/10.1016/S0960-0779(02)00029-2} {\bibfield  {journal}
  {\bibinfo  {journal} {Chaos, Solitons \& Fractals}\ }\textbf {\bibinfo
  {volume} {14}},\ \bibinfo {pages} {839} (\bibinfo {year} {2002})}\BibitemShut
  {NoStop}%
\bibitem [{\citenamefont {Bai}(2009)}]{bai_dqmc_notes}%
  \BibitemOpen
  \bibfield  {author} {\bibinfo {author} {\bibfnamefont {Z.}~\bibnamefont
  {Bai}},\ }\href {https://www.cs.ucdavis.edu/~bai/lnqmc.pdf} {\bibinfo {title}
  {Lecture notes on determinant quantum monte carlo (dqmc)}},\ \bibinfo
  {howpublished} {University of California, Davis, Department of Computer
  Science} (\bibinfo {year} {2009}),\ \bibinfo {note} {accessed:
  2023-10-27}\BibitemShut {NoStop}%
\bibitem [{\citenamefont {LOH}\ and\ \citenamefont
  {GUBERNATIS}(1992)}]{LOH1992177}%
  \BibitemOpen
  \bibfield  {author} {\bibinfo {author} {\bibfnamefont {E.}~\bibnamefont
  {LOH}}\ and\ \bibinfo {author} {\bibfnamefont {J.}~\bibnamefont
  {GUBERNATIS}},\ }\bibfield  {title} {\bibinfo {title} {Chapter 4 - stable
  numerical simulations of models of interacting electrons in condensed-matter
  physics},\ }in\ \href
  {https://doi.org/https://doi.org/10.1016/B978-0-444-88885-3.50009-3} {\emph
  {\bibinfo {booktitle} {Electronic Phase Transitions}}},\ \bibinfo {series}
  {Modern Problems in Condensed Matter Sciences}, Vol.~\bibinfo {volume} {32},\
  \bibinfo {editor} {edited by\ \bibinfo {editor} {\bibfnamefont
  {W.}~\bibnamefont {HANKE}}\ and\ \bibinfo {editor} {\bibfnamefont
  {Y.}~\bibnamefont {KOPAEV}}}\ (\bibinfo  {publisher} {Elsevier},\ \bibinfo
  {year} {1992})\ pp.\ \bibinfo {pages} {177--235}\BibitemShut {NoStop}%
\end{thebibliography}%
%TC:endignore

\end{document}

\setlength{\parindent}{0pt}
\setlength{\parskip}{1em} % Adjust the value as needed
%\preprint{APS}

\title{Supplemental Information of Effective Field  Neural Networks }
\maketitle

In this supplementary information, we present the complete workflow of EFNN with symmetrization for the quantum double exchange model in Section~\ref{sec_qt_conv}. We introduce the basics of Pad\'{e} approximation, continued fractions, and continued functions in Section~\ref{sec_pade}, and demonstrate that EFNN implements a continued function, while ResNet, DenseNet and DNN are not continued functions. In Section~\ref{sec_rg}, we derive the structure of EFNN from the renormalization group (RG) equation. In Section~\ref{sec_symmetrization}, we show that the convolution layer ensures the system's invariance under the point group $\operatorname{O}(3)$. In Section~\ref{sec_adapted_structures}, we present the adapted architectures of ResNet, DenseNet, and DNN for energy fitting of the quantum double exchange model. In Section~\ref{sec_DQMC_details}, we break down the details of DQMC, and demonstrate how EFNN solves the computational bottleneck problem in DQMC for quantum double exchange model. We report the speed-up of EFNN against ED. In Section~\ref{sec_data_distribution}, we report the distributions of EFNN's predictions, and the ground truth data, for quantum double exchange model, demonstrating that the accuracy of EFNN.

\section{Computational workflow of  EFNNs with symmetrization}
\label{sec_qt_conv}

\begin{figure}[hbtp!]
    \centering
    \includegraphics[width=1\linewidth]{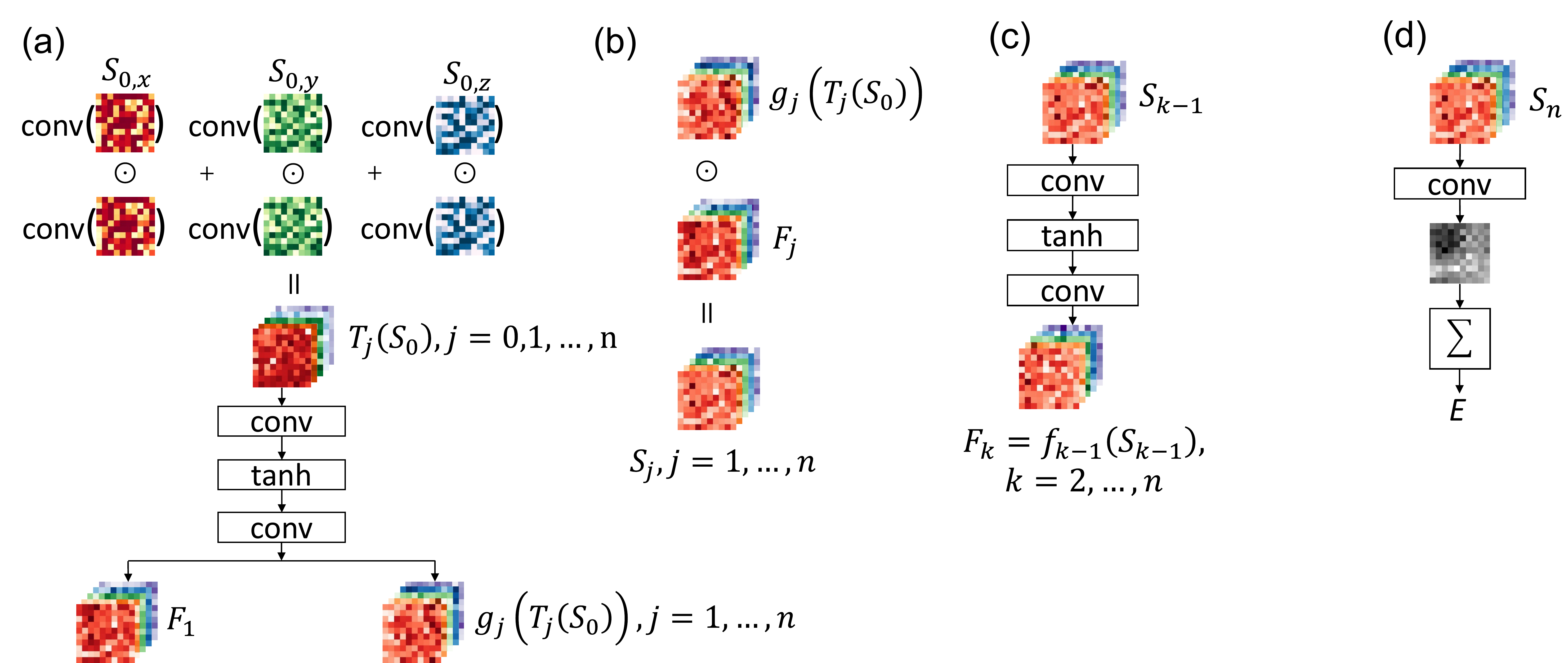}
    \caption{Computational workflow of  EFNNs with symmetrization}
    \label{fig_conv_details}
\end{figure}

In this section, we present the complete workflow of EFNN with symmetrization for the quantum double exchange model.
Fig.\ref{fig_conv_details}(a) illustrates the symmetrization process: each of the three channels of $S_{0}$ ($S_{0,x}$, $S_{0,y}$, $S_{0,z}$) undergoes convolution and element-wise multiplication within the same channel. The resulting products are summed to form the symmetrization layer $T_{j}(S_{0})$. Subsequently, $T_{j}(S_{0})$ is passed through a sequence of layers---convolution, $\tanh$ activation, and convolution---to produce $F_{1}$ for $j=0$ or $g_{j}(T_{j}(S_{0}))$ for $j=1,\ldots,n$. Fig.~\ref{fig_conv_details}(a) is Fig.~3 in the main text. Fig.~\ref{fig_conv_details}(b) shows the generation of quasi-particle layers: the transformed symmetrization layer $g_{j}(T_{j}(S_{0}))$ is element-wise multiplied with the effective field layer $F_{j}$, yielding the quasi-particle layer $S_{j}$ for $j=1,\ldots,n$.  Fig.~\ref{fig_conv_details}(c) depicts the transformation to effective field layers: each quasi-particle layer $S_{k-1}$ is processed through convolution, $\tanh$ activation, and convolution to generate the effective field layer $F_{k}$ for $k=2,\ldots,n$. Fig.~\ref{fig_conv_details}(d) illustrates the energy evaluation: the final quasi-particle layer $S_{n}$ is transformed into a single-channel matrix, and an element-wise summation of this matrix yields the scalar energy prediction $E$. These visualizations are generated after training the model with $C=5$ channels and $n=2$ layers, using the 576th sample from the test set as input $S_{0}$. 

\section{Mathematical details of DenseNet, ResNet, DNN}
\label{sec_pade}
In this section, we describe the mathematical structures of EFNN, ResNet, DenseNet, and DNN.

We denote the input spins as $S_{0}\in\mathbb{R}^{N}$.

\begin{figure}[hbtp!]
    \centering
    \includegraphics[width=0.7\linewidth]{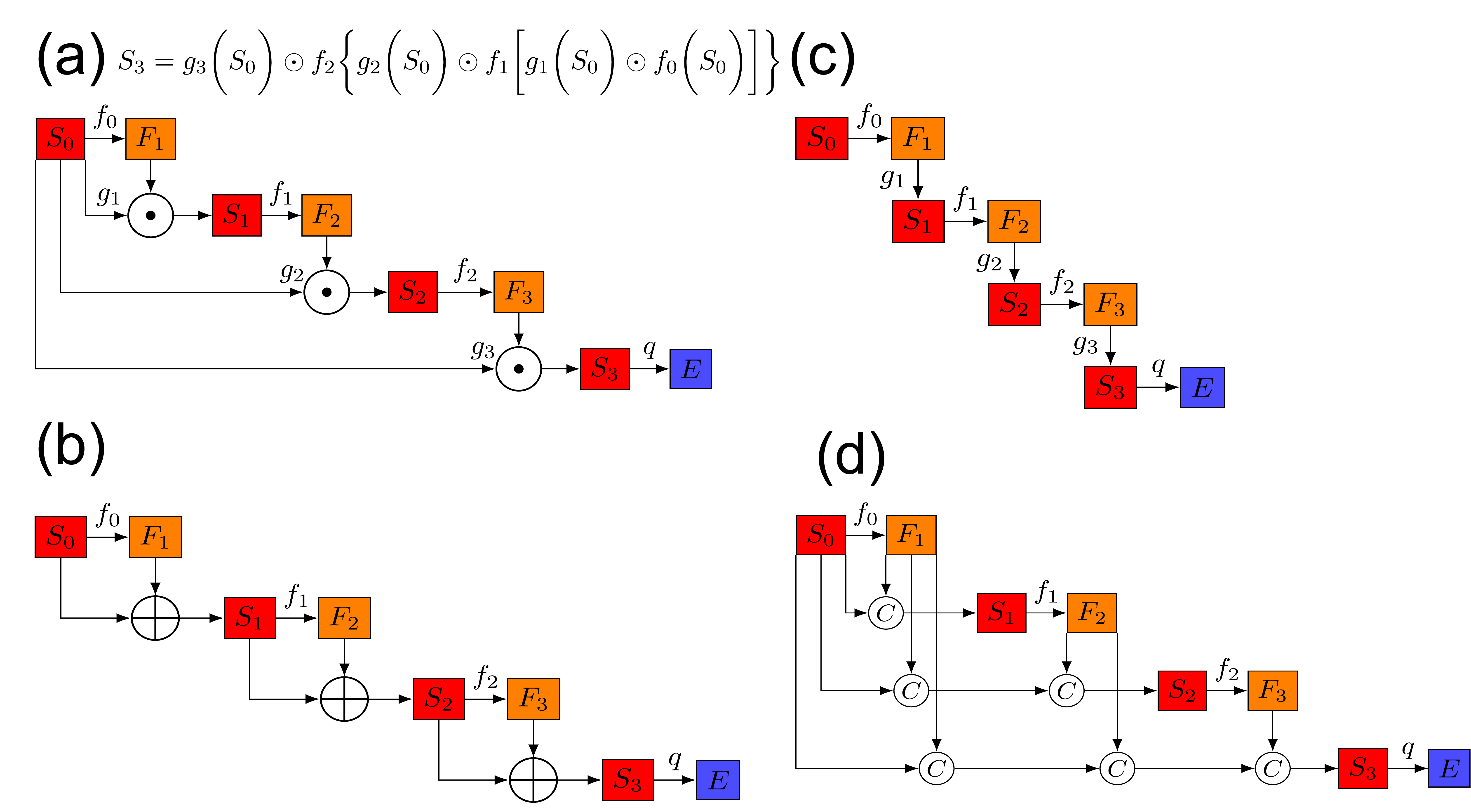}
    \caption{Comparison of structures of (a) EFNN, (b) ResNet, (c) DNN, and (d) DenseNet}
    \label{fig_comp_nets}
\end{figure}

\noindent\textbf{Why EFNN is a continued function}

In Fig.~\ref{fig_comp_nets}(a), the functions $g_{k}$, $k=1,2,3$, $f_{j}$, $j=0,1,2$, all have the forms of a composition of a linear mapping $\circ$ tanh $\circ$ linear mapping. We denote
\begin{align}
    g_{k}(z)&=\beta_{k}\tanh(\gamma_{k}z+\delta_{k})+\varphi_{k},\,\,k=1,2,3,\\
    f_{j}(z)&=d_{j}+c_{j}\tanh(A_{j}z+b_{j}),\,\,j=0,1,2. \label{eqn_EFNN_fj}
\end{align}

The quasi-particle layers in EFNN form a nested structure that implements a continued function. For instance, the third quasi-particle layer is expressed as:
\begin{align}
     S_3 = g_3\bigg(S_0\bigg) \odot f_2\bigg\{g_2\bigg(S_0\bigg) \odot f_1\bigg[g_1\bigg(S_0\bigg) \odot f_0\bigg(S_0\bigg)\bigg]\bigg\},\label{eqn_EFNN_S3}
\end{align}
where $\odot$ denotes the Hadamard (element-wise) product.

To illustrate why this structure constitutes a continued function, we draw an analogy with Pad\'{e} approximants, which can be expressed as continued fractions---a special case of continued functions.

In physics and applied mathematics, perturbative series naturally express physical interactions to arbitrary order. However, these series have a fundamental dilemma: truncating at low order fails to capture the complete physics, while including higher-order terms typically leads to divergence. Renormalization provides the solution to this problem, with Pad\'{e} approximants being the most widely used numerical scheme for transforming divergent series into convergent expressions. To match a perturbative series $f(x)\sim c_{0}+c_{1}x+c_{2}x^{2}+\cdots$ with a Pad\'{e} approximant $\frac{P_{M}(x)}{Q_{M}(x)}$, one equates truncated $c_{0}+c_{1}x+c_{2}x^{2}+\cdots$ with the rational function $\frac{P_{M}(x)}{Q_{M}(x)}$, then deals with the algebraic equation $(c_{0}+c_{1}x+c_{2}x^{2}+\cdots)Q_{M}(x)=P_{M}(x)$. $M$ is the order of the polynomial. The approximation accuracy typically improves as $M\to\infty$. Examples of continued fraction:

\begin{itemize}
    \item Exponential function:
    \begin{align}
        e^{x}&\sim 1+\frac{1}{1!}x+\frac{1}{2!}x^{2}+\frac{1}{3!}x^{3}+\frac{1}{4!}x^{4}+\cdots= 1 + \cfrac{x}{1 - \cfrac{x}{2 + x - \cfrac{2x}{3 + x - \cfrac{3x}{4 + x - \ddots}}}}\label{eqn_ex}.
    \end{align}
    \item Natural logarithm:
    \begin{align}
        \log(1+x)&\sim x-\frac{1}{2}x^{2}+\frac{1}{3}x^{3}-\frac{1}{4}x^{4}+\cdots= \cfrac{x}{1 + \cfrac{1^2 x}{2 + \cfrac{1^2 x}{3 + \cfrac{2^2 x}{4 + \ddots}}}} \label{eqn_log}.
    \end{align}
\end{itemize}

The Pad\'{e} approximant for $e^{x}$ in Eq.~(\ref{eqn_ex}) has everyday application throughout numerical analysis, since this is how matrix exponential is computed in various programming languages~\cite{expm_Higham_2010}, for example, the \texttt{expm} in Python, Mathematica, and MATLAB. In Eq.~(\ref{eqn_log}), the perturbative series diverges for $x\geq 1$, however, the Pad\'{e} approximant always converges and captures well the value of $\log(1+x)$. The continued fraction can be naturally generalized, by replacing the inverse operation with other nonlinear functions. Examples of continued function~\cite{bender1999advanced}: 

\begin{itemize}
    \item $C_{N}(z)=c_{0}\exp(c_{1}z \exp(c_{2}z \cdots \exp(c_{N}z)))$,
    \item $C_{N}(z)=c_{0}\sqrt{1+c_{1}z\sqrt{1+c_{2}z\sqrt{1+\cdots c_{N-1}z\sqrt{1+c_{N}z}}}}$,
    \item $C_{N}(z)=\log(c_{0}+z\log(c_{1}+z\log(c_{2}+\cdots z\log(c_{N}+z))))$. 
\end{itemize}

To see why EFNN is a continued function, we first turn to a simpler case by requiring $S_{0}\in\mathbb{R}$. We let $f_{j}(z)=d_{j}+\frac{c_{j}}{b_{j}+A_{j}z}$, then $S_{3}$ in Eq.~(\ref{eqn_EFNN_S3}) becomes:

\begin{align}
    S_{3}&=g_3(S_0)d_2 + \cfrac{g_3(S_0)c_2}{b_2 + A_2 g_2(S_0)d_1 + \cfrac{A_2 g_2(S_0)c_1}{b_1 + A_1 g_1(S_0)d_0 + \cfrac{A_1 g_1(S_0)c_0}{b_0 + A_0 S_0}}}. 
\end{align}

Since $S_{0}$ is actually a vector, and rational functions always introduce fictitious poles, we would like to replace the inverse function with tanh, using Eq.~(\ref{eqn_EFNN_fj}). 
 Then the neural network implementation with tanh then gives us:

\begin{align}
    S_{3} &= g_3(S_0) \odot \Bigg[d_2 + c_2 \tanh\Bigg(b_2 + A_2 g_2(S_0) \odot\bigg[d_1 + c_1 \tanh\bigg(b_1 + A_1 g_1(S_0) \odot  \Big[d_0 + c_0 \tanh\Big(b_0 + A_0 S_0\Big)\Big]\bigg)\bigg]\Bigg)\Bigg].\label{eqn_S3_tanh}
\end{align}
And we have fully demonstrated that $S_{3}$ is a continued function.

Note that in Eq.~(\ref{eqn_S3_tanh}), $S_{3}$ is vector-valued, each component of $S_{3}$, which is a finite value, contains infinite number of perturbation terms, as shown in Eq.~(\ref{eqn_S0_all_components_perturb}).

\begin{align}
    \sum_{n_{1},n_{2},\cdots n_{N}=0}^{\infty}a_{n_{1}n_{2}\cdots n_{N}}[S_{0}^{(1)}]^{n_{1}}[S_{0}^{(2)}]^{n_{2}}\cdots[S_{0}^{(N)}]^{n_{N}}. \label{eqn_S0_all_components_perturb}
\end{align}

\noindent\textbf{Why ResNet, DenseNet, and DNN are not continued function}

For DNN, as shown in Fig.~\ref{fig_comp_nets}(c), the mapping $S_{0}\mapsto F_{1}\mapsto S_{1}\mapsto F_{2}\mapsto \cdots \mapsto S_{3}=g_{3}(f_{2}(g_{2}(f_{1}(g_{1}f_{0}(S_{0}))))) $, does not contain any skip connections, as required by continued function. Therefore it is apparent that DNN is not a continued function. 

For ResNet, 
\begin{align}
    F_{j}&=f_{j}(S_{j-1}),\\
    S_{j}&=S_{j-1}\oplus F_{j},j=1,2,3,
\end{align}
$\oplus$ is element-wise addition in Fig.~\ref{fig_comp_nets}(b). Then

\begin{align}
    S_{3}&=S_{0}+f_{0}(S_{0})+ f_{1}\Big[S_{0}+f_{0}(S_{0})\Big]+f_{2}\Bigg\{S_{0}+f_{0}(S_{0})+f_{1}\Big[S_{0}+f_{0}(S_{0})\Big] \Bigg\}. \label{eqn_resnet}
\end{align}
$f_{j}$ is nonlinear.
Clearly Eq.~(\ref{eqn_resnet}) is not a correct continued function. The correct form would be: 
\begin{align}
    S_{3}&=S_{0}+f_{2}[S_{0}+f_{1}(S_{0}+f_{0}(S_{0}))]. \label{eqn_resnet_correct}
\end{align}
There are too many redundant terms in Eq.~(\ref{eqn_resnet}). The expression in Eq.~(\ref{eqn_resnet_correct}) has the structure of Fig.~\ref{fig_comp_nets}(a), with $\odot$ replaced by $\oplus$, and $g_{j}$ set to identity.

For DenseNet, as shown in Fig.~\ref{fig_comp_nets}(d), \circled{$C$} is vector concatenation: for $x\in\mathbb{R}^{a}$, $y\in\mathbb{R}^{b}$, $x\textrm{ \circled{$C$} }y=[x^{\intercal},y^{\intercal}]^{\intercal}$. We take nonlinear functions $f_{j}$, the subsequent layer are:
\begin{align}
    F_{1}&=f_{1}(W_{0}S_{0}+b_{1}),\\
    S_{1}&=\begin{bmatrix}
        S_{0}\\
        F_{1}
    \end{bmatrix},\\
    F_{2}&=f_{2}\bigg([W_{10},W_{11}]\begin{bmatrix}
        S_{0}\\
        F_{1}
    \end{bmatrix}+b_{1}\bigg),\\
    S_{2}&=\begin{bmatrix}
        S_{0}\\
        F_{1}\\
        F_{2}
    \end{bmatrix},\\
    F_{3}&=f_{3}\bigg([W_{20},W_{21},W_{22}]\begin{bmatrix}
        S_{0}\\
        F_{1}\\
        F_{2}
    \end{bmatrix}+b_{2}\bigg),\\
    S_{3}&=\begin{bmatrix}
        S_{0}\\
        F_{1}\\
        F_{2}\\
        F_{3}\label{eqn_densenet_S3}
    \end{bmatrix}. 
\end{align}
The explicit expressions for $F_{1}, F_{2}, F_{3}$ read
\begin{align}
   F_{1}&=f_{0}(W_{0}S_{0}+b_{0}),\\
F_{2}&=f_{1}\bigg[W_{10}S_{0}+W_{11}f_{0}\bigg(W_{0}S_{0}+b_{0}\bigg)+b_{1}\bigg],\\
    F_{3}&=f_{2}\bigg\{W_{20}S_{0}+W_{21}f_{0}\bigg(W_{0}S_{0}+b_{0}\bigg)+W_{22}f_{1}\bigg[W_{10}S_{0}+W_{11}f_{0}\bigg(W_{0}S_{0}+b_{0}\bigg)+b_{1}\bigg]+b_{2} \bigg\}. \label{eqn_densenet}
\end{align}
The expression in Eq.~(\ref{eqn_densenet_S3}) is not a correct continued function. The $S_{3}$ layer concatenates previous layers, a summation of $S_{3}$'s components gives energy, such complicated function cannot be interpreted, and $S_{0}$ is not properly renormalized.

\section{From renormalization group equation to EFNN}
\label{sec_rg}
In the previous section, we have explained that EFNN is a continued function. In this section, we show that the EFNN  structure can be derived from a renormalization group (RG) equation, as all continued functions can be. 

We follow the arguments in Refs.~\cite{YUKALOV1994553_higher_orders,PhysRevE.55.6552_bootstrap,PhysRevE.58.1359_exponential}, to derive the EFNN structure from RG equation. The derivation proceeds in 4 major steps: Step 1, we transform the perturbative series using an algebraic prefactor, and obtain  exponential renormalization through RG equation. Step 2, we apply nested renormalization to create the exponential self-similar approximant with hierarchical structure. Step 3, we replace unbounded functions with bounded tanh functions while preserving asymptotic properties. Step 4, we generalize from scalar to vector inputs using Hadamard products, yielding the final EFNN architecture.

\noindent\textbf{Step 1: using algebraic transform to renormalize perturbative series through RG equation}

We first use scalar $S_{0}\in\mathbb{R}$, target function is $E(S_{0})$, we know its perturbative series up to $k$-th order:
\begin{align}
    p_{0}(S_{0})&=a_{0},\\
    p_{1}(S_{0})&=a_{0}+a_{1}S_{0},\\
    \vdots\\
    p_{k-1}(S_{0})&=a_{0}+a_{1}S_{0}+\cdots +a_{k-1}S_{0}^{k-1},\\
    p_{k}(S_{0})&=a_{0}+a_{1}S_{0}+\cdots +a_{k-1}S_{0}^{k-1}+a_{k}S_{0}^{k}. 
\end{align}
We apply an algebraic transform to the above polynomials:
\begin{align}
    u_{0}&=S_{0}^{\theta}p_{0}(S_{0})=a_{0}S_{0}^{\theta},\\
    u_{1}&=S_{0}^{\theta}p_{1}(S_{0})=a_{0}S_{0}^{\theta}+a_{1}S_{0}^{1+\theta},  \\
    \vdots\\
    u_{k-1}(S_{0})&=S_{0}^{\theta}p_{k-1}(S_{0})=a_{0}S_{0}^{\theta}+a_{1}S_{0}^{1+\theta}+\cdots+a_{k-1}S_{0}^{\theta+k-1} \\
    u_{k}(S_{0})&=S_{0}^{\theta}p_{k}(S_{0})=a_{0}S_{0}^{\theta}+a_{1}S_{0}^{1+\theta}+\cdots+a_{k-1}S_{0}^{\theta+k-1}+a_{k}S_{0}^{\theta+k}. 
\end{align}

We embed the discrete functions $\{u_{j}(S_{0}): j\in\mathbb{N}\}$ into $\{u(t,S_{0}): t\geq 0\}$, with $u_{j}(S_{0})=u(j,S_{0})$, since continuous variable is easier to deal with. $t$ is interpreted as time. We let $u(0,S_{0})=f$, and further denote $u(t,S_{0})$ as $u(t,S_{0},f)$, to show that the initial value of the flow is $f$, while the variable in physical space is $S_{0}$. We assume that $u$ satisfies the self-similar equation:
\begin{align}
    u(t_{2}+t_{1},S_{0},f)&=u(t_{2},S_{0},u(t_{1},S_{0},f)),\label{eqn_u_self_similar}
\end{align}
because we hope to find the fixed point solution $u^{*}$:
\begin{align}
    u(t,S_{0},u^{*}(S_{0}))&=u^{*}(S_{0}),
\end{align}

to guarantee the best convergence when using variable $S_{0}$~\cite{YUKALOV2002839_fractal}, i.e., in the physical space. $u^{*}(S_{0})$ does not depend on $t$, because it is already a fixed point solution. $u^{*}(S_{0})$ does not depend on initial value $f$, because we hope that $u^{*}$ is an attractor.  Eq.~(\ref{eqn_u_self_similar}) leads to the Lie equation:
\begin{align}
    \partial_{t}u(t,S_{0},f)&=v(u(t,S_{0},f)),\label{eqn_Lie}
\end{align}
the velocity is given by
\begin{align}
    v(u(t,S_{0},f))&=\partial_{\epsilon}u(\epsilon,S_{0},u(t,S_{0},f))\bigg|_{\epsilon=0}. 
\end{align}

Lie equation is exactly RG equation, when $t$ is scaled as $t=\log \tau$: 
\begin{align}
    \frac{\partial u}{\partial\log \tau}&=v(u). 
\end{align}

We can't obtain the fixed point solution $u^{*}$ exactly, and need to compute the so called $k$-th order quasi-fixed point solution. We use the Euler scheme to approximate the velocity at $t=k$:
\begin{align}
    v(u(k,S_{0},f))&\simeq u(k,S_{0},f)-u(k-1,S_{0},f)=a_{k}S_{0}^{\theta+k}. 
\end{align}
We assume that starting from $u_{k-1}$, it takes time $T_{k}$ to reach the $k$-th order quasi-fixed point $u_{k}^{*}$, integrating the Lie equation Eq.~(\ref{eqn_Lie}), we obtain:
\begin{align}
    \int_{u_{k-1}}^{u_{k}^{*}}\frac{du}{a_{k}\xi^{\theta+k}}&=T_{k}. 
\end{align}
The dummy variable $\xi$ is in physical space, starting with $S_{0}$ for lower limit $u_{k-1}(S_{0},f)$. We choose the renormalization map from physical space to function space as algebraic:
\begin{align}
    u&=b\xi^{\theta},
\end{align}

then
\begin{align}
    u_{k}^{*}&=(u_{k-1}^{-\frac{k}{\theta}}-\frac{k}{\theta}\frac{a_{k}}{b^{1+\frac{k}{\theta}}}T_{k})^{-\frac{\theta}{k}}.
\end{align}
We multiply the above equation by $S_{0}^{-\theta}$ to scale back, and obtain a renormalized value for $E(S_{0})$:
\begin{align}
    E_{k}^{*}(S_{0})&=S_{0}^{-\theta}u_{k}^{*}\nonumber\\
    {}&=p_{k-1}(S_{0})[1-\frac{k}{\theta}\frac{a_{k}}{b^{1+\frac{k}{\theta}}}T_{k}S_{0}^{k}p_{k-1}^{\frac{k}{\theta}}(S_{0})]^{-\frac{\theta}{k}}. \label{eqn_Ek_star}
\end{align}

Starting from Eq.~(\ref{eqn_Ek_star}), by tuning values of $\theta$, $b$, $T_{k}$, we can arrive at several continued function approximants, including Pad\'{e}. Here, we let $\theta\to\infty$, and obtain

\begin{align}
    E_{k}^{*}(S_{0})&=p_{k-1}(S_{0})\exp(\frac{a_{k}}{b}T_{k}S_{0}^{k}). 
\end{align}

This is to say, this one-step renormalization is a map:
\begin{align}
    p_{k}(S_{0})&=a_{0}+a_{1}S_{0}+\cdots+ a_{k}S_{0}^{k}\mapsto E_{k}^{*}(S_{0})=(a_{0}+a_{1}S_{0}+\cdots a_{k-1}S_{0}^{k-1})\exp(\frac{a_{k}}{b}T_{k}S_{0}^{k}). 
\end{align}

We repeat the renormalization process for the polynomial prefactor until only the affine term is left:
\begin{align}
    p_{k}(S_{0})&\mapsto (a_{0}+a_{1}S_{0})\exp(\frac{a_{2}}{b}T_{2}S_{0}^{2}+\frac{a_{3}}{b}T_{3}S_{0}^{3}+\cdots +\frac{a_{k}}{b}T_{k}S_{0}^{k}). \label{eqn_affine_pk}
\end{align}

\noindent\textbf{Step 2: using nested renormalization to obtain exponential self-similar approximant}

We use the exponential renormalization to renormalize the polynomial in the exponential part in Eq.~(\ref{eqn_affine_pk}), and always leave a linear term as a prefactor, we obtain the exponential self-similar approximant:
\begin{align}
    p_{k}(S_{0})&\mapsto (a_{0}+a_{1}S_{0})\exp(b_{1}S_{0}\exp(b_{2}S_{0}\cdots \exp(b_{n_{k}}S_{0}))).\label{eqn_exp_continued_function}
\end{align}
$n_{k}$ is an integer depending on $k$. One often performs a perturbation analysis for the exponential self-similar approximant with the perturbative series to obtain the unknown coefficients $b_{1}$, $b_{2}$, \ldots, $b_{n_{k}}$. Note that one may also include the linear term $a_{1}S_{0}$ in the exponential, we have the freedom of not taking this approach, and  we have numerically tested that this approach does not work well.

\noindent\textbf{Step 3: further renormalization for exponential self-similar approximant}

Since linear functions and exponential functions are not bounded, to achieve better convergence, we need to further renormalize the nested exponential in  Eq.~(\ref{eqn_exp_continued_function}). Since we do not use order by order matching to determine the unknown coefficients, we can set $n_{k}$ with a given integer. We take $n_{k}=3$ as an example. To further renormalize, we proceed in 5 substeps. We denote the exponential continued function as

\begin{align}
    \Phi_{1}(S_{0})&=(\alpha_{3}S_{0}+\kappa_{3})\exp(\alpha_{2}S_{0}\exp(\alpha_{1}S_{0}\exp(A_{0}S_{0}))).
\end{align}

First, we scale and shift the exp function in $\Phi_{1}(S_{0})$:
\begin{align}
\Phi_{1}(S_{0})&\mapsto \Phi_{2}(S_{0})\nonumber\\
{}&=(\alpha_{3}S_{0}+\kappa_{3})\Bigg[d_{2}^{'}+c_{2}\exp\Bigg(b_{2}+\alpha_{2}S_{0}\bigg[d_{1}^{'}+c_{1}\exp\bigg(b_{1}+\alpha_{1}S_{0}\Big[d_{0}^{'}+c_{0}\exp\Big(A_{0}S_{0}\Big)\Big]\bigg)\bigg]\Bigg)\Bigg].
\end{align}

Second, we shift the origin of $S_{0}$ in $\Phi_{2}(S_{0})$:
\begin{align}
\Phi_{2}(S_{0})&\mapsto \Phi_{3}(S_{0})\nonumber\\
{}&=(\alpha_{3}S_{0}+\theta_{3})\Bigg[d_{2}^{'}+c_{2}\exp\Bigg(b_{2}+(\alpha_{2}S_{0}+\theta_{2})\bigg[d_{1}^{'}+c_{1}\exp\bigg(b_{1}+(\alpha_{1}S_{0}+\theta_{1})\Big[d_{0}^{'}+c_{0}\exp\Big(b_{0}+A_{0}S_{0}\Big)\Big]\bigg)\bigg]\Bigg)\Bigg].
\end{align}

Third, renormalize exp by tanh. We know that $1+\tanh(y)\sim \exp(y)$, when $|y|$ is small, and $\tanh(y)$ is bounded on $\mathbb{R}$. Therefore we replace $\exp(y)$ by $1+\tanh(y)$:
\begin{align}
\Phi_{3}(S_{0})&\mapsto \Phi_{4}(S_{0})\nonumber\\
{}&= (\alpha_{3}S_{0}+\theta_{3})\Bigg[d_{2}+c_{2}\tanh\Bigg(b_{2}+(\alpha_{2}S_{0}+\theta_{2})\bigg[d_{1}+c_{1}\tanh\bigg(b_{1}+(\alpha_{1}S_{0}+\theta_{1})\Big[d_{0}+c_{0}\tanh\Big(b_{0}+A_{0}S_{0}\Big)\Big]\bigg)\bigg]\Bigg)\Bigg].
\end{align}

Fourth, renormalize the linear prefactors before tanh. Since $y\sim \tanh(y)$, when $|y|$ is small, we perform the following renormalizations:
\begin{align}
    \alpha_{j}S_{0}+\theta_{j}&\mapsto g_{j}(S_{0})=\beta_{j}\tanh(\gamma_{j}S_{0}+\delta_{j})+\varphi_{j},\,\,j=1,2,3.
\end{align}
Then
\begin{align}
\Phi_{4}(S_{0})&\mapsto \Phi_{5}(S_{0})\nonumber\\
{}&=g_{3}(S_{0})\Bigg[d_{2}+c_{2}\tanh\Bigg(b_{2}+g_{2}(S_{0})\bigg[d_{1}+c_{1}\tanh\bigg(b_{1}+g_{1}(S_{0})\Big[d_{0}+c_{0}\tanh\Big(b_{0}+A_{0}S_{0}\Big)\Big]\bigg)\bigg]\Bigg)\Bigg]. \label{eqn_Phi5}
\end{align}

Fifth, rescale $g_{1}(S_{0})$, $g_{2}(S_{0})$:
\begin{align}
\Phi_{5}(S_{0})&\mapsto \Phi_{6}(S_{0})\nonumber\\
{}&=g_{3}(S_{0})\Bigg[d_{2}+c_{2}\tanh\Bigg(b_{2}+A_{2}g_{2}(S_{0})\bigg[d_{1}+c_{1}\tanh\bigg(b_{1}+A_{1}g_{1}(S_{0})\Big[d_{0}+c_{0}\tanh\Big(b_{0}+A_{0}S_{0}\Big)\Big]\bigg)\bigg]\Bigg)\Bigg].
\end{align}

\noindent\textbf{Step 4: generalization to vector case}

The above derivations take $S_{0}\in\mathbb{R}$. For a spin system, $S_{0}$ is vector-valued. We generalize $\Phi_{6}(S_{0})$, let $S_{0}\in\mathbb{R}^{N}$, and transform scalar product to Hadamard product $\odot$, and let $b_{k}\in\mathbb{R}^{N}$, $d_{k}\in\mathbb{R}^{N}$, $A_{k}\in\mathbb{R}^{N\times N}$ , $c_{k}\in\mathbb{R}^{N\times N}$, $k=0,1,2$. For $g_{j}(S_{0})$, $j=1,2,3$, we let $\gamma_{j}\in\mathbb{R}^{N\times N}$, $\beta_{j}\in\mathbb{R}^{N\times N}$, $\delta_{j}\in\mathbb{R}^{N}$, $\varphi_{j}\in\mathbb{R}^{N}$. Then
\begin{align}
\Phi_{6}(S_{0})&=g_{3}(S_{0})\odot \Bigg[d_{2}+c_{2}\tanh\Bigg(b_{2}+A_{2}g_{2}(S_{0})\odot \bigg[d_{1}+c_{1}\tanh\bigg(b_{1}+A_{1}g_{1}(S_{0})\odot\Big[d_{0}+c_{0}\tanh\Big(b_{0}+A_{0}S_{0}\Big)\Big]\bigg)\bigg]\Bigg)\Bigg].
\end{align}

The above $\Phi_{6}(S_{0})$ is just the $S_{3}$ in Eq.~(\ref{eqn_S3_tanh}). We need one more step to obtain the energy, i.e., summing up all components of $\Phi_{6}(S_{0})$ to get a real scalar as the predicted energy.

\section{Symmetrization}
\label{sec_symmetrization}
In the main text, the symmetrization layer is given by
\begin{align}
    T_{j}(S_{0})&=\sum_{k=x,y,z}\operatorname{conv}(S_{0,k})\odot \operatorname{conv}(S_{0,k}),\,\,j=0,1,\ldots,n.
\end{align}
We choose to use 2 different matrices as the filters for the 2 convolutional operations in $T_{j}(S_{0})$,  and rewrite $T_{j}(S_{0})$ as
\begin{align}
    T_{j}(S_{0})&=\sum_{k=x,y,z}\operatorname{conv}_{2j}(S_{0,k})\odot \operatorname{conv}_{2j+1}(S_{0,k}).
\end{align}

For $\operatorname{conv}_{q}$, the convolution filter is  $W_{q}\in\mathbb{R}^{C\times R\times S}$, $C$ is the output channel number. The $[a,b,c]$-th element of the convolution result is
\begin{align}
    \operatorname{conv}_{q}(S_{0,k})[a,b,c]&=\sum_{\alpha=0}^{R-1}\sum_{\beta=0}^{S-1}S_{0,k}[b+\alpha,c+\beta]W_{q}[a,\alpha,\beta],
\end{align}
where we assume  a proper padding scheme is chosen in the above equation.  Then 
\begin{align}
    \operatorname{conv}_{2j}[a,b,c]&=\sum_{\alpha=0}^{R-1}\sum_{\beta=0}^{S-1}S_{0,k}[b+\alpha,c+\beta]W_{2j}[a,\alpha,\beta],\\
    \operatorname{conv}_{2j+1}[a,b,c]&=\sum_{\gamma=0}^{R-1}\sum_{\delta=0}^{S-1}S_{0,k}[b+\gamma,c+\delta]W_{2j+1}[a,\gamma,\delta], 
\end{align}

the $[a,b,c]$-th element of the  symmetrization layer reads
\begin{align}
    T_{j}(S_{0})[a,b,c]&=\sum_{k=x,y,z}\operatorname{conv}_{2j}(S_{0,k})[a,b,c]\cdot \operatorname{conv}_{2j+1}(S_{0,k})[a,b,c] \nonumber\\
    {}&=\sum_{\alpha,\beta,\gamma,\delta}W_{2j}[a,\alpha,\beta]W_{2j+1}[a,\gamma,\delta] \sum_{k=x,y,z} S_{0,k}[b+\alpha,c+\beta]S_{0,k}[b+\gamma,c+\delta].
\end{align}

We let
\begin{align}
    s_{1}&=\begin{bmatrix}
        S_{0,x}[b+\alpha,c+\beta]\\
        S_{0,y}[b+\alpha,c+\beta]\\
        S_{0,z}[b+\alpha,c+\beta]
    \end{bmatrix},
\end{align}

\begin{align}
    s_{2}&=\begin{bmatrix}
        S_{0,x}[b+\gamma,c+\delta]\\
        S_{0,y}[b+\gamma,c+\delta]\\
        S_{0,z}[b+\gamma,c+\delta]
    \end{bmatrix}.
\end{align}

Then
\begin{align}
    s_{1}^{\intercal} s_{2}&=\sum_{k=x,y,z}S_{0,k}[b+\alpha,c+\beta]S_{0,k}[b+\gamma,c+\delta].
\end{align}

For a  matrix $U\in\operatorname{O}(3)$,
\begin{align}
    (Us_{1})^{\intercal}Us_{2}&=s_{1}^{\intercal}U^{\intercal}Us_{2} \nonumber\\
    {}&=s_{1}^{\intercal}s_{2}. 
\end{align}
As a result, under any operation $U\in O(3)$, the value $T_{j}[a,b,c]$ remains invariant.

\section{ResNet, DenseNet, DNN adapted for Monte Carlo energy fitting of quantum double exchange model}
\label{sec_adapted_structures}
In this section, we present the structures of ResNet, DenseNet, and DNN, adapted for  Monte Carlo energy fitting  of quantum double exchange model. 
\begin{figure}
    \centering
    \includegraphics[width=0.7\linewidth]{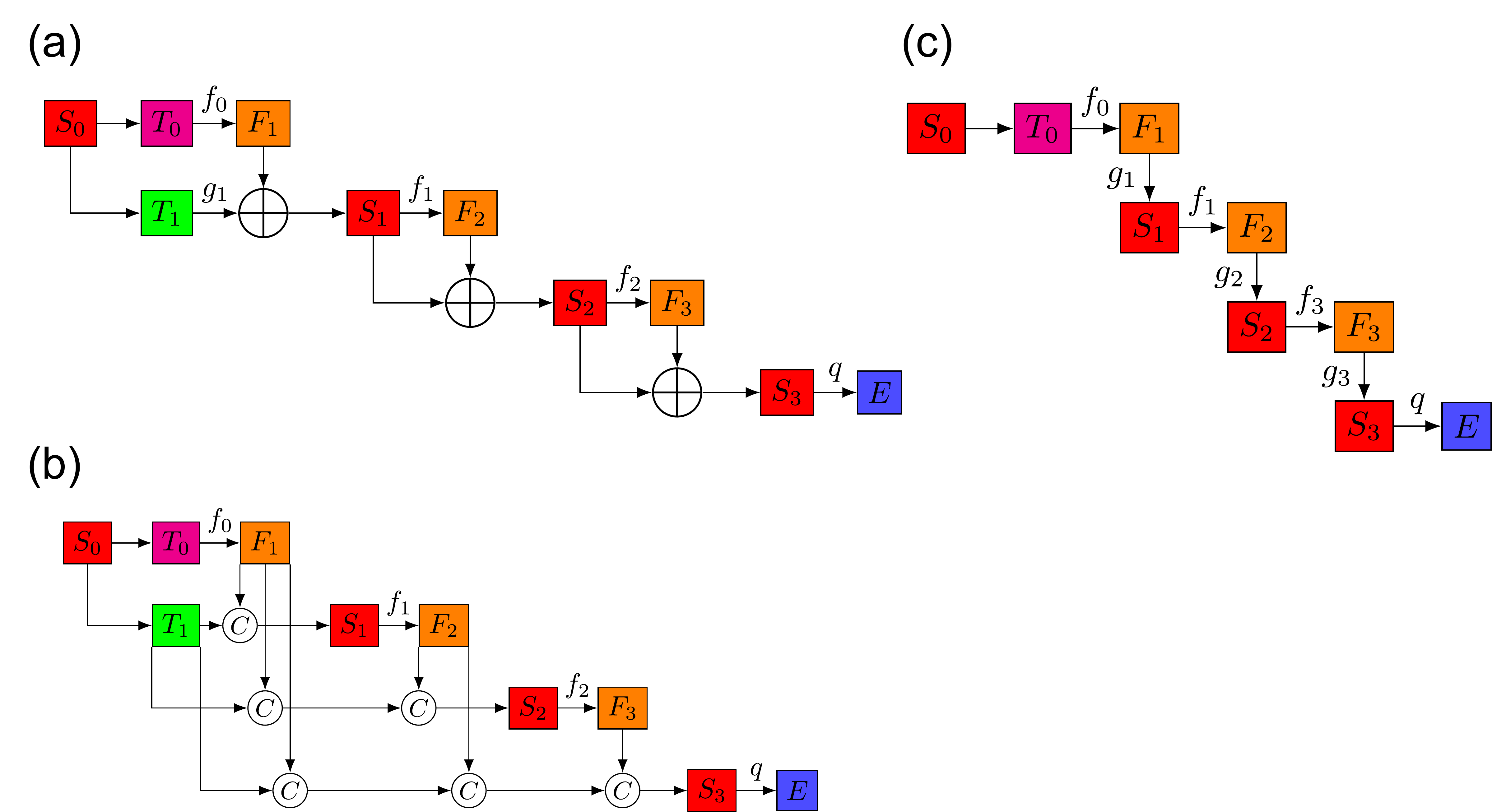}
    \caption{(a) ResNet, (b) DenseNet, and (c) DNN adapted for Monte Carlo energy fitting  of quantum double exchange model. $\oplus$ is element-wise addition, while \protect\circled{$C$} is concatenation. }
    \label{fig_qt_adapted}
\end{figure}

As shown in Fig.~\ref{fig_qt_adapted}(a) and Fig.~\ref{fig_qt_adapted}(b), for ResNet and DenseNet, we add 2 symmetrization layers $T_{0}$ and $T_{1}$ to obtain $O(3)$ invariant results. For DNN in Fig.~\ref{fig_qt_adapted}(c), we use 1 symmetrization layer $T_{0}$, as no skip connection is used for DNN. How the nonlinear layers are computed, how the symmetrization is performed, and how the final energy is computed are similar to Fig.~3 in the main text.

\section{DQMC and the speed-up by EFNN}
\label{sec_DQMC_details}

In this section, we explain the computational bottleneck problem in Determinant Quantum Monte Carlo (DQMC), and report the computational speed-up achieved by EFNN. 

\textbf{Outline of DQMC computation in 2D quantum double exchange model}

For this quantum double exchange model, the DQMC computation has 3 major steps~\cite{bai_dqmc_notes,LOH1992177}.

\textbf{1. Converting the Hamiltonian to quadratic form}

The 2D quantum double exchange model is a lattice system describing electronic hopping interactions and on-site coupling between itinerant electrons and localized classical spins. In this framework, the classical spins act as a random field. The Hamiltonian reads (with chemical potential part)

\begin{align}
    H(S)&=-t\sum_{\braket{i,j},\alpha}(c_{i\alpha}^{\dagger}c_{j\alpha}+c_{j\alpha}^{\dagger}c_{i\alpha})-\frac{J}{2}\sum_{i,\alpha,\beta}\vec{s}_{i}\cdot c_{i\alpha}^{\dagger}\vec{\sigma}_{\alpha\beta}c_{i\beta}-\mu\sum_{i\alpha}n_{i\alpha},
    \label{eqn_H_qt_double_exchange}
\end{align}
$-t\in\mathbb{R}$ is the hopping amplitude, $-\frac{J}{2}$ is the on-site interaction strength, $\vec{s}_{i}\in\mathbb{S}^{2}$ is a classical spin vector. $\sigma^{x}$, $\sigma^{y}$, $\sigma^{z}$ are Pauli matrices. $\alpha,\beta$ denote spin components $\uparrow$ and $\downarrow$.

We first need to expand the site index $i$ to double index, the double index read $[a,b]$, $a,b=0,1,\ldots,N-1$, denoting a lattice point on a square. Now a creation operator reads $c_{ab\uparrow}^{\dagger}$ or $c_{ab\downarrow}^{\dagger}$. 

We group the creation operators into a row vector. The row vector has two parts,
\begin{align}
    c^{\dagger}&=[c_{\uparrow}^{\dagger},c_{\downarrow}^{\dagger}], \label{eqn_c_dagger_combined}
\end{align}

the first part $c_{\uparrow}^{\dagger}$ is a row of creation operators with spin $\uparrow$, the explicit expression reads

\begin{align}
    c_{\uparrow}^{\dagger}&=[c_{00\uparrow}^{\dagger},c_{01\uparrow}^{\dagger}, \ldots,c_{0,N-1,\uparrow}^{\dagger}, c_{10\uparrow}^{\dagger},c_{11\uparrow}^{\dagger}, \ldots,c_{1,N-1,\uparrow}^{\dagger}, \ldots, c_{N-1,0,\uparrow}^{\dagger},c_{N-1,1,\uparrow}^{\dagger},\ldots,c_{N-1,N-1,\uparrow}^{\dagger}].\label{eqn_c_dagger_up}
\end{align}

The creation operators in Eq.~(\ref{eqn_c_dagger_up}) are arranged in row-major order. Similarly, we arrange $\downarrow$ creation operators in row-major order:
\begin{align}
    c_{\downarrow}^{\dagger}&=[c_{00\downarrow}^{\dagger},c_{01\downarrow}^{\dagger},\ldots,c_{0,N-1,\downarrow}^{\dagger},c_{10\downarrow}^{\dagger},c_{11\downarrow}^{\dagger},\ldots, c_{1,N-1,\downarrow}^{\dagger}, \ldots,c_{N-1,0,\downarrow}^{\dagger},c_{N-1,1,\downarrow}^{\dagger},\ldots,c_{N-1,N-1,\downarrow}^{\dagger}]. \label{eqn_c_dagger_down}
\end{align}

The Hamiltonian in Eq.~(\ref{eqn_H_qt_double_exchange}) is expressed in a quadratic form:
\begin{align}
    H(S)&=c^{\dagger}\Gamma(S) c,\label{eqn_H_Gamma_S}
\end{align}

Hermitian matrix $\Gamma(S)\in\mathbb{C}^{2N^{2}\times 2N^{2}}$, 
\begin{align}
    \Gamma(S)&=\begin{bmatrix}
        T_{x}+T_{y} & 0\\
        0& T_{x}+T_{y}
    \end{bmatrix}
    -\frac{J}{2}
    \begin{bmatrix}
        S^{z} & S^{x}-iS^{y}\\
        S^{x}+iS^{y} & -S^{z}
    \end{bmatrix}
    +
    \begin{bmatrix}
        -\mu I &0\\
        0 & -\mu I
    \end{bmatrix}.
\end{align}

$T_{x}$ and $T_{y}$ are hopping matrices in  $x$ and $y$ directions. For $T_{x}\in\mathbb{R}^{N^{2}\times N^{2}}$, the $[rN+s,(r+1)N+s]$-th element is $-t$, the $[(r+1)N+s,rN+s]$-th element is $-t$, the rest are 0. The boundary condition is PBC, the $r$, $s$ indices are understood as $r\mod N$, $s\mod N$.  For hopping matrix $T_{y}\in\mathbb{R}^{N^{2}\times N^{2}}$, the $[rN+s,rN+(s+1)]$-th element and the $[rN+(s+1),rN+s]$-th element are $-t$, the rest are 0.   The boundary condition is PBC, the $r$, $s$ indices are understood as $r\mod N$, $s\mod N$.  

Matrix $S^{x}$ contains all the $s_{ab}^{x}$ components along the diagonal:
\begin{align}
    S^{x}&=\operatorname{diag}(s_{00}^{x},s_{01}^{x},\ldots,s_{0,N-1}^{x},s_{10}^{x},s_{11}^{x},\ldots,s_{1,N-1}^{x},\ldots,s_{N-1,0}^{x},s_{N-1,1}^{x},\ldots,s_{N-1,N-1}^{x}). \label{eqn_S_x_all}
\end{align}

Matrix $S^{y}$ contains all the $s_{ab}^{y}$ components along the diagonal:
\begin{align}
    S^{y}&=\operatorname{diag}(s_{00}^{y},s_{01}^{y},\ldots,s_{0,N-1}^{y},s_{10}^{y},s_{11}^{y},\ldots,s_{1,N-1}^{y},\ldots,s_{N-1,0}^{y},s_{N-1,1}^{y},\ldots,s_{N-1,N-1}^{y}).\label{eqn_S_y_all} 
\end{align}

Matrix $S^{z}$ contains all the $s_{ab}^{z}$ components along the diagonal:
\begin{align}
    S^{z}&=\operatorname{diag}(s_{00}^{z},s_{01}^{z},\ldots,s_{0,N-1}^{z},s_{10}^{z},s_{11}^{z},\ldots,s_{1,N-1}^{z},\ldots,s_{N-1,0}^{z},s_{N-1,1}^{z},\ldots,s_{N-1,N-1}^{z}). 
    \label{eqn_S_z_all}
\end{align}

\textbf{2. Convert tr to det}

In this step, with the quadratic form in Eq.~(\ref{eqn_H_Gamma_S}), the trace is converted to det, following the DQMC scheme:

\begin{align}
    Z(S)&=\sum_{S}\operatorname{tr}e^{-\beta H(S)}\nonumber\\
    {}&=\sum_{S}\operatorname{tr}e^{-\beta c^{\dagger}\Gamma(S)c}\nonumber\\
    {}&=\sum_{S}\det[I+e^{-\beta \Gamma(S)}].\label{eqn_Z_S_qt_double_exchange}
\end{align}
In this step, the trace of a $4^{N^{2}}\times 4^{N^{2}}$ matrix is converted to the determinant of a $2N^{2}\times 2N^{2}$ matrix, this is the quantum-to-classical mapping in DQMC.  

More precisely, the partition function is
\begin{align}
    Z(S)&=\int_{\{\vec{s}_{ab}\in\mathbb{S}^{2}:a,b=0,1,\ldots,N-1\}} \operatorname{tr} e^{-\beta H(S)}dS,\label{eqn_Z_S_integral}
\end{align}

since all the classical continuous spins have norm 1. Eq.~(\ref{eqn_Z_S_qt_double_exchange}) can replace Eq.~(\ref{eqn_Z_S_integral}), only differing by a constant factor when the data points are dense and even enough. 

\textbf{3. Classical Monte Carlo using determinants}

We use the determinants in Eq.~(\ref{eqn_Z_S_qt_double_exchange}) as statistical weight for Monte Carlo sampling. Determinants in Eq.~(\ref{eqn_Z_S_qt_double_exchange}) have the nice property that they are all strictly positive, because they can be written as 
\begin{align}
    \det[I+ e^{-\beta \Gamma (S)}]&=\prod_{j=0}^{2N^{2}-1}[1+e^{-\beta E_{j}(S)}]>0,\label{eqn_tr_2_det_qt_double_exchange}
\end{align}

$\{E_{j}(S)\}$ are the eigenvalues of Hermitian matrix $\Gamma (S)$. In each Monte Carlo step, we perform update on \textbf{one} component of \textbf{one} classical spin vector, among $N^{2}$ such 3D vectors in $\mathbb{S}^{2}$ . Evaluating Eq.~(\ref{eqn_tr_2_det_qt_double_exchange}) requires Exact Diagonalization (ED), for such $2N^{2}\times 2N^{2}$ matrix, one evaluation requires complexity $O((2N^{2})^{3})=O(8N^{6})$. Note that evaluating the det in left hand side of Eq.~(\ref{eqn_tr_2_det_qt_double_exchange}) has exactly the same complexity. There are $3N^{2}$ scalar components in $S$ in total, a sweep to update every component has complexity $3N^{2}\cdot O(8N^{6})\sim O(24N^{8})$. In order to obtain convergent results, we typically need 500 \textbf{effective samples}, then the total computational cost is
\begin{align}
    500\times (\textrm{autocorrelation length})\times (\textrm{sweep number})\times O(24N^{8}).\label{eqn_cost_qt_double_excchange}
\end{align}

As $N$ increases, with computational complexity in Eq.~(\ref{eqn_cost_qt_double_excchange}), the computation becomes prohibitive for classical Monte Carlo very quickly. We use EFNN to address the computational bottleneck problem by training EFNN using exact value of Eq.~(\ref{eqn_tr_2_det_qt_double_exchange}), given $S$. We make one more transformation, and define a Monte Carlo energy $E_{\operatorname{MC}}(S)$,
\begin{align}
    e^{-\beta E_{\operatorname{MC}}(S)}&= \det[I+ e^{-\beta \Gamma (S)}]\nonumber\\
    {}&=\prod_{j=0}^{2N^{2}-1}[1+e^{-\beta E_{j}(S)}],
\end{align}
then 
\begin{align}
    E_{\operatorname{MC}}(S)&=-T\sum_{j=0}^{2N^{2}-1}\log[1+e^{-\frac{1}{T}E_{j}(S)}]. 
\end{align}

$ E_{\operatorname{MC}}(S)$ is formally a free energy. We use ED (Exact Diagonalization) to generate $[S,E_{\operatorname{MC}}(S)]$ data as training set to EFNN. ED has the same complexity as determinant computation. 

\begin{table}[hbtp!]
\centering
\caption{Average computational time for EFNN inference, $C=3$, 3 layers, and ED, quantum double exchange model} % The Title
\label{tab_efnn_ed_time} % The Label for referencing
\begin{tabular}{|l|l|l|l|l|l|l|l|}
\hline
$N$         & 10                   & 15                   & 20                   & 25                   & 30                   & 35                   & 40                   \\ \hline
EFNN time/s & $4.56\times 10^{-3}$ & $1.6\times 10^{-2}$  & $1.62\times 10^{-2}$ & $1.40\times 10^{-2}$ & $7.08\times 10^{-3}$ & $9.38\times 10^{-3}$ & $7.59\times 10^{-3}$ \\ \hline
ED time/s   & $6.24\times 10^{-3}$ & $3.11\times 10^{-2}$ & $1.17\times 10^{-1}$ & $4.82\times 10^{-1}$ & $1.46$               & $3.56$               & $7.76$               \\ \hline
\end{tabular}
\end{table}

\begin{figure}[hbtp!]
    \centering
    \includegraphics[width=0.5\linewidth]{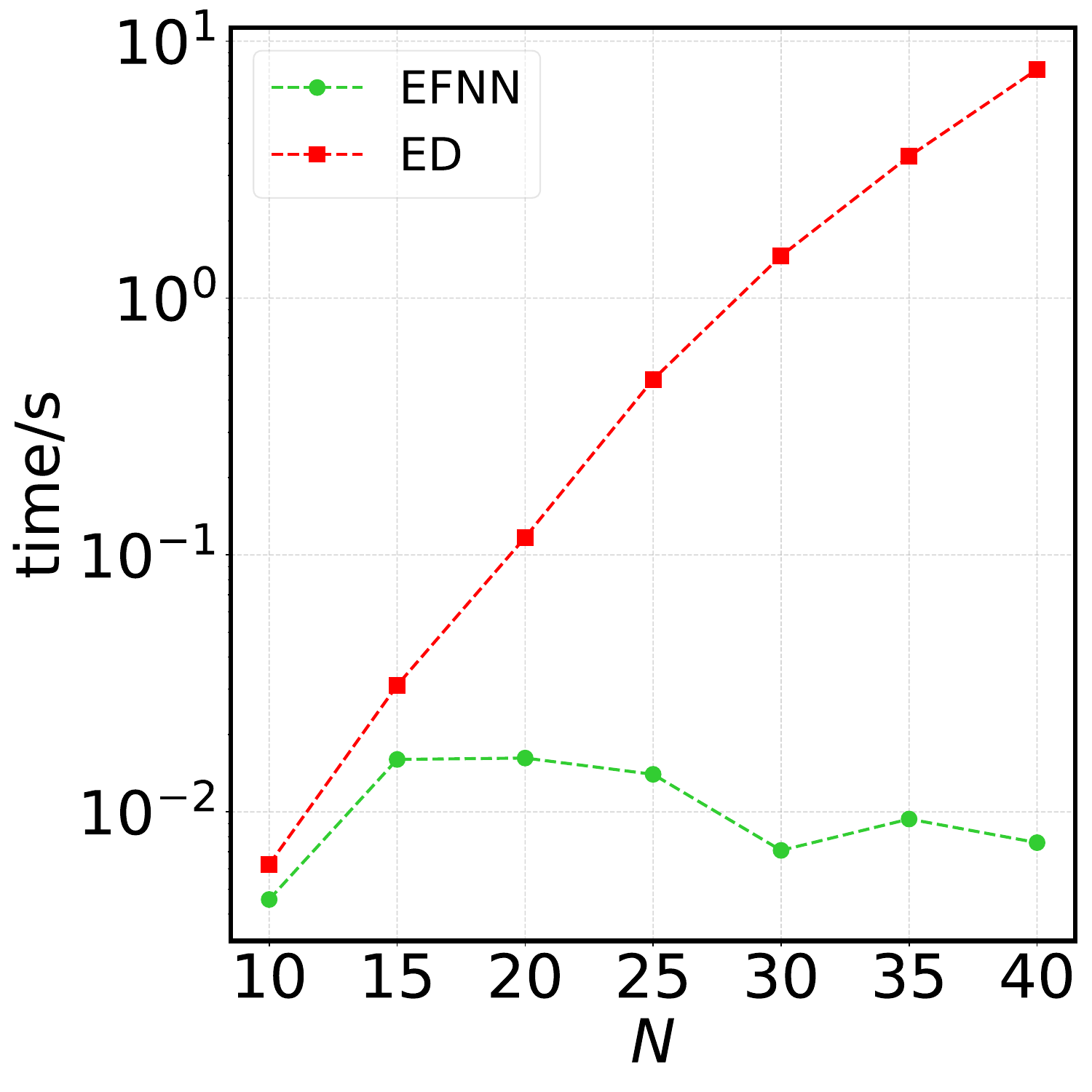}
    \caption{Average computational time for EFNN inference, $C=3$, 3 layers, and ED. The y-axis has log scale. }
    \label{fig_efnn_ed_time}
\end{figure}

We report the average computation times  of EFNN and ED in Table.~\ref{tab_efnn_ed_time}, and make a plot in Fig.~\ref{fig_efnn_ed_time}, for comparison. While the EFNN model was trained on a supercomputer, this benchmark compares the inference time for a single $E_{\textrm{MC}}(S)$ value given a configuration $S$ against the corresponding ED calculation, both performed on an Alienware m15 R12 laptop. As shown in
Table.~\ref{tab_efnn_ed_time} and the green line in Fig.~\ref{fig_efnn_ed_time}, the EFNN inference time remains approximately constant at order $10^{-2}$ s when the lattice size $N\times N$ grows from $10\times 10$ to $40\times 40$. Conversely, as shown by the red line, the ED computational time scales polynomially with $N$, rapidly reaching approximately 10 s for $N=40$. This scaling behavior makes ED computationally prohibitive for Monte Carlo sampling. Notably, for $N=40$, the EFNN achieves a speedup of approximately $10^{3}$.

\section{Details of inference error and data distribution, quantum double exchange model}
\label{sec_data_distribution}

In this section, we report the error of EFNN prediction, plot the distributions of the test  data (ground truth) and the EFNN predictions, for the quantum double exchange model.

We report the absolute value of mean of $E_{\operatorname{MC}}$ for test set, RMSE of prediction by EFNN, and relative error in Table.~\ref{tab_performance_metrics_EFNN}, with $C=3$, layer number = 3, in quantum double exchange model. The relative error is defined as 
\begin{align}
    \frac{\textrm{RMSE}}{|\textrm{mean of } E_{\operatorname{MC}}|}.
\end{align}

As shown in Table.~\ref{tab_performance_metrics_EFNN}, even though the model is trained on a $10\times 10$ lattice under boundary condition PBC, when applied to larger lattices, the RMSE  increases only marginally. Moreover, the $|\textrm{mean of }E_{\operatorname{MC}}|$ increases very quickly with respect to $N$. As a result, the difference of the prediction value by EFNN and the ground truth value of $E_{\operatorname{MC}}(S)$, becomes smaller and smaller perturbations as $N$ increases, demonstrating the accuracy of EFNN.

\begin{figure}[hbtp!]
    \centering
    \includegraphics[width=1\linewidth]{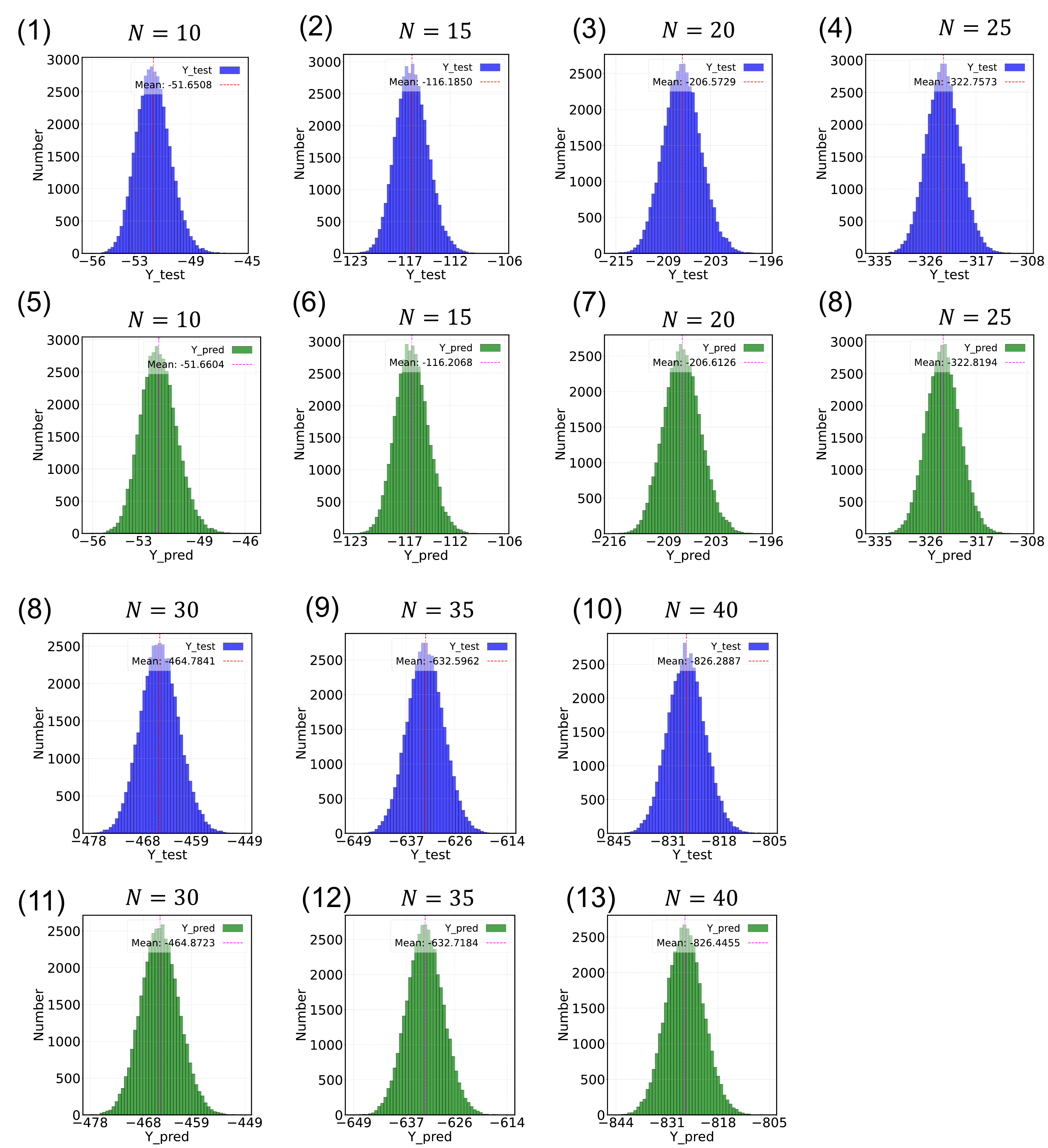}
    \caption{EFNN's prediction results of quantum double exchange model, with $C=3$ and 3 layers:  histogram of Y\_test and Y\_pred, for $N=10,15,20,25,30,35,40$. }
    \label{fig_all_Y_test_Y_pred_efnn}
\end{figure}

We present the histograms of test set data (Y\_test, ground truth value of $E_{\operatorname{MC}}(S)$) and prediction results computed by EFNN (Y\_pred, predicted value of $E_{\operatorname{MC}}(S)$) for the quantum double exchange model in Fig.~\ref{fig_all_Y_test_Y_pred_efnn}, for $N=10,15,20,25,30,35,40$. The blue histograms are for ground truth data Y\_test, and green histograms are for predictions Y\_pred. Note that as lattice size increases, the model trained on $10\times 10$ lattice still correctly predicts $E_{\operatorname{MC}}(S)$ for larger lattices, up to $N=40$, as shown by the overall histogram shapes, and the mean values of Y\_test and Y\_pred. Note that as $N$ increases, the absolute value of mean Y\_test increases.

\begin{table}[hbtp!]
\centering
\caption{Performance metrics for EFNN, $C=3$, 3 layers, quantum double exchange model} % The Title
\label{tab_performance_metrics_EFNN} % The Label for referencing
\begin{tabular}{|l|l|l|l|l|l|l|l|}
\hline
$N$               & 10                   & 15                   & 20                   & 25                   & 30                   & 35                   & 40                   \\ \hline
$|\textrm{mean of $E_{\operatorname{MC}}$}|$ & 51.6508              & 116.1850             & 206.5729             & 322.7573             & 464.7841             & 632.5962             & 826.2887             \\ \hline
RMSE         & 0.0584               & 0.0989               & 0.1340               & 0.1714               & 0.2118               & 0.2549               & 0.3003               \\ \hline
relative error    & $1.13\times 10^{-3}$ & $8.51\times 10^{-4}$ & $6.49\times 10^{-4}$ & $5.31\times 10^{-4}$ & $4.56\times 10^{-4}$ & $4.03\times 10^{-4}$ & $3.63\times 10^{-4}$ \\ \hline
\end{tabular}
\end{table}

\newpage
\bibliography{reference}